\documentclass[reprint,superscriptaddress,preprintnumbers,amsmath,amssymb,aps,prd,tightenlines,longbibliography,nofootinbib,balancelastpage]{revtex4-2}

\usepackage{graphicx}
\usepackage[dvipsnames]{xcolor}
\usepackage[sort&compress]{natbib}
\usepackage{amsmath,amssymb,bm,bbm,slashed,subdepth}
\usepackage{xr-hyper}
\usepackage[colorlinks=true
,urlcolor=blue
,anchorcolor=blue
,citecolor=blue
,filecolor=blue
,linkcolor=red
,menucolor=blue
,linktocpage=true
,pdfproducer=medialab
,pdfa=true
]{hyperref}
\usepackage{cleveref}
\usepackage{enumerate}
\usepackage{epsfig, subfigure}
\usepackage{setspace}
\usepackage{booktabs, tabularx}
\usepackage{units}
\usepackage{placeins}
\usepackage{multirow}
\usepackage{mathtools}
\usepackage[normalem]{ulem}
\usepackage[page,toc,titletoc,title]{appendix}
\usepackage{float}

\begin{document}

\title{On the Energy Distribution of the Galactic Center Excess' Sources}

\author{Florian List}
\affiliation{Department of Astrophysics, University of Vienna, Türkenschanzstraße 17, 1180 Vienna, Austria}

\author{Yujin~Park}
\affiliation{Theory Group, Lawrence Berkeley National Laboratory, Berkeley, CA 94720, USA}
\affiliation{Berkeley Center for Theoretical Physics, University of California, Berkeley, CA 94720, USA}

\author{Nicholas~L.~Rodd}
\affiliation{Theory Group, Lawrence Berkeley National Laboratory, Berkeley, CA 94720, USA}
\affiliation{Berkeley Center for Theoretical Physics, University of California, Berkeley, CA 94720, USA}

\author{Eve~Schoen}
\affiliation{Theory Group, Lawrence Berkeley National Laboratory, Berkeley, CA 94720, USA}
\affiliation{Berkeley Center for Theoretical Physics, University of California, Berkeley, CA 94720, USA}

\author{Florian Wolf}
\affiliation{Department of Astrophysics, University of Vienna, Türkenschanzstraße 17, 1180 Vienna, Austria}

\begin{abstract}
The Galactic Center Excess (GCE) may yet herald the discovery of annihilating dark matter.
Weighing against that conclusion are analyses showing evidence for dim point sources within the spatial structure of the emission.
Due to technical limitations these analyses are purely spatial with all spectral information that could disentangle the excess from astrophysical backgrounds discarded.
Here, we demonstrate that a neural network simulation-based inference approach can jointly analyze the spatial and spectra data.
The addition is profound: energy information drives the putative point sources to be significantly dimmer, indicating either the GCE is truly diffuse in nature or made of an exceptionally large number of sources.
Quantitatively, for our best fit background model, the excess is essentially consistent with Poisson emission as predicted by dark matter.
If due to point sources, our median prediction is ${\cal O}(10^5)$ sources, or more than 35,000 at 90\% confidence---both orders of magnitude larger than the hundreds preferred by earlier point-source analyses of the GCE, although variations allowed by background systematics could reduce the required number of sources by roughly an order of magnitude.
\end{abstract}

\maketitle

\begin{figure*}[!t]
\centering
\includegraphics[width=0.45\textwidth]{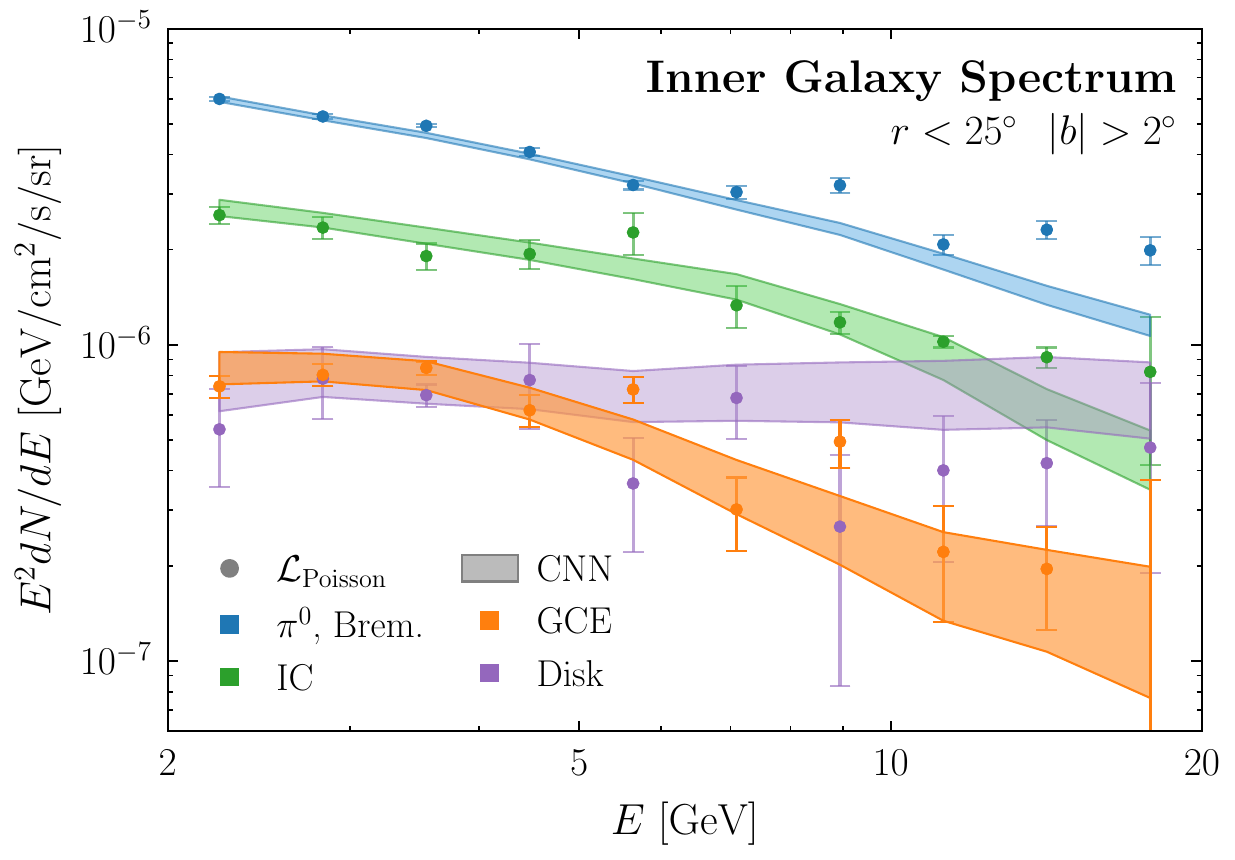}
\hspace{0.5cm}
\includegraphics[width=0.45\textwidth]{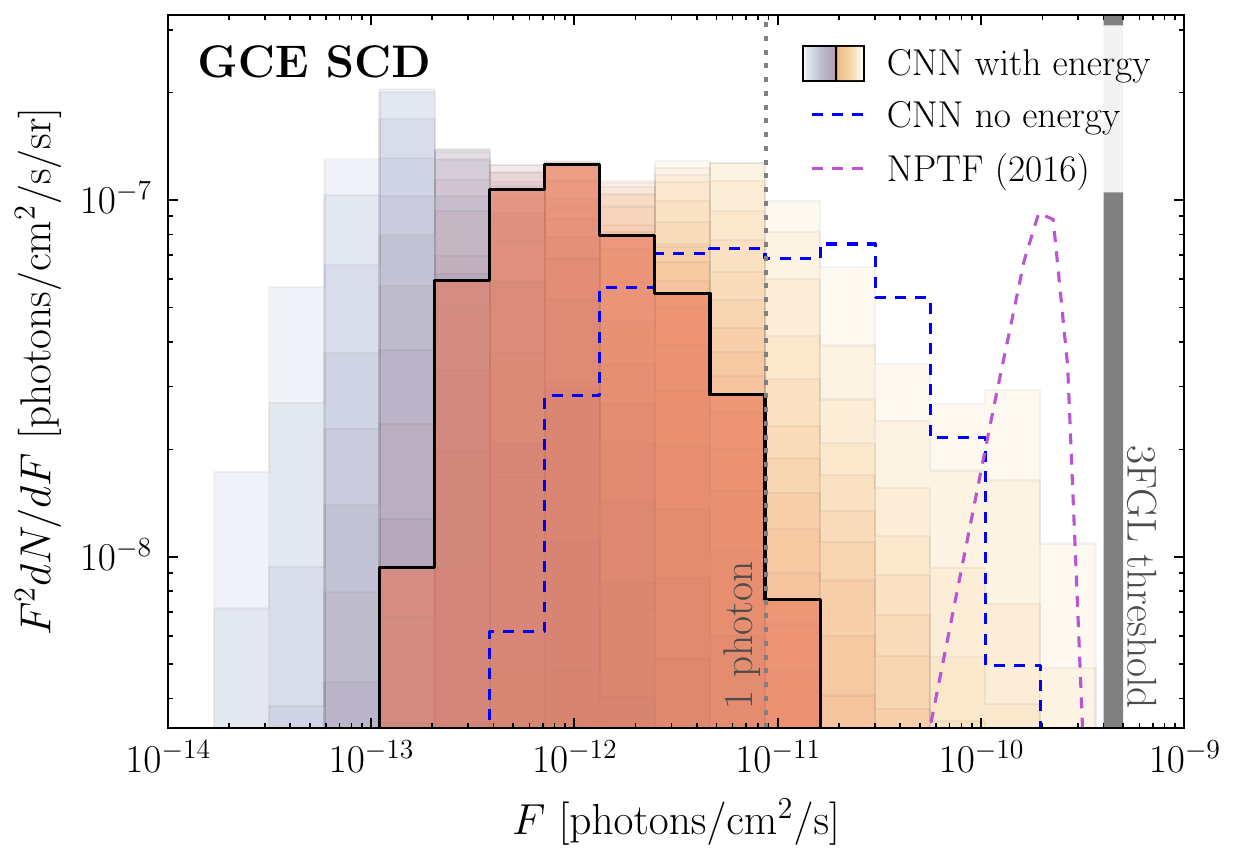}
\vspace{-0.2cm}
\caption{(Left) The spectrum and $1\sigma$ errors for different contributions to the $\gamma$-ray flux in the Inner Galaxy as determined by a conventional Poisson template fit (${\cal L}_{\rm Poisson}$, markers and error bars) and the convolutional neural network (CNN, bands) developed in this work.
The CNN explicitly accounts for the presence of point sources in its determination of the GCE and disk fluxes; the Poisson likelihood does not.
(An isotropic and Fermi bubbles template are also included in the fit, with spectra shown in the SM.)
(Right) The source-count distribution (SCD) extracted for the GCE via three different approaches.
The NPTF (dashed purple) shows the result of likelihood based approaches in 2016 which suggested that the excess was made of point sources just below the Fermi 3FGL detection threshold~\cite{Mishra-Sharma:2016gis}.
An energy independent CNN (dashed blue) deployed on the latest Fermi data and background models prefers a dimmer SCD, peaked at fluxes below where sources produce 1 photon on average.
The addition of energy drives the SCD considerably dimmer (median black, quantiles shown from purple to orange) to a level where it is almost indistinguishable from Poisson emission: at 95\% confidence only 3\% of the emission can be excluded as being consistent with Poisson.
This divergence between methods is not observed for the disk SCD, cf. Fig.~\ref{fig:SCD-Disk}.
}
\vspace{-0.5cm}
\label{fig:Results}
\end{figure*}

What is the experiment required to discover dark matter (DM)?
In many scenarios where DM is an ${\cal O}($GeV-TeV$)$ mass thermal WIMP the answer to this question is the Fermi Large Area Telescope.
Such models predict that DM should annihilate today with a similar strength to which its relic abundance was set in the early Universe and thereby generate an observable and largely spherically symmetric emission of $\gamma$-rays peaking at the Galactic Center.
This insight was one of the core motivations for the construction and launch of the Fermi telescope~\cite{Baltz:2008wd,Fermi-LAT:2009ihh}.
Remarkably, Fermi observed exactly such emission shortly after its launch~\cite{Goodenough:2009gk,Hooper:2010mq}.

The origin of this Galactic Center Excess (GCE) has been debated since its discovery; see Refs.~\cite{Murgia:2020dzu,Leane:2022bfm} for a review.
Whilst a compelling case can be made in favor of DM~\cite{Hooper:2011ti,Hooper:2013nhl,Daylan:2014rsa,Zhou:2014lva,Calore:2014xka}, it has not been possible to exclude an astrophysical origin for the emission.
In particular, it was recognized early on that the energy spectrum of the GCE was well described not only by DM annihilation but also millisecond pulsars~\cite{Hooper:2010mq,Abazajian:2012pn,Hooper:2011ti}.
If the excess originates from a large population of astrophysical point sources it might be expected that the excess would correlate more with the plane of the galaxy rather than being spherically symmetric; this question has been intensely studied although a clear consensus has not yet emerged~\cite{2016PhRvL.116e1102B, Macias:2016nev,Macias:2019omb,2021PhRvD.103f3029D,Song:2024iup,Bartels:2017vsx,Cholis:2021rpp,McDermott:2022zmq,Zhong:2024vyi,Ramirez:2024oiw}.

An alternative path to resolving the nature of the GCE is to exploit the defining feature of a point source: the generation of more than one photon from a single location.
The pixelized photon count map predicted by DM annihilation follows the Poisson distribution.
Although controlled by the underlying and uncertain Inner Galaxy DM distribution (see e.g. Ref.~\cite{Hussein:2025xwm}), a detector observes a relatively smooth spatial distribution punctuated by Poisson fluctuations.
Point sources, with their ability to produce multiple photons from the same position, inject further pixel-to-pixel variation into the map.
Even for a population of sources that are each too dim to detect, collectively they can induce an observable modification to the photon map away from the Poisson prediction.
This intuition can be encoded in an analytic likelihood to distinguish the scenarios -- as done for the 1pPDF~(1-point probability distribution function, \cite{Malyshev:2011zi,Zechlin:2015wdz,Zechlin:2016pme,Zechlin:2017uzo,Manconi:2019ynl,Calore:2021jvg}), NPTF~(Non-Poissonian Template Fitting, \cite{Lee:2014mza,Lee:2015fea,Mishra-Sharma:2016gis}), and CPG~(Compound Poisson Generator, \cite{Collin:2021ufc}) -- and underpinned the NPTF based claim in \textcite{Lee:2015fea} that the GCE had a point source origin.
The robustness of these methods has been debated~\cite{Leane:2019xiy,Chang:2019ars,Buschmann:2020adf,Leane:2020nmi,Leane:2020pfc}.
What is indisputable is that computational tractability mandates the likelihood be evaluated with two significant approximations: 1. pixel-to-pixel correlations are neglected; and 2. photon energies are discarded.
Machine learning approaches have been successfully leveraged to move beyond the first approximation~\cite{Caron:2017udl,List:2020mzd,List:2021aer,Mishra-Sharma:2021oxe,Amerio:2023uet,Christy:2024gsl,Eckner:2025waa}.
In particular, a likelihood-free analysis based on convolutional neural networks (CNNs~\cite{List:2021aer}) concluded that whilst the GCE may have a point-source origin, the sources would need to be considerably dimmer than concluded in the earlier likelihood studies~\cite{Lee:2015fea, Mishra-Sharma:2016gis}.

In the present Letter we build on the CNN methodology to demonstrate that it can further remove the second assumption and provide the first energy-dependent analysis of the point-source nature of the GCE.
Conceptually, this is a significant step forward in studies of the excess.
It is known that the GCE has a distinct spectrum to the dominant backgrounds, so energy information provides a powerful lever arm through which methods could disentangle the nature of the excess from background mismodeling.
We find that this lever arm plays an important role in practice: as shown in Fig.~\ref{fig:Results}, our method returns a similar spectrum to the excess as earlier studies, but a far dimmer population of sources, the median prediction of which is consistent with purely Poisson emission as expected for a DM origin of the Fermi excess.

The remainder of the article outlines the CNN based framework we develop to achieve an energy-dependent analysis of the GCE.
We further discuss the robustness of our findings, especially to the ever-present systematic uncertainty associated with background mismodeling: we must worry that our best models of the $\gamma$-ray sky have significant imperfections.
Beyond this, a far broader study of systematic uncertainties can be found in the Supplemental Material (SM) where we mention Refs.~\cite{Ballet:2023qzs,Malyshev:2024obk,stossi2024space,zhao2020individual,papamakarios2019sequential,anau2024scalable,Steinwart2011,Gautam2021,Ploeg2020,Fermi-LAT:2012edv,Ulyanov2016,Ioffe2015,Linden:2016rcf,1965TrAlm...5...87E,Burkert:1995yz}.

\vspace{0.2cm}
\noindent {\bf Exploiting Energy with a CNN.}
%
We begin with an outline of the CNN approach and a review of the machine learning pipeline of Refs.~\cite{List:2020mzd,List:2021aer}, which form the foundation of the energy extension developed in this Letter.

In Ref.~\cite{List:2020mzd}, CNNs were established as a powerful template fitting method for $\gamma$-ray photon-count maps.
Specifically, a CNN was trained on simulated Fermi mock maps to predict what fraction of the total flux each component generated.
The simulated maps were determined using spatial templates for the expected emission components in the Inner Galaxy, including astrophysical diffuse emission and a model for the GCE.
In determining the flux and associated uncertainties for each, the CNN learned template fitting.
Compared to conventional template fitting, however, the CNN could model the flux as arising from point sources or Poisson emission, not simply the latter.
This flexibility was exploited to model the GCE as a linear combination of both.
While the CNN generally recovered the flux fractions at percent level accuracy in simulated maps, the results on the Fermi data regarding the nature of the GCE were inconclusive. 
Apart from the modeling uncertainties -- most crucially of the Galactic foreground emission -- the fundamental degeneracy between faint point sources and Poissonian emission proved a major obstacle: with decreasing brightness of the individual GCE sources, the CNN's confusion between the point source and Poisson hypotheses increased until the method could no longer make any distinction between the two.
The confusion is physical.
Although not fully appreciated in Ref.~\cite{List:2020mzd} or more broadly in the early GCE literature, point source versus Poisson emission is a nested hypothesis: an infinite number of infinitely dim sources is exactly Poisson emission (see e.g. Refs.~\cite{Collin:2021ufc,List:2021aer}).
To make progress, the CNN needs to be asked a more physical question.

This shortcoming was addressed in Ref.~\cite{List:2021aer} with the use of an extended CNN-based framework that inferred not only template flux fractions but also the brightness distribution for the point-source emission components.
This was achieved by using two distinct CNNs.
The first estimated the template fluxes; however, with the important improvement over Ref.~\cite{List:2020mzd} that the entire GCE emission was modeled as originating from point sources, and the CNN was tasked with estimating a \emph{single} GCE flux fraction, rather than two (Poisson and point source).
This approach explicitly avoids the ambiguity emphasized above; Poisson emission can be reconstructed as dim point sources, so long as one ensures that the training data contains maps with a GCE faint enough to be indistinguishable from ``true'' Poissonian emission as judged by the CNN.
Point source emission can be distinguished from Poisson only through sources that generate more than one detected photon.
The final output of the first CNN was a prediction for the total flux fraction in the Inner Galaxy for each emission component.

The second CNN in Ref.~\cite{List:2021aer} was used to infer the underlying brightness distribution of the sources for the GCE and disk templates.
In particular, the CNN learned the source-count distribution (SCD) for each component, which is commonly stated as a flux distribution for the sources, $dN/dF$ (cf. Fig.~\ref{fig:Results}).
For the GCE, the method found an SCD peaked at a flux corresponding to only 3-4 photons per source, from which it was further deduced that at least a third of the GCE flux was preferred to have a point-source origin at 95\% confidence.
(In Ref.~\cite{Mishra-Sharma:2021oxe} a different machine learning approach derived a comparable result.)
We note that Ref.~\cite{List:2021aer} demonstrated that the CNN was robust against systematics such as misreconstructing a north-south asymmetry as evidence for point sources~\cite{Leane:2020nmi,Leane:2020pfc}.

In this work, we closely follow Ref.~\cite{List:2021aer} in regard to the template fitting and the estimation of the SCDs.
The key extension is that the CNN is now trained on mock Fermi maps with ten logarithmically spaced bins from 2-20 GeV as opposed to a single energy bin over that range as used previously.
(A preliminary version of this extension was shown in Ref.~\cite{Wolf:2024oqb}.)
Exploiting spectral information allows us to improve on existing approaches in two crucial aspects.
\begin{enumerate}
\item \emph{Faithfulness of the modeling:} the CNN can capture that distinct emission components are predicted to have distinct spectra; moreover, we account for the energy dependence of the Fermi instrument response, especially its point-spread function (PSF);
\item \emph{More expressive results:} we simultaneously reconstruct the spectra of all templates and the SCD for point-source models.
\end{enumerate}

\vspace{0.2cm}
\noindent {\bf CNN Architecture.}
%
Although our CNN architecture is similar to Ref.~\cite{List:2021aer}, let us briefly describe it here.
For the recovery of the spectra and SCD we use two independent networks, whose structure is comparable, apart from their final layers and the loss functions used for their training.
We use the \texttt{DeepSphere} framework~\cite{Perraudin2019a, Defferrard2020}, which provides a spherical counterpart of standard CNN operations.
\texttt{DeepSphere} leverages the \texttt{HEALPix} tessellation of the sphere~\cite{Gorski2005} to build a graph where each pixel center represents a vertex and neighboring pixels are connected by an edge.
The inputs of our CNNs are given by photon-count maps of shape $n_{\mathrm{pix}} \times n_{\mathrm{bins}}$, where $n_{\mathrm{pix}}$ denotes the number of pixels in our region and $n_{\mathrm{bins}}$ represents the energy bins that are treated as channels, analogous to color channels in standard CNNs.

A series of (graph-)convolutional blocks -- each consisting of a convolution, maximum pooling, a ReLU activation function, and batch normalization -- is then used to gradually transform the localized information provided by the input map into a more abstract representation.
Each layer reduces the spatial resolution while increasing the number of feature channels, thereby capturing progressively higher-level patterns.
Notably, after the first convolution, all channels already integrate information across all energy bins.
This occurs because each convolution operates across all input channels and spatial neighbors simultaneously, applying a learned filter that mixes information along the spatial and the energy axes right from the start.
The convolutional blocks are followed by fully-connected blocks, which interpret the extracted information and produce predictions for either $dN/dE$ or $dN/dF$s, depending on the network.

For the spectra, we train the CNN to estimate the fractional contribution of each template to the flux in each energy bin.
As a loss function, we use the negative log of a Gaussian distribution with a diagonal covariance matrix, which enables uncertainty estimation on the spectra.
To enable full non-parametric posterior inference of the SCDs, we use a simulation-based inference approach rooted in quantile regression.
The network is trained to predict binned histograms of the cumulative $F^2 dN/dF$ using the ``Earth mover's pinball loss''~\cite{list2021earth}, which allows us to estimate arbitrary quantile levels $\tau \in (0, 1)$ consistent with the data.
During training, the quantile level is randomly sampled, and at evaluation time, it can be systematically scanned to reconstruct the full posterior across all flux bins.
We infer the distribution of $F^2 dN/dF \propto F dN/d\!\log_{10}\!F$ as the histogram is defined with logarithmic flux bins, and this result captures the amount of flux generated by point sources in each bin.

\vspace{0.2cm}
\noindent {\bf Data and Modeling.}
%
We use 812 weeks of Fermi data as collected from 4 August 2008 to 23 February 2024, selecting the p8R3\_ULTRACLEANVETO events (evclass=1024) rated as having the best angular reconstruction (evtype=32).
The resulting list of photons is divided into ten logarithmically spaced energy bins from 2 to 20 GeV and further subdivided spatially using a \texttt{HEALPix} grid with \texttt{nside=256}, resulting in pixels of width $\simeq$$0.2^{\circ}$.
We then study the Inner Galaxy, defined as the region within $25^{\circ}$ of the Galactic Center, and mask locations within $2^{\circ}$ of the Galactic plane or the energy-dependent 95\% containment radius of sources in the 3FGL catalog~\cite{Fermi-LAT:2015bhf}.
(Applying the same procedure to the newer 4FGL catalog~\cite{Fermi-LAT:2022byn} would result in an unacceptably large fraction of the region being masked.)

The spatial templates used to decompose the region are divided into four Poisson-like and two point-source maps, with the latter flexible enough to capture Poisson emission. 
We use energy-dependent versions of the templates employed in Ref.~\cite{List:2021aer}.
For the Poisson-like, two templates describe the Galactic diffuse emission, divided into a combined model for $\pi^0$ plus bremsstrahlung flux and separately inverse Compton (IC), both determined using ``Model O''~\cite{Buschmann:2020adf,Macias:2016nev,Macias:2019omb}.
The final two Poisson templates model the isotropic emission and the Fermi bubbles~\cite{Su:2010qj,Fermi-LAT:2014sfa}.
All Poisson templates are spatially smoothed with the Fermi PSF at the relevant energy.
The point-source components are modeled by a generalized NFW profile~\cite{Navarro:1996gj} with $\gamma=1.2$ for the GCE, as well as a template to capture point sources below the 3FGL threshold in the Milky Way disk treated with a double exponential with scale radius $R_s = 5\,$kpc and height $z_s = 0.3\,$kpc.
The impact of varying these templates is studied in the SM.
(We emphasize that although the isotropic and Fermi bubbles templates do not have spectra depicted in Fig.~\ref{fig:Results}, they are included in the fit and results are shown in the SM.)

The mock datasets for training the CNNs are generated using the same Fermi parameters and templates outlined above.
All simulated point-source counts
are stochastically smeared 
with an energy dependent PSF.
A key choice is the space of priors adopted for the mock data generation.
These are detailed in the SM.
We emphasize here that the simulated SCDs, which are drawn from a random skew normal distribution for $dN/d\!\log_{10}\!F$, are chosen to ensure there are ample maps that are indistinguishable from Poisson emission.
In particular, our priors allow the mean of the SCD to be as low as $F \simeq 3 \times 10^{-14}$ photons/cm$^2$/s.
The energy spectrum for each template is taken to be a skew normal distribution, with the location $\xi$, scale $\omega^2$, and skewness $a$ drawn randomly.
The parameters are chosen to ensure the simulated spectra cover a broad range of possibilities.
(Specifically, $\xi$, $\omega^2$, and $a$ are drawn from a log-uniform, a chi-squared distribution, and log uniform, with parameters given in the SM.)
We multiply the normalized energy spectra to the spatial template for the additional energy information without altering the total flux attributed to the respective model components.
The wide priors for the parameters allows the simulated spectra to cover the space of viable spectral shapes, and we assume a common spectrum for all sources (cf. Ref.~\cite{Christy:2024gsl}).

Using the above procedure we generate one million simulated Fermi maps; 790,000 were used for training, 200,000 for validation, and 10,000 for testing.
Using the test maps we show in the SM that the CNN is producing reliable predictions with calibrated errors.

\begin{figure}[!t]
\centering
\includegraphics[width=0.45\textwidth]{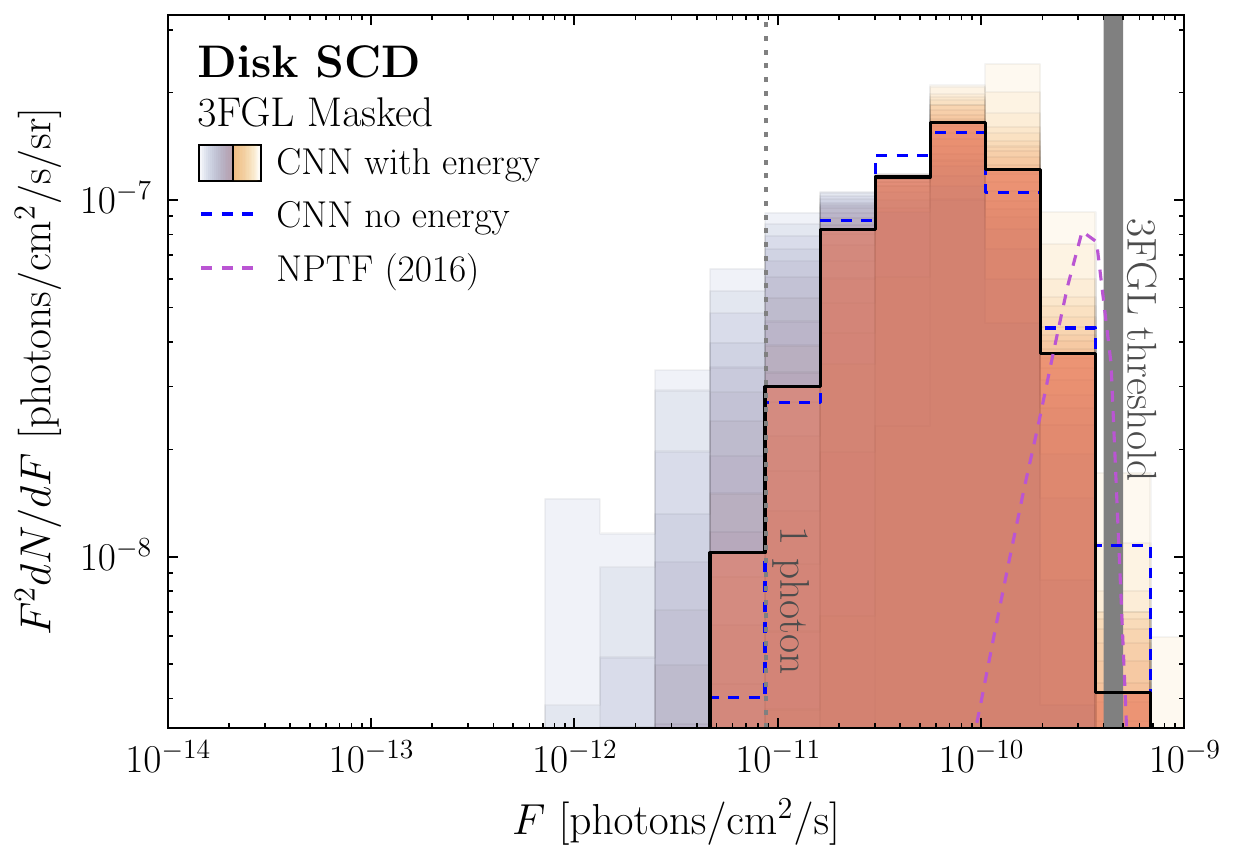}
\vspace{-0.2cm}
\caption{The inferred SCD for the disk displayed in the same format as for the GCE in Fig.~\ref{fig:Results}.
Unlike for the GCE SCD, the addition of energy does not significantly alter the CNN prediction of the disk SCD.}
\vspace{-0.5cm}
\label{fig:SCD-Disk}
\end{figure}

\vspace{0.2cm}
\noindent {\bf $\boldsymbol{dN/dE}$ and $\boldsymbol{dN/dF}$ for the GCE.}
%
The application of our energy dependent CNN to the actual Fermi data leads to the results shown in Figs.~\ref{fig:Results} and \ref{fig:SCD-Disk}.
Consider the spectrum.
Broad agreement is observed between the CNN prediction and that obtained by applying a conventional Poisson likelihood template analysis.
We comment on two aspects of the results.
Firstly, we emphasize that unlike the CNN, the Poisson fit makes no accounting for the presence of unmasked point sources.
In spite of this, the two methods make broadly consistent predictions; the largest discrepancy occurs for the disk above 8\,GeV, and confusion with the spatially similar IC template is undoubtedly playing a role.
Further, the analysis region has the brightest 3FGL sources masked and so only relatively dim sources remain in the data.
For such dim sources, a purely Poissonian fit -- even if based on the wrong statistics -- can still reliably infer the flux fractions.
Secondly, the CNN spectra show a far smaller degree of bin-to-bin variability than the Poisson spectra, indicating a correlation in the results across energy.
As we show in the SM, this correlation can be traced to the fact the CNN is trained only on smooth spectra, and can be reduced if the training includes spectra with random variations.
In particular, the CNN produces the entire spectrum simultaneously, whereas the Poisson spectrum is built bin by bin.

Consider next the SCDs.
The most striking feature is the significantly reduced brightness returned by the energy dependent CNN.
The difference with NPTF results derived in 2016~\cite{Mishra-Sharma:2016gis} is especially pronounced.
(As shown in the SM, applying the NPTF with updated data and background models leads to a dimmer SCD.)
CNN results using a single energy bin for the data generation and training, as in Ref.~\cite{List:2021aer}, already suggested a dimmer SCD, as shown in the blue dashed curve.
The addition of energy leads to a result where the $F^2 dN/dF$ is located almost entirely below the 1 photon threshold.
As the relative consistency in the inferences for the disk SCD demonstrate, where there are clearly expected to be sources there is greater stability between methods.

We end this section with two comments on the GCE SCD.
If interpreted as a distribution of sources, the median SCD prediction requires roughly 200,000 sources to produce this, whereas the 90\% upper quantile prediction is closer to 35,000.
In either event a large number of sources would be required, which forms an essential input to understanding any point-source interpretation of the GCE, see e.g. Refs.~\cite{Dinsmore:2021nip,Holst:2024fvb}.
(For comparison, the NPTF distribution shown in Fig.~\ref{fig:Results} predicts $\simeq$200 sources.)
Secondly, such a dim SCD can also be interpreted as Poisson emission.
Indeed, following Ref.~\cite{List:2021aer}, we trained a separate neural network to use the inferred SCDs to distinguish between Poisson and point-source emission, to the extent possible, and to exclude a fraction of Poisson emission when the map was clearly bright sources.
At 95\% confidence and for the median SCD, the network was only able to exclude 3\% of the map being from Poisson emission.

\vspace{0.2cm}
\noindent {\bf Systematic Studies.}
%
In the SM we explore the robustness of our conclusions under two systematic variations to our analysis: changes to the spatial templates and changes to the CNN.

Firstly, Fig.~\ref{fig:Results} demonstrates that the addition of energy impacts the SCD.
Yet energy is added in two ways: both as a set of input channels for the CNN and also in the simulation of more realistic Fermi maps which model the energy dependence of the templates and the underlying instrument response.
One could ask if the latter alone is driving the difference.
It is not.
If we train a CNN on data simulated in a single energy bin and one trained on data generated in multiple energy bins but then summed back to a single bin, the results for the Fermi GCE SCD are largely indistinguishable and always brighter than with the energy dependent CNN.

Secondly, do the energy channels make the CNN more robust to background mismodeling?
Results here are less conclusive.
To test this, we took an energy dependent and independent CNN trained with Model O.
We then evaluated the performance of both networks on test maps generated with a different diffuse model.
Specifically we use ``Model A'' as introduced in Ref.~\cite{Calore:2014xka}.
Although Model A provides a worse fit to the actual Fermi data~\cite{Buschmann:2020adf} (see their Fig. 3), that goodness-of-fit is irrelevant to tests on purely simulated data.
As demonstrated in the SM, the two CNNs perform similarly across various metrics in the presence of diffuse mismodeling. 
In terms of SCD recovery both methods perform well, although they exhibit a slight bias towards recovering the SCD as too dim, which is particularly interesting given Fig.~\ref{fig:Results}.

Lastly, as a further test of robustness we retrained our full energy dependent CNN on data generated exclusively with Model A.
The results are broadly consistent with Model O with the single key difference that the median GCE SCD prediction is closer to the energy independent result in Fig.~\ref{fig:Results}, although it remains well below the 3FGL threshold.
Although Model A provides a considerably worse fit to the Fermi sky than Model O (the $2 \Delta \log \mathcal{L}$ between the models is $\simeq648.3$), this result exemplifies that our median energy dependent CNN prediction in Fig.~\ref{fig:Results} is not independent of the background diffuse model.
For example, the median number of sources in this case is reduced from ${\cal O}(10^5)$ to ${\cal O}(10^4)$.
This remains the key systematic that deserves further study.
Beyond updated diffuse models, the CNN could be trained on linear combinations of different background models (see e.g. Ref.~\cite{List:2020mzd}), include adaptive template fitting methods such as SkyFACT~\cite{Storm:2017arh,Bartels:2017vsx,Manconi:2024tgh,2025arXiv251103350M}, or even models that have been weighted by polynomials or harmonics (see e.g. Ref.~\cite{Buschmann:2020adf}).
This represents an important benchmark for future work.

\vspace{0.2cm}
\noindent {\bf Conclusions.}
%
What powers the GCE remains a key question of the Fermi era.
Although a notoriously difficult problem, progress remains possible, and machine learning is proving a powerful tool for pushing past long standing limitations.
Our work exemplifies this: we have presented the first simultaneous inference of the GCE spectrum and SCD, revealing that the inclusion of energy shifts the SCD to become indistinguishable from the Poisson emission predicted by DM, thereby undercutting one of the key pieces of evidence in favor of the point-source hypothesis.

At present, all conclusions regarding the GCE are accompanied by the caveat of background systematics.
Whilst this highlights the importance of independent attempts to test the nature of the GCE -- including looking for a millisecond pulsar population in radio~\cite{Calore:2015bsx,Frail:2024vwn} or CTAO~\cite{CTAConsortium:2010umy,CTAConsortium:2017dvg,Gautam:2021wqn} observations or Dwarf searches for a DM signal~\cite{McDaniel:2023bju,DiMauro:2022hue} (cf. Ref.~\cite{Berlin:2025fwx}) -- any reliable inference from the Fermi Galactic Center dataset requires reliable methods.
To that end, there are straightforward extensions to the work presented here that should be studied.
The method could be extended to a wider energy range, especially lower energies where the unique spectrum of the GCE could be more apparent.
Further, given the inferred SCD has significant support in the regime beyond the 1 photon line where the emission becomes Poisson-like, fundamentally building the Poisson degeneracy into the model would improve the interpretability of results.
More broadly, it is likely the CNN based approach we have adopted, while performing well, is not yet the optimal use of machine learning for this problem.
Exploring how these methods can be pushed even further is an important milestone for the future and in determining whether the discovery of DM has already been made with the GCE.

\vspace{0.2cm}
\noindent {\it Acknowledgments.}
%
We thank Dan Hooper and Joshua Foster for discussions, and the latter in particular for assistance in generating the latest Fermi sky maps.
FL thanks Christopher Eckner, Noemi Anau Montel, and Francesca Calore for many discussions on machine learning techniques for $\gamma$-ray data.
We further acknowledge the constructive feedback on a draft version of this work from Gabriel Collin, Dan Hooper, Tim Linden, Dmitry Malyshev, Silvia Manconi, and Tracy Slatyer.
This work was supported in part by the Office of High Energy Physics of the U.S. Department of Energy under contract DE-AC02-05CH11231 and DESC0025293.

\bibliographystyle{apsrev4-1}
\bibliography{refs}

\clearpage
\newpage
\maketitle
\onecolumngrid
\begin{center}
\textbf{\large On the Energy Distribution of Galactic Center Excess' Point Sources} \\ 
\vspace{0.05in}
{ \it \large Supplemental Material}\\ 
\vspace{0.05in}
{}
{Florian List, Yujin Park, Nicholas L. Rodd, Eve Schoen, and Florian Wolf}

\end{center}
\setcounter{equation}{0}
\setcounter{figure}{0}
\setcounter{table}{0}
\setcounter{section}{0}
\renewcommand{\theequation}{S\arabic{equation}}
\renewcommand{\thefigure}{S\arabic{figure}}
\renewcommand{\thetable}{S\arabic{table}}
\renewcommand*{\thesection}{S.\Roman{section}}
\interfootnotelinepenalty=10000 

\setstretch{1.1}

The purpose of this Supplemental Material is twofold.
Firstly, in Sec.~\ref{sec:furtherdetails} we provide the non-essential details of our CNN to provide additional context for the results shown in Figs.~\ref{fig:Results} and \ref{fig:SCD-Disk} and also to demonstrate that those results were derived from a validated network.
Secondly, in Secs.~\ref{sec:templatevariations} and \ref{sec:CNNvariations} we explore the robustness of the results to systematic variations.
In particular, Sec.~\ref{sec:templatevariations} studies the impact of the most concerning systematic: background modeling uncertainties.
After this, in Sec.~\ref{sec:CNNvariations} we demonstrate the robustness of our results to systematic variations in the CNN itself.
We begin, however, with a summary of the various systematic tests performed throughout the Supplemental Material.

\section{Summary of Systematic Tests}
\label{sec:overview}

We present first a high-level summary of the tests performed in this work.
These tests fall broadly into two categories.
The first are those where a new CNN is created on data different from the baseline training dataset in some way.
For these tests, we can obtain a prediction on the Fermi data and compare how different it is from the baseline results, with our findings quantified in Tab.~\ref{tab:summary1}.
For the second category, we take the baseline CNN and have it make predictions on maps outside its training data.
In these cases, we test the accuracy of the CNN when the (simulated) test maps are systematically different from the training data, and show results in Tab.~\ref{tab:summary2}.

Our results are quantified in terms of the impact on our inference for the GCE $dN/dF$ as measured by \emph{calibration}, \emph{sharpness}, and \emph{signed Earth mover's distance} (SEMD), each described in detail in Sec.~\ref{sec:furtherdetails}.
Briefly, however, the SEMD measures how much the GCE $dN/dF$ has shifted compared to the baseline CNN, whereas calibration quantifies the frequency with which the truth fall within the CNN's predicted error bars, while sharpness tracks the size of those error bars.

\begin{table}[h]
\centering
\begin{tabular}{lll}
\toprule
\textbf{Test Variation} & \textbf{SEMD} \hspace{10mm} & \textbf{Figure} \\
\midrule
Subset in time & 1.60 & Fig.~\ref{fig:old-edep-results}\\
Unimodal GCE SCD          & $-$0.78 & Fig.~\ref{fig:unimodal}\\
Alternative Diffuse Model A          & 3.15 & Fig.~\ref{fig:ModAResults}  \\
Brighter GCE SCD prior                 & 2.01     & Fig.~\ref{fig:brighterGCE} \\
\bottomrule
\end{tabular}
\caption{Summary table for experiments with new CNNs.
SEMD is calculated as the (cumulative) baseline CNN median prediction $-$ the varied CNN's median prediction for the GCE's $dN/dF$ in difference of number of flux bins.
For example, the value SEMD $= 1.6$ in the first row ``Subset in time'' means that the CNN found a \textit{brighter} SCD in that variation as compared to the baseline, whose magnitude is characterized by a net shift of the median by 1.6 flux bins to the right.
For reference, the $[0.16,\,0.84]$ quantile range of SEMD for the baseline CNN on test maps is $[-0.90,\,1.15]$.
}
\vspace{-0.3cm}
\label{tab:summary1}
\end{table}

\begin{table}[h]
\centering
\begin{tabular}{llll}
\toprule
\textbf{Test Variation} & \textbf{GCE Calibration [\%]} \hspace{2mm} 
& \textbf{GCE Mean 95\%-IQR [bins]} \hspace{2mm} 
&\textbf{Figure} \\
\midrule
Baseline & 1.60 & 0.223 & Fig.~\ref{fig:sharpcalibration_hist} \\
Alternative Diffuse Model A & 8.14 & 0.226 & Fig.~\ref{fig:mismodeling_hist} -- row 1 \\
Thick Disk           & 0.86 & 0.222 & Fig.~\ref{fig:mismodeling_hist} -- row 2 \\
GCE NFW $\gamma=1$          & 7.16 & 0.210 & Fig.~\ref{fig:mismodeling_hist} -- row 3 \\ 
\bottomrule
\end{tabular}
\caption{Summary table for experiments with our baseline CNN applied to out-of-training-distribution data.
The GCE's $dN/dF$ calibration is summarized by the Mean Calibration Error as a percentage, $(1/T) \sum_\tau^T |\hat{c}(\alpha_\tau) - \alpha_\tau| \times 100$, where $\hat{c}(\alpha_\tau)$ is the measured CNN calibration at confidence level $\alpha_\tau$. 
The baseline calibration measures the performance of the CNN on test maps within its priors, while in all other tests, the calibration is based on CNN predictions on maps systematically different from its training data.
The mean 95\%-IQR describes the sharpness of the predictions and is measured as the IQR between the 2.5 and 97.5 percentiles.
Both quantities are averaged over the flux bins.
}
\vspace{-0.3cm}
\label{tab:summary2}
\end{table}

\section{Validation and Further Results for the default CNN}
\label{sec:furtherdetails}

\begin{figure*}[!b]
\centering
\includegraphics[width=0.45\textwidth]{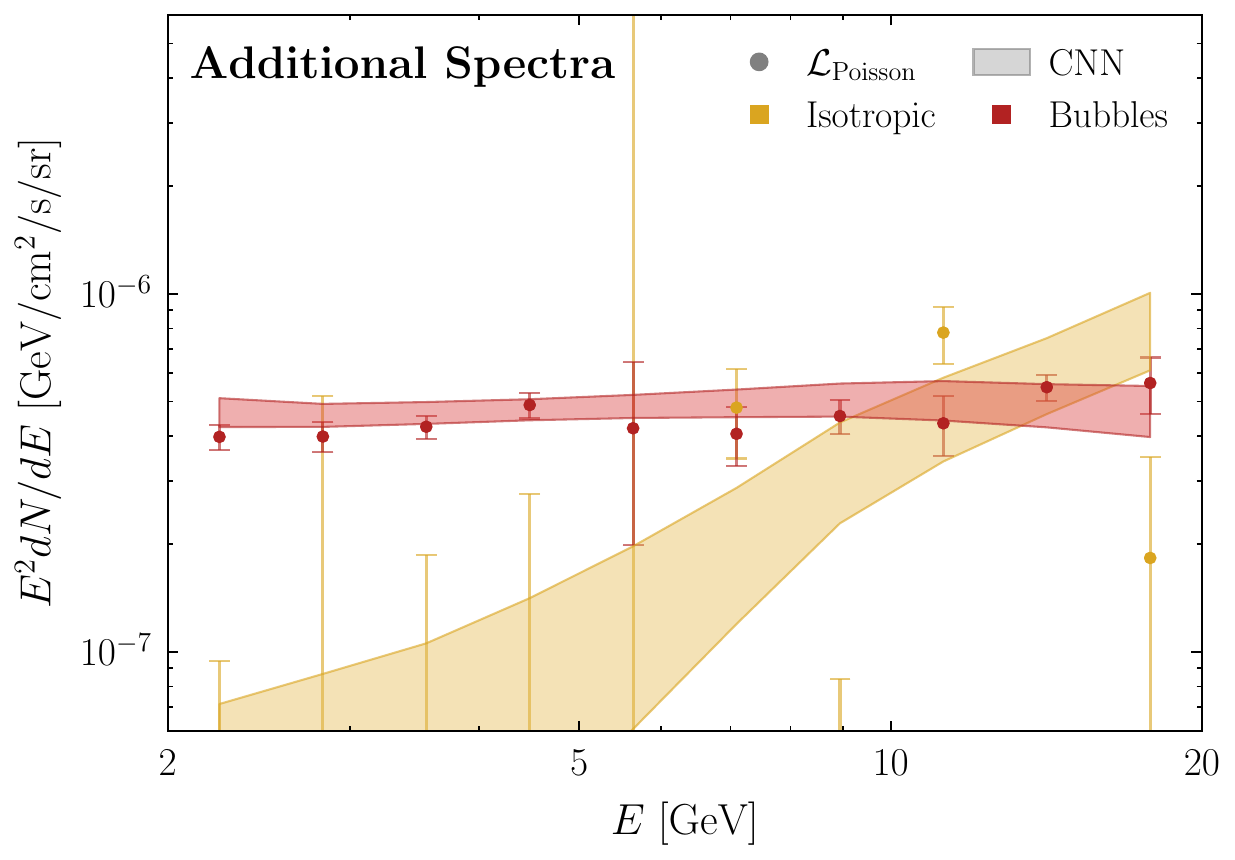}
\vspace{-0.2cm}
\caption{As in Fig.~\ref{fig:Results} but for the spectrum of the isotropic and Fermi bubbles templates.
}
\vspace{-0.5cm}
\label{fig:IsoBub}
\end{figure*}

In the main text we outlined our two step CNN procedure where energy was used as a set of additional input channels.
In the present section we provide further details of that approach and demonstrate the validation of the network.
Seven topics are covered.
Firstly, we provide the full details of the priors used to train the network.
From here we discuss extensively the different representations one can choose for the SCD, showing four different representations of the GCE SCD provided in Fig.~\ref{fig:Results}.
This discussion is followed by a study of how truly Poisson emission is reconstructed within our framework.
Building on the SCD comparison, we then discuss the comparison between our SCD results and those obtained with conventional likelihood techniques.
We then turn to validation, showing the CNN performance on test data, demonstrating both that the method is able to reconstruct a very dim SCD and also that the error bands returned are sensible.
A brief digression on the interpretation of a key metric of the CNN performance, namely \emph{sharpness}, is provided, before we end with the details of the third neural network we implemented to make a robust statement regarding the amount of Poisson emission associated with the GCE.

Before we begin, in Fig.~\ref{fig:IsoBub} we show the recovered spectrum for the isotropic and Fermi bubbles templates using a Poisson likelihood fit and our energy dependent CNN.
These results supplement those on the left of Fig.~\ref{fig:Results} and were neglected there simply for clarity of presentation.
(We model the Fermi bubbles using the template of Ref.~\cite{Su:2010qj}, assuming uniform emission within the bubbles.
For additional details about our templates, we refer the interested reader to e.g. Ref.~\cite[Supp. Mat. Sec. 2]{List:2020mzd}.
We note that the templates used in the present work are energy-dependent versions updated to reflect the longer observation time and correspondingly higher exposure.)
The results show good agreement between the two methods for the Fermi bubbles predictions.
For the isotropic component the agreement is reasonable yet generally worse, although it is subdominant to other components over most of the energy range.
Regarding the isotropic, as noted in the main text the CNN can only return a smooth distribution as that is what it is trained on; the Poisson likelihood, by comparison, can accommodate large bin-to-bin fluctuations.
The rising (in $E^2 dN/dE$) isotropic flux at higher energies is likely a result of residual background mismodeling.
Nonetheless, as the flux is only significant at the highest energies it is unlikely to have a large impact on our results.

Further, although we have not modeled them, there must be an isotropic point-source contribution within our region arising from extragalactic sources.
We have not included these sources as they are hard to constrain in our small analysis region (indeed the Poisson isotropic component is poorly constrained).
It is also unlikely their presence would impact our results: the extragalactic sources should peak below the 3FGL threshold, away from where we find the GCE SCD, and previous studies have found their effect to be small in the inner Galaxy region \cite[e.g.][]{Lee:2015fea}.
Nevertheless they are there and one could imagine including these sources but only allow them to float within a tight prior as determined away from the Galactic Center.
We leave such an analysis to future work.

\subsection{Priors Adopted for Network Training}
\label{ssec:priors}

\begin{table}[h]
\centering
\begin{tabular}{lll}
\toprule
\textbf{Template} & \textbf{Parameter} & \textbf{Priors} \\
\midrule
Diffuse $\pi^0$\,+\,Brem. & $A$ & $U([1.75, 3.5])$ \\
Diffuse IC          & $A$ & $U([1, 2.25])$ \\
Isotropic           & $A$ & $U([0, 0.5])$ \\
Fermi Bubbles & $A$ & $U([0, 0.5])$ \\
GCE                 & $\xi$     & $U([-13, -9])$ \\
                    & $\omega^2$ & $0.25\, \chi^2_1$ \\
                    & $a$        & $\mathcal{N}(0, 3)$ \\
                    & $F_{\text{tot}}$ & $U([0, 1.4 \times 10^{-7}])$ \\
Disk                & $\xi$     & $U([-12, -9])$ \\
                    & $\omega^2$ & $0.25\, \chi^2_1$ \\
                    & $a$        & $\mathcal{N}(0, 3)$ \\
                    & $F_{\text{tot}}$ & $U([0, 2.8 \times 10^{-7}])$ \\
\bottomrule
\end{tabular}
\caption{The set of priors used for generating the training data for all parameters except for the spectrum, where priors are presented in Tab.~\ref{tab:spec_priors}.
This table specifically outlines those priors which determine the flux associated with each template.
For the four Poisson templates, we specify the range over which we draw their template normalization, $A$, which in each case is taken from a uniform distribution $U([a, b])$ that runs from $a$ to $b$.
(All templates are unit normalized in the region within 30 degrees of the galactic center, masking within 2 degrees of the plane.)
For the point sources, first we draw the total flux for the population, specified by $F_{\rm tot}$ (in units of photons/cm$^2$/s), and then determine how that flux is divided amongst point sources.
For this we use the skew normal distribution of Eq.~\eqref{eq:skewnormal} (over the logarithmic fluxes) defined by $\xi$ (location), $\omega^2$ (scale), and $a$ (related to skewness).
These three parameters are drawn from a uniform distribution, a chi-squared distribution with one degree of freedom, $\chi^2_1$, and a normal distribution ${\cal N}(\mu,\sigma)$, respectively.
Note that for the GCE, all simulated maps involve two separate point source contributions added together so that the GCE SCD can be bimodal and have a maximum total flux of 2.8$\times 10^{-7}$\,photons/cm$^2$/s.}
\vspace{-0.3cm}
\label{tab:priors}
\end{table}

\begin{table}[h]
\centering
\begin{tabular}{lll}
\toprule
\textbf{Parameter} & \textbf{Priors} \\
\midrule
$\xi$      & $U([\log_{10}2,\, \log_{10}20])$ \\
$\omega^2$ & $4\, \chi^2_1$ \\
$a$        & $\mathcal{N}(0, 5)$ \\
\bottomrule
\end{tabular}
\caption{Priors for the spectrum for each template, which is taken to be a skew normal distribution (for the logarithmic energies) with random parameters.}
\vspace{-0.3cm}
\label{tab:spec_priors}
\end{table}

A key ingredient in specifying the CNN is the set of priors that determine the simulated maps on which it is trained and validated.
Our choices are specified in Tab.~\ref{tab:priors} and \ref{tab:spec_priors}.
For the templates that are described by Poisson emission, we randomly determine a normalization for the templates which fixes the total expected flux and then determine the counts in each pixel randomly according to the Poisson distribution.
The counts are distributed randomly among the ten energy bins (logarithmically spaced between 2 and 20 GeV) accordingly to a skew normal distribution,
\begin{equation}
f(x) = \frac{2}{\omega} \, \phi\!\left[ \frac{x - \xi}{\omega} \right] \, \Phi\!\left[ a \left( \frac{x - \xi}{\omega} \right) \right]\!,
\label{eq:skewnormal}
\end{equation}
where $\phi(x)$ and $\Phi(x)$ represent the probability distribution function (PDF) and cumulative distribution function (CDF), respectively, of a symmetric standard normal distribution with $\mu=0$ and $\sigma=1$.
The parameters $\xi$, $\omega$, and $a$ indicate the location, scale, and skewness of the skew-normal distribution, and are chosen randomly for each template.

\begin{figure*}[!t]
\centering
\includegraphics[width=0.45\textwidth]{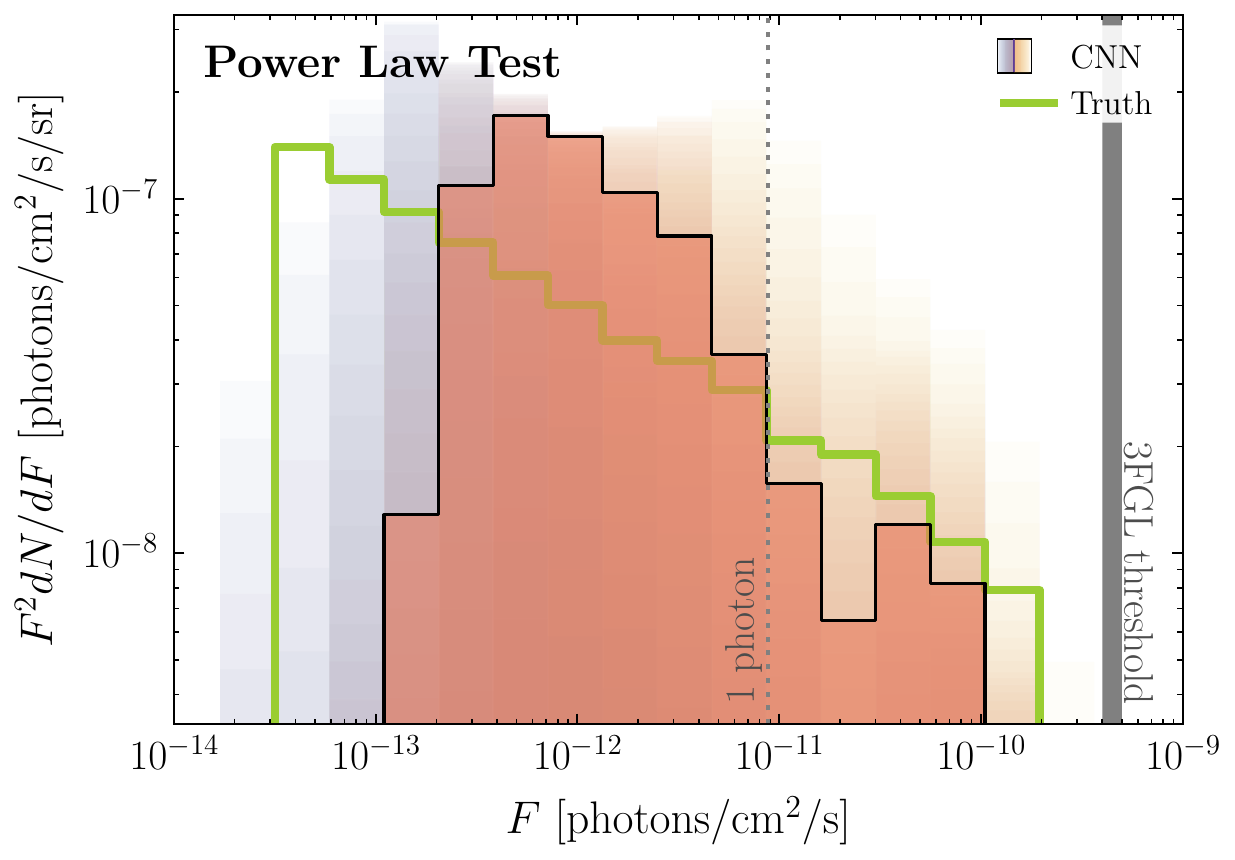}
\hspace{0.5cm}
\includegraphics[width=0.45\textwidth]{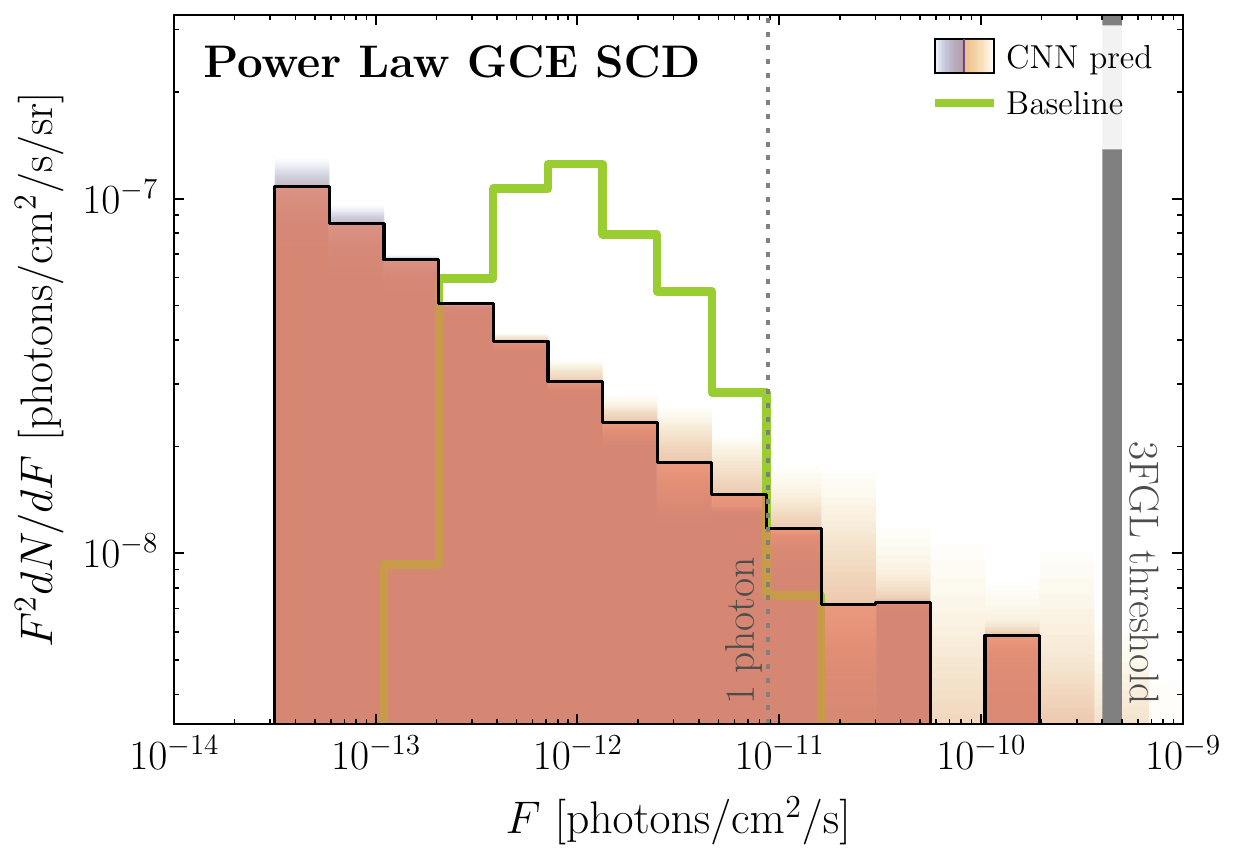}
\vspace{-0.2cm}
\caption{(Left) The performance of our baseline CNN to recover an injected power law SCD.
As can be seen, although the recovery is not perfect, the broad features of the power law are recovered.
We emphasize that well below the 1 photon line the flux bins become degenerate as the emission becomes indistinguishable from Poisson; therefore the poorer agreement at the very lowest fluxes is less concerning.
(Right) The results out our baseline CNN retrained on power law SCDs.
The recovered SCD continues to place the majority of the flux below the 1 photon threshold.
}
\vspace{-0.5cm}
\label{fig:SCD-PowerLaw}
\end{figure*}

For the GCE and disk templates, which are described with point-source statistics, we need instead to distribute their flux between various point sources that are drawn from a randomly determined SCD.
We again specify the SCD as a skew normal distribution, with parameters chosen such that particularly in the GCE case the network is shown maps where the emission is so dim as to be fully indistinguishable from Poisson.
One important distinction between the GCE and disk component is that for the GCE we generate two separate populations of point sources and add them together; in other words, the GCE will generally have a bimodal SCD, whereas the disk SCD is unimodal.
The rationale for this choice is to ensure the network is flexible enough to account for the possibility the GCE could be a fraction of truly dim emission, as predicted by DM, and a truly point-source-like component.
In spite of this, the central prediction in Fig.~\ref{fig:Results} is suggestively unimodal, although we show in Sec.~\ref{sec:CNNvariations} that a network trained on purely unimodal GCE distributions returns an SCD that has a similar mean to that shown in the main text, but that is noticeably narrower.
Of course, it is always possible that the space of priors chosen is not sufficient to include all types of emission that could be present in the actual Fermi data.
As a test to see if our method can capture a broader range of emission, on the left of Fig.~\ref{fig:SCD-PowerLaw} we show that the priors are sufficiently flexible to reasonably recover an injected power law SCD spanning a wide flux range.
For this test, the dataset was generated using a random draw from our default priors in Tab.~\ref{tab:priors}, with the exception that the SCD was forced to have the depicted power law shape.
The results of this test suggest that if the true SCD was a soft power law with the majority of its flux placed below the 1 photon line, then the inferred SCD by our baseline analysis would not look too distinct 
To test this further, we retrained our baseline network with the GCE SCD now described by a power law in $F^2 dN/dF$ with an index between -0.5 to 0.5 (although always cutting sources outside $3\times 10^{-14}$ and $1\times 10^{-9}$ counts$/$cm$^2/$s to ensure the bright sources that are masked in the real data are not included).
The result of this is shown on the right of Fig.~\ref{fig:SCD-PowerLaw}, and consistent with the above the results prefer a soft power law.

\subsection{Interpretations and Representations of the Source Count Distribution}
\label{ssec:SCD4}

A key result of the present work is the inference of the SCD for the GCE.
The SCD, written as $dN/dF$, specifies the number of sources that have a flux in the range $[F,\,F+dF]$ and encodes information on the population and how it is expected to impact the Fermi dataset.
Throughout this work, ``flux'' refers to the integrated photon flux over the energy range considered in our analysis, i.e.\ 2-20 GeV.
With information about the spatial distribution of the source population the SCD can be converted to a more physical luminosity distribution, but as $\gamma$-ray fluxes are what impacts Fermi observations, we focus our attention on the flux distribution.
The goal of the present section is to discuss different ways to interpret the SCD, both to add further context for interpreting our results, but also to clarify how the CNN infers the distribution.

Given an SCD we can compute the expected number of total sources from $N_{\rm tot} = \int_0^{\infty} dF\,dN/dF$ and also the expected total flux produced from the population, $F_{\rm tot} = \int_0^{\infty}dF\, FdN/dF$.
Even if each underlying source had the same luminosity, once a realistic spatial model is included, the distribution of fluxes observed at the Earth is generally spread over many orders of magnitude.
As such, it is convenient to make inferences and present results in logarithmic fluxes, and in particular by default we take the variable of our distributions to be $\log_{10}\!F$ rather than $F$. 
In this space, we now have $F_{\rm tot} = \int_{-\infty}^{\infty} d\!\log_{10}\!F\,FdN/d\!\log_{10}\!F$ and similarly for $N_{\rm tot}$.
Note, however, that this implies that when plotted on logarithmic axes, the area under the following quantity determines the total flux,
\begin{equation}
F\frac{dN}{d\!\log_{10}\!F} = \ln(10) F^2\frac{dN}{dF} \propto F^2\frac{dN}{dF}.
\end{equation}
This is the rationale for showing $F^2 dN/dF$ by default in the figures of the main text as they directly inform a comparison for the total flux associated with various models.
In particular, we can see that the energy dependent and independent results recover a similar flux fraction of $\simeq 7\%$, whereas the 2016 NPTF results -- with a far narrower distribution -- are more than a factor of 4 smaller.
We emphasize that this is the flux fraction attributed to the median NPTF SCD and this value varies somewhat between analyses, see Refs.~\cite{Lee:2015fea,Mishra-Sharma:2016gis,Leane:2019xiy,Buschmann:2020adf}.
What is more consistent between the classic NPTF analyses is the presence of an SCD peaking below the 3FGL threshold.
We expand on the comparison with NPTF in the next section.
Note also that when showing $F^2 dN/dF$, we further divide by the size of our region of interest, $\Omega \simeq 0.49\,{\rm sr}$, to make the results more representative of the density of sources.

\begin{figure*}[!t]
\centering
\includegraphics[width=.79\linewidth]{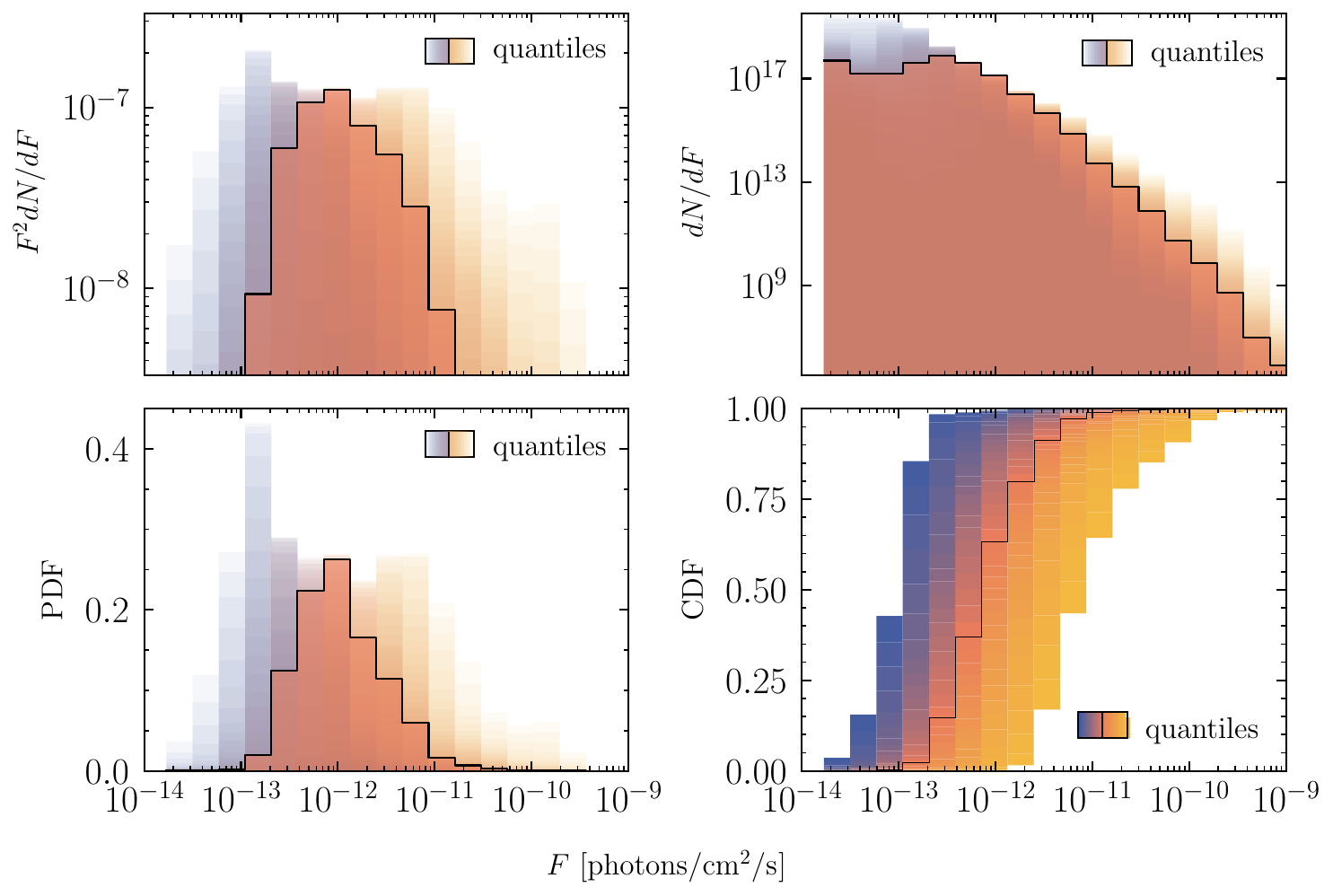}
\vspace{-0.2cm}
\caption{Here we show four representations of the GCE SCD.
Starting from the top left and working clockwise: 1. $F^2dN/dF$ in units of [photons/cm$^2$/s/sr] as in Fig.~\ref{fig:Results}; 2. $dN/dF$, so similar but divided by $F^2$; 3. the CDF, which is the direct output of the CNN; and 4. the associated PDF, normalized so that it sums to unity over all flux bins.
In all cases we show quantiles -- defined in terms of the CDF value in each bin -- ranging from purple (2.5\%) to yellow (97.5\%) progressing between in steps of 2.5\%, with the median shown in black.
}
\vspace{-0.5cm}
\label{fig:SCDs_4_ways}
\end{figure*}

We emphasize, however, that it is $dN/dF$ rather than $F^2 dN/dF$ that controls the probability of seeing sources at a given brightness.
Accordingly, caution is required when comparing $F^2dN/dF$ to flux thresholds as the distribution is weighted to larger fluxes (this effect is smaller for narrower distributions).
In particular, in several figures depicting an SCD, we show two reference lines, labeled ``1 photon'' and ``3FGL threshold''.
The 3FGL threshold~\cite{Fermi-LAT:2015bhf}, shown between $(4-5)\times 10^{-10}$\,photons/cm$^2$/s, is taken as an approximate threshold above which sources should begin to be individually resolved.
We emphasize the approximate nature of this association: the 3FGL catalog was determined a decade ago and newer catalogs have since been produced~\cite{Ballet:2023qzs}.
For our default analysis we prefer the comparison with the older 3FGL as if one masks all sources in the newer catalogs a large fraction of the ROI is removed.
Moving forward, the best approach may be not to use a point-source mask and instead infer unresolved and resolved sources simultaneously; agreement between the inferred $dN/dF$ and the catalog sources at large fluxes could then serve as a key validation of the method’s accuracy.
See e.g. Ref.~\cite{Malyshev:2024obk} for a machine learning approach to resolved sources in the Galactic Center.
The 1 photon line, at $\simeq 8.8 \times 10^{-12}$\,photons/cm$^2$/s, is the flux where a source is expected to generate a single photon, which as discussed in the main text is conceptually important as it sets the scale below which the flux begins to become indistinguishable from Poisson emission (see also Refs.~\cite{Collin:2021ufc,List:2021aer}).
We return to this point in Sec.~\ref{ssec:Poisson}.

So far, we have outlined two different representations of the SCD: $dN/dF$ and $F^2dN/dF$.
For our central results for the GCE, both of these are shown in Fig.~\ref{fig:SCDs_4_ways}.
Neither of these correspond to what the CNN actually infers.
As discussed, $dN/dF$ can be used to compute the total expected flux generated by the point-source population.
In the CNN framework we adopt, however, the flux (or truthfully the flux fraction) is inferred by the first network, this value is then fixed when the SCD is inferred at the second step.
Instead, the CNN infers a probability distribution associated with the sources, or more accurately, a binned cumulative probability distribution.
As discussed in Ref.~\cite{List:2021aer}, this quantity makes the incorporation of quantiles on the probability distribution straightforward through the use of the Earth mover's pinball loss~\cite{list2021earth}.
As mentioned in the main text, the probability distribution we infer is related to $F^2dN/dF$, or as we work with logarithmic binning, $FdN/d\!\log_{10}\!F$: the CNN is trained to infer the amount of flux the point-source population has in each log-spaced flux bin (not the number of sources).
The Earth mover's loss ensures that the method is penalized for placing the flux too far from its true location (i.e. placing the flux two bins away from its true location receives a stronger penalty than a single bin).
The ``pinball'' aspect of the loss means that the penalty for over- or underestimating the true CDF in each bin is \textit{asymmetric}, with a tilt that depends on the queried quantile level $\tau$.
The loss function used for training is therefore $\tau$ dependent---in such a way that its expectation for any given $\tau$ is minimized by the $\tau$-th quantile.

With this in mind, in Fig.~\ref{fig:SCDs_4_ways} we also show the CDF -- the distribution the CNN actually infers -- as well as the derived PDF for the GCE.
In all cases, the quantile levels shown range from 2.5\% (purple) to 97.5\% (orange) in steps of 2.5\%.
The ordering of our quantiles is defined such that lower quantiles correspond to a dimmer PDF or $dN/dF$; this implies that for the CDF the quantiles appear reversed, as a CDF that rises earlier corresponds to a dimmer PDF.
Note also that for the CDF we show the quantiles in full color as they can be distinguished due to their monotonicity, whereas they overlap when converting them to PDFs so we show at various levels of transparency.
By default, all distributions have 26 flux bins: 24 equally spaced log bins between $[10^{-13.5},\,10^{-7}]\,$photons/cm$^2$/s, and an overflow bin on either side.
Within this space, the histogram of the PDF is normalized such that $\sum_i p(F_i)=1$, and we can therefore compute in each bin $F_i^2\,(dN/dF)_i = [F_{\rm tot}/\Delta \ln (10)]\, p(F_i)$, where $\Delta$ is the width of the logarithmic bins.

\subsection{Reconstruction of Poisson Emission}
\label{ssec:Poisson}

\begin{figure*}[!t]
\centering
\includegraphics[width=.45\linewidth]{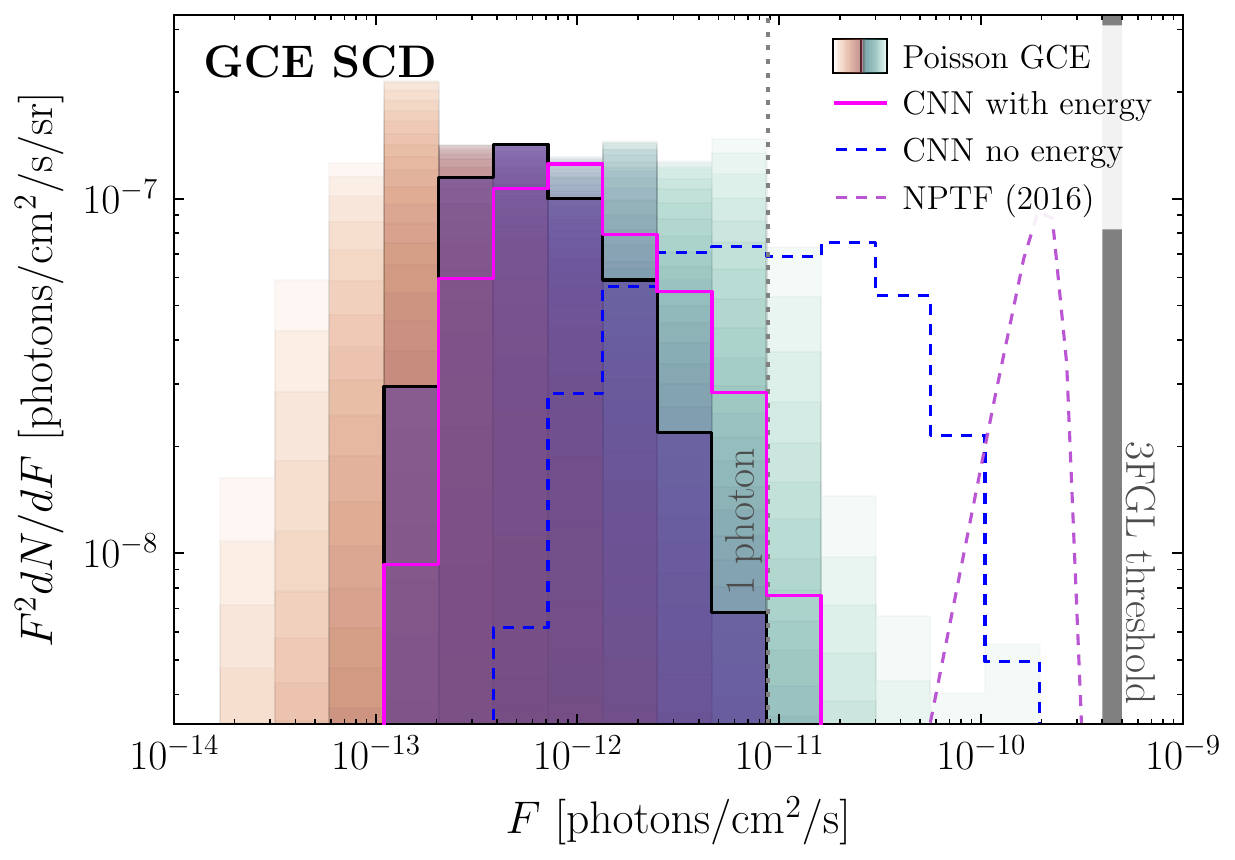}
\vspace{-0.2cm}
\caption{How truly Poisson emission is recovered by our energy dependent network, compared to the CNN predictions for Fermi.
We show the CNN one hundred simulated Fermi maps where the GCE flux is completely Poisson and then display the median result, quantile by quantile, for the resulting SCD prediction.
In particular, the solid black line corresponds to the median of the one hundred predicted SCD medians.
As can be seen, the range over which Poisson emission is recovered is fully inclusive of the median GCE prediction on the actual Fermi data.}
\vspace{-0.5cm}
\label{fig:poissonGCE}
\end{figure*}

Given its importance, let us expand the discussion of how Poisson emission is reconstructed in our framework.
We reiterate the reason Poisson emission is interesting to study is that a population of point sources that each produce on average far less than 1 photon is mathematically indistinguishable from Poisson emission.
Therefore, all SCDs shown in this work contain an intrinsic degeneracy in the sources well below the 1 photon line.

The goal of the present section is to quantify the behavior of our network when presented with truly Poisson emission, although formally this was never shown to the network during training.
In fact, as emphasized in the main text, we show the CNN maps that include point sources that are sufficiently dim to be indistinguishable from Poisson emission during the training process.
As such, even though it is never shown exact Poisson emission for the GCE during training, it should be able to reconstruct Poisson emission at the dim end of the SCD.
We demonstrate this in Fig.~\ref{fig:poissonGCE} where we show how the network recovers literal Poisson emission; in other words, this shows how the network would recover the DM prediction or equivalently a population of point sources where no source ever produced more than one photon.
In particular, the flux is on average recovered below the 1 photon line; again, we emphasize that as we show $F^2 dN/dF$ the distribution is shifted to higher fluxes as compared to $dN/dF$, although we choose to show the result as such in order to facilitate a comparison with Fig.~\ref{fig:Results}.
As can be seen, the recovered Poisson flux is close to what is found in the Fermi data---in Sec.~\ref{ssec:nn3} we return to this point and quantify this intuition into an exclusion on the amount of Poisson emission consistent with the recovered SCD.
To be more explicit as to what is shown in the plot, we simulated one hundred maps where the GCE is generated from exclusively Poisson emission and all other spectral and SCD parameters are taken to be the values that fit the Fermi data best.
For each map we extract the SCD prediction by our CNN and then for each quantile compute the median over the one hundred simulated maps, which is what is finally shown.

\subsection{Comparison with Analytic Likelihood Techniques}
\label{ssec:NPTFCompare}

\begin{figure*}[!t]
\centering
\includegraphics[width=0.45\textwidth]{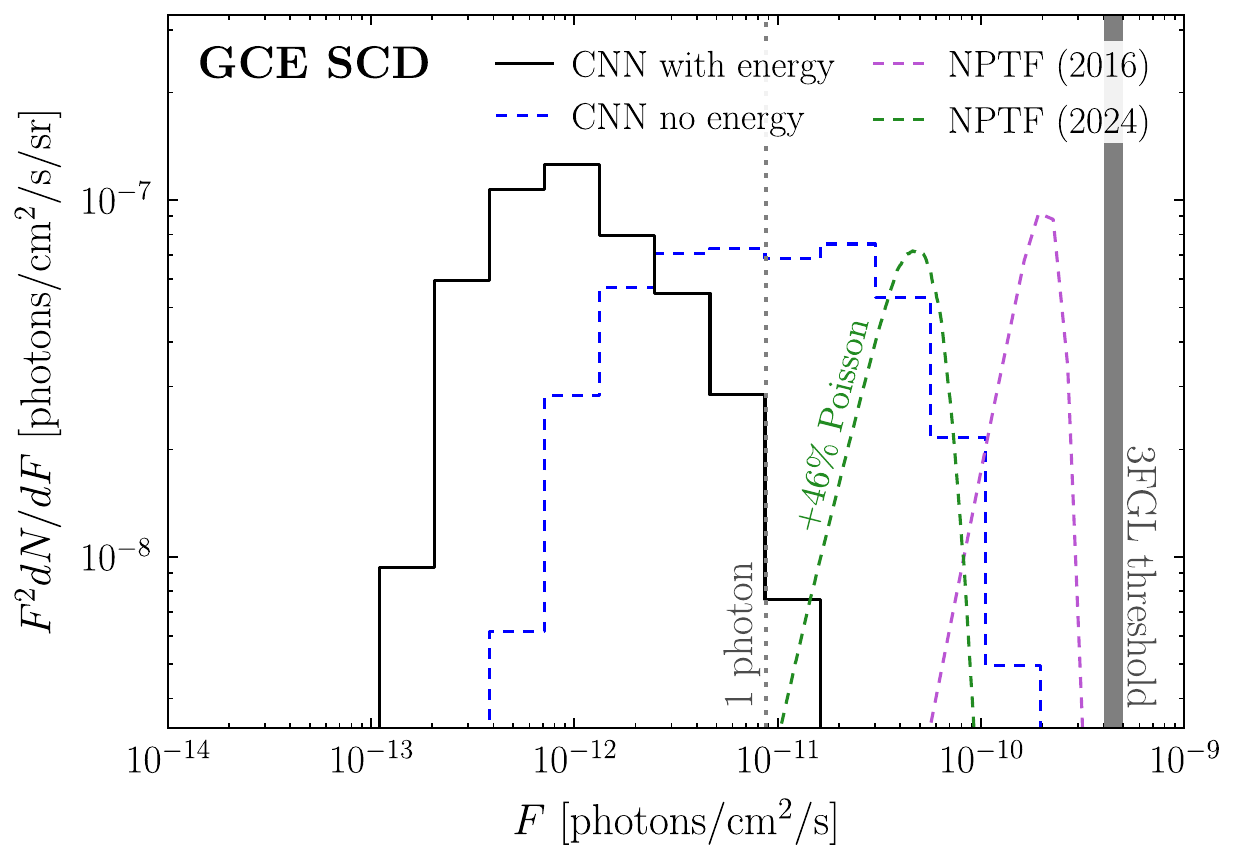}
\hspace{0.5cm}
\includegraphics[width=0.45\textwidth]{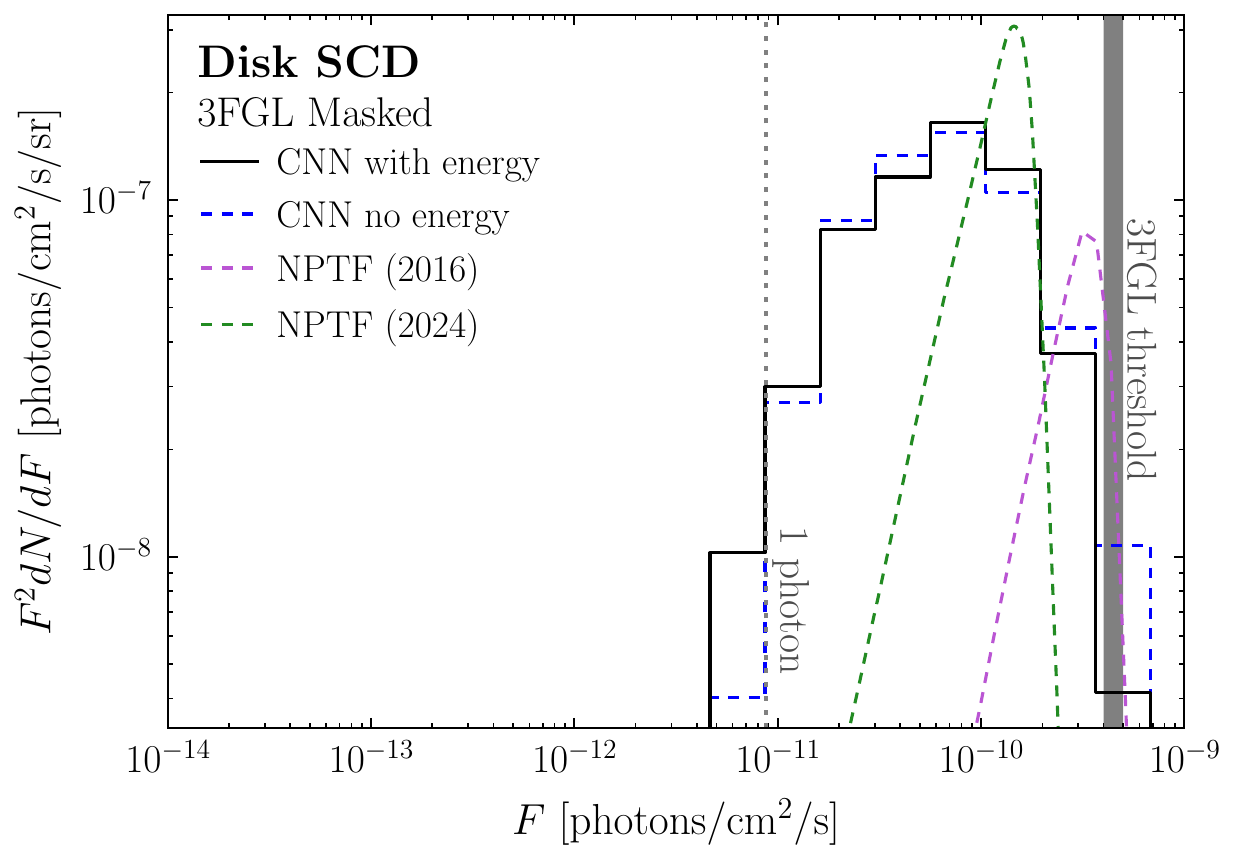}
\vspace{-0.2cm}
\caption{As in Figs.~\ref{fig:Results} and \ref{fig:SCD-Disk}, but with a more direct comparison to the analytic likelihood approach, represented by the NPTF.
In particular, the ``NPTF (2016)'' curves are designed to represent the state of the art predictions a decade ago.
By contrast, the ``NPTF (2024)'' results in green depict what analytic methods recover on otherwise identical analysis choices to what we use by default for our CNN.
Importantly, the green curve only accounts for 54\% of the GCE flux recovered by the NPTF; the remaining 46\% is attributed to genuinely Poisson emission.
(For the 2016 results, the fit preferred effectively zero Poisson emission.)
In all cases the curves show the median predictions of each method.}
\vspace{-0.5cm}
\label{fig:NPTF-Comparison}
\end{figure*}

As discussed in the main text, the classic results regarding the GCE SCD were derived using analytic likelihood based methods such as NPTF~\cite{Lee:2015fea,Mishra-Sharma:2016gis}.
In particular, in Figs.~\ref{fig:Results} and \ref{fig:SCD-Disk}, the results labeled ``NPTF (2016)'' are taken directly from Ref.~\cite{Mishra-Sharma:2016gis}, as this work used an identical energy range to what we consider, although in a single bin.
There were also several other differences in analysis in that work, including a larger region of interest was used and an older diffuse model (technically \texttt{gll\_iem\_v02\_P6\_V11\_DIFFUSE} or \texttt{p6v11}).

The inclusion of the ``NPTF (2016)'' results was intended to provide a comparison for what was the state of the art a decade ago, rather than a direct comparison with likelihood based methods.
In Fig.~\ref{fig:NPTF-Comparison} we show what is designed to be a more direct comparison between our results and NPTF.
In particular, we show the same curves as in the main text, but add to these ``NPTF (2024).''
Those results make use of the public code \texttt{NPTFit}~\cite{Mishra-Sharma:2016gis} to analyze the same dataset and region as for our default CNN and we replace \texttt{p6v11} with Model O to describe the diffuse emission and with equivalent priors to the 2016 analysis. For more details on how Model O has been constructed, we refer the reader to Ref.~\cite{Buschmann:2020adf}.
As can be seen, the move to a newer (and better fitting) background model drives the inferred GCE SCD dimmer, a point that was noted already in Ref.~\cite{List:2021aer}, although still brighter than either prediction of the CNN.
We emphasize that the CNN results are describing the total GCE emission, whereas for the NPTF that flux is artificially divided into a point-source component (shown) and a Poisson contribution that cannot be directly mapped into this space.
Indeed, with the SCD becoming dimmer in the 2024 results, the confusion between the point source and Poisson components becomes significant: the median prediction for the NPTF is that 54\% of the flux is due to point sources and 46\% Poisson emission (cf. the 2016 prediction where almost the entire flux was attributed to point sources).
Given this, it is interesting to observe the rough consistency between the NPTF (2024) GCE SCD and the upper end of the energy independent CNN.

\subsection{Network Validation}

\begin{figure}[!t]
\centering
\includegraphics[width=.47\linewidth]{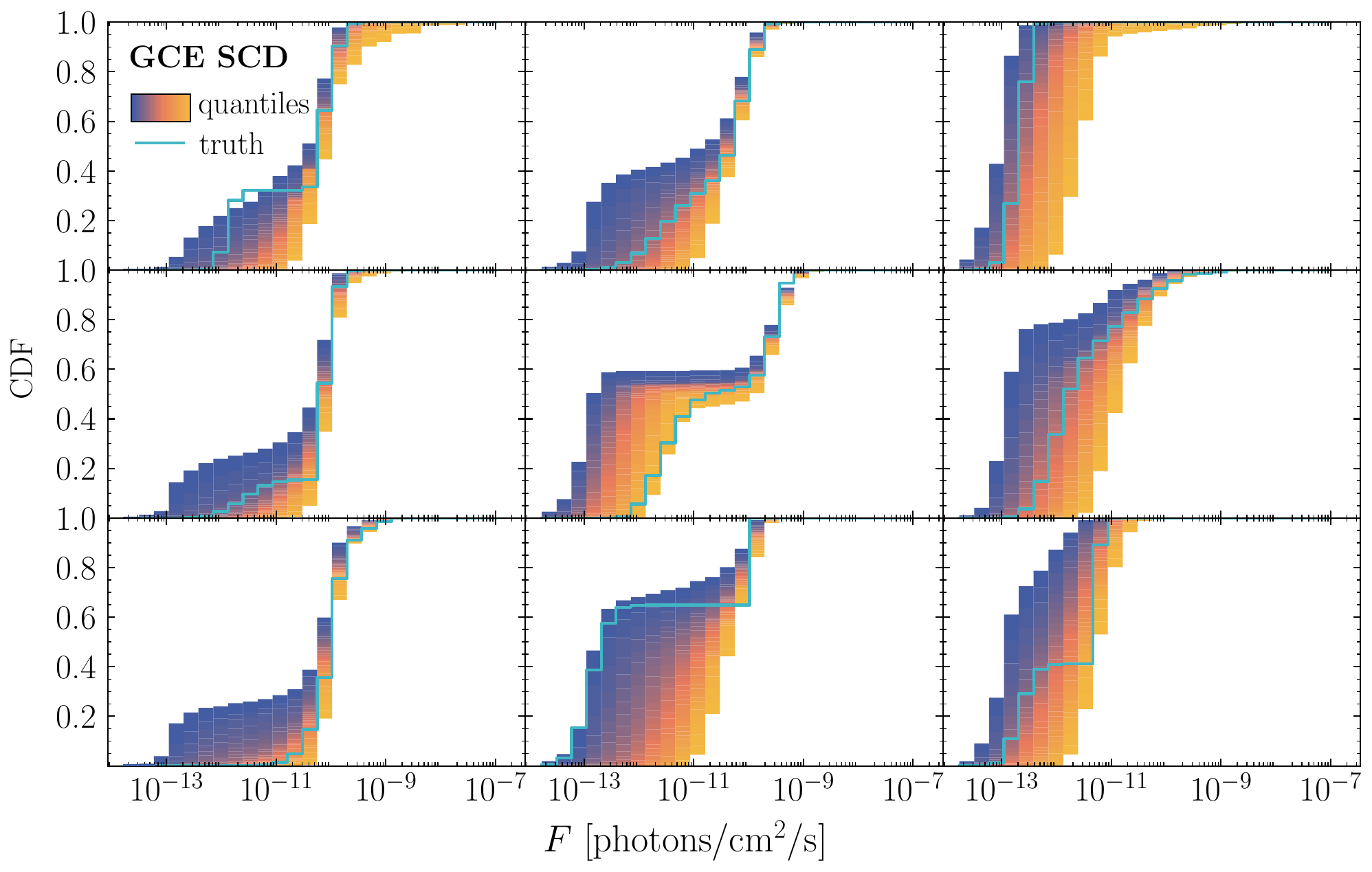}
\hspace{0.5cm}
\includegraphics[width=.47\linewidth]{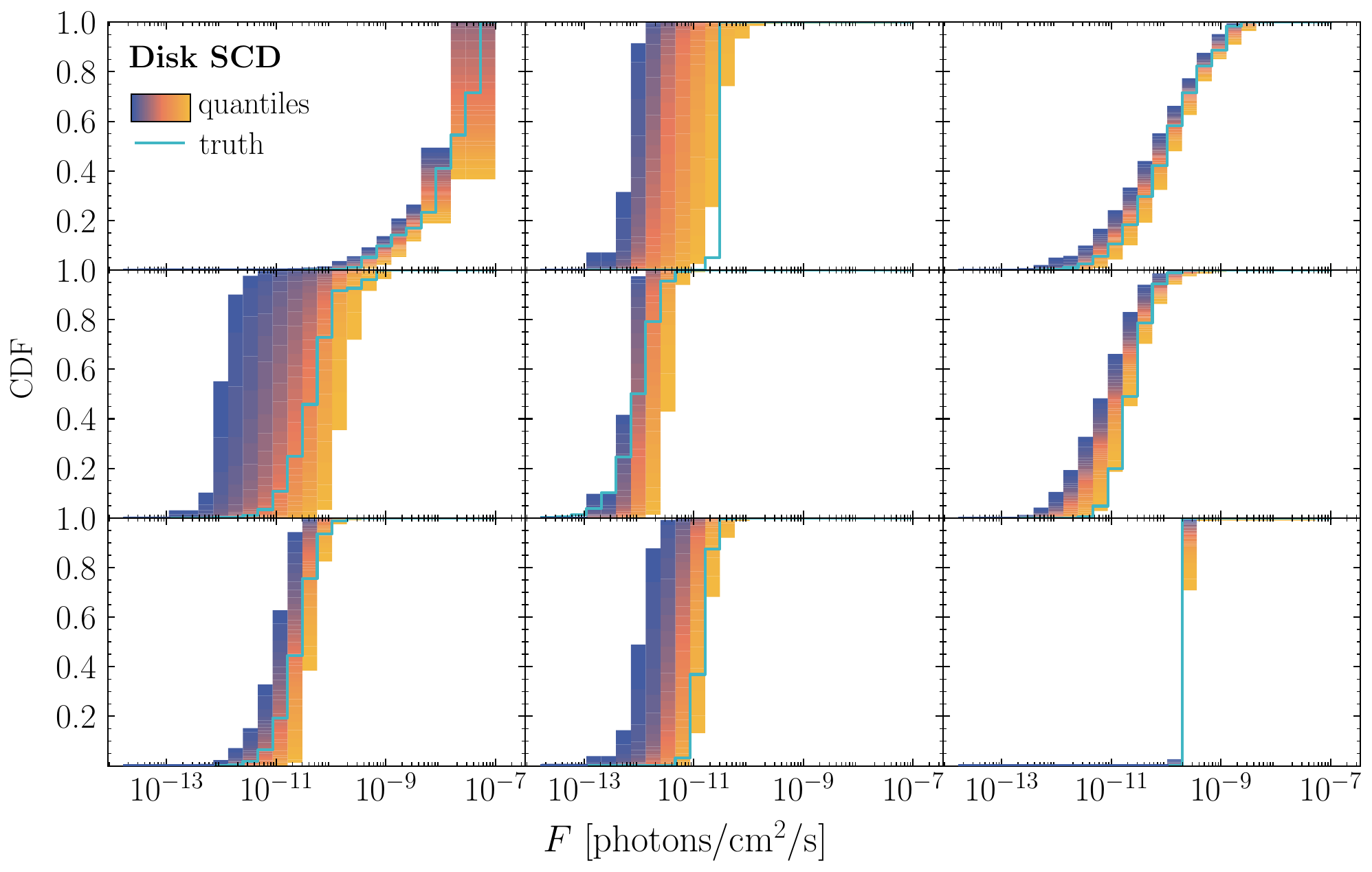}
\vspace{-0.2cm}
\caption{The SCDs predicted by the energy dependent CNN on nine randomly chosen test maps as compared to the true SCDs shown in light blue.
All SCDs are shown as CDFs which correspond to the space in which the CNN is trained.
Quantiles which fall above the median are dimmer, whereas those which fall below are brighter.
For several of the maps (e.g. bottom center), the bimodality in the GCE SCD can be clearly seen, resulting in two sharp increases in the CDF that are separated by a nearly horizontal stretch.}
\vspace{-0.5cm}
\label{fig:exCDFs}
\end{figure}

\begin{figure}[!b]
\centering
\includegraphics[width=.64\linewidth]{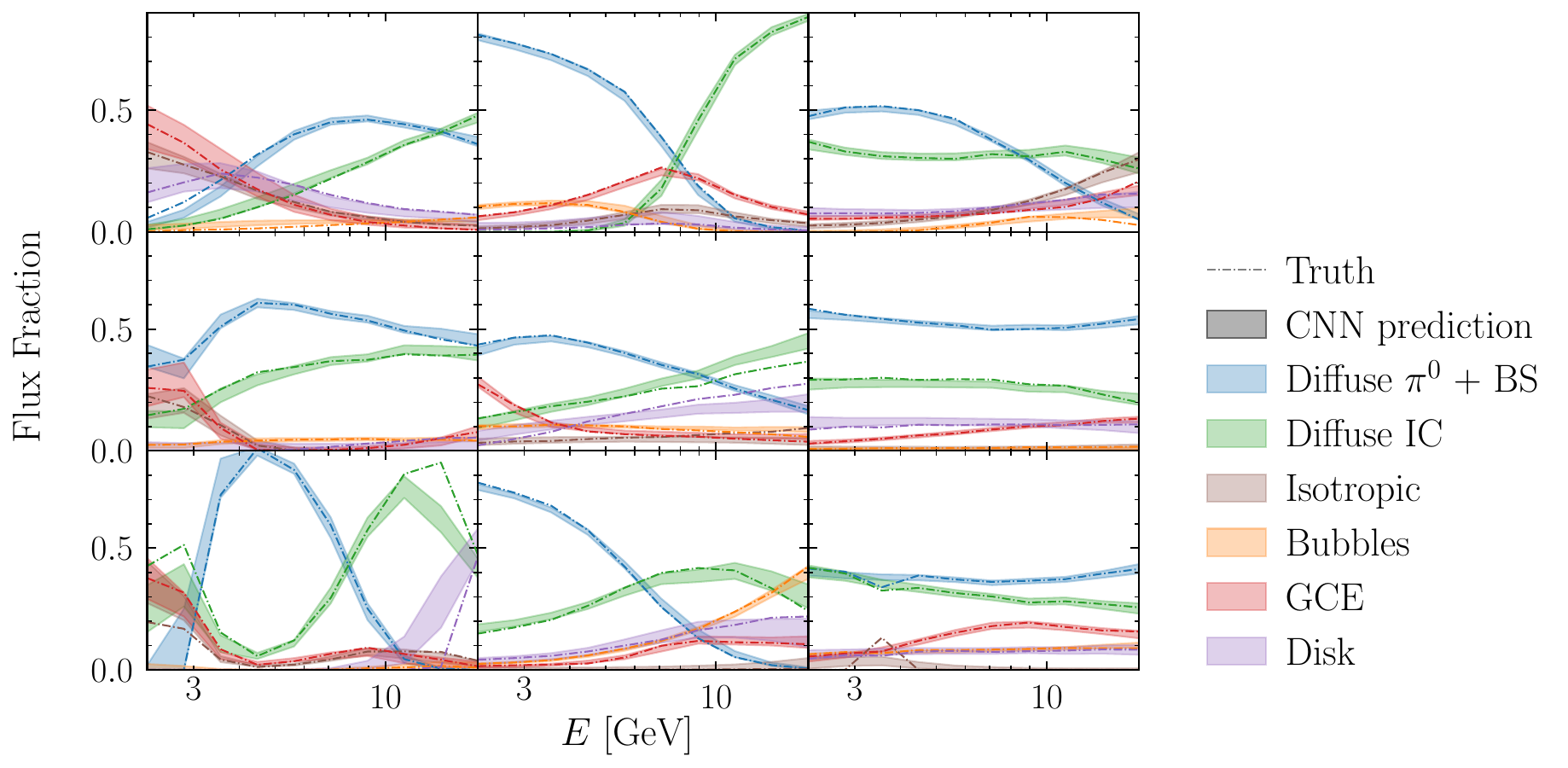}
\vspace{-0.2cm}
\caption{The 1$\sigma$ bands on the energy spectra predicted by the CNN, compared to the true spectra for nine test maps. 
The energy spectra correspond to the same maps as shown in Fig.~\ref{fig:exCDFs}.
The fraction of the flux of the total map per template is plotted per energy bin as this is the space in which the CNN trains.
}
\vspace{-0.5cm}
\label{fig:exSpectras}
\end{figure}

Once the CNN is trained, before turning to real data it is important to validate the network performance on datasets that are drawn from the same space of maps used to train the CNN.
(The performance on maps outside the training space is studied in Sec.~\ref{sec:templatevariations}.)
In particular, we study the performance on 500 example Fermi sky maps, which are drawn from the priors used to generate the training dataset, but crucially not shown to the network during training.
Focusing on the SCD recovery, in Fig.~\ref{fig:exCDFs} we show nine examples for the recovery of the GCE and disk SCD.
Although chosen randomly, these results exemplify the general performance: the CNN can recover the SCD within its uncertainty bands and when the sources are well above the one photon threshold ($\simeq 10^{-11}$\,photons/cm$^2$/s) the uncertainty bands tend to be smaller.
Notably, the uncertainty bands predicted by the neural network are \textit{pointwise} rather than simultaneous.
Intuitively, this means that one should not expect the true SCD to lie within the predicted 95\% interquantile range across \textit{all} flux bins for 95\% of the maps -- as would be the case for simultaneous uncertainty bands, which are typically wider.
Instead, the predicted bands are expected to be well calibrated when considering each flux bin individually, as analyzed in what follows.
Similarly, in Fig.~\ref{fig:exSpectras} we show the performance of the spectra recovery for nine example maps.
Again in general the performance is good and where there tends to be mismodeling in these examples it is a misattribution of the disk and IC flux, particularly at high energies.

\begin{figure*}[!t]
\centering
\includegraphics[width=.32\linewidth]{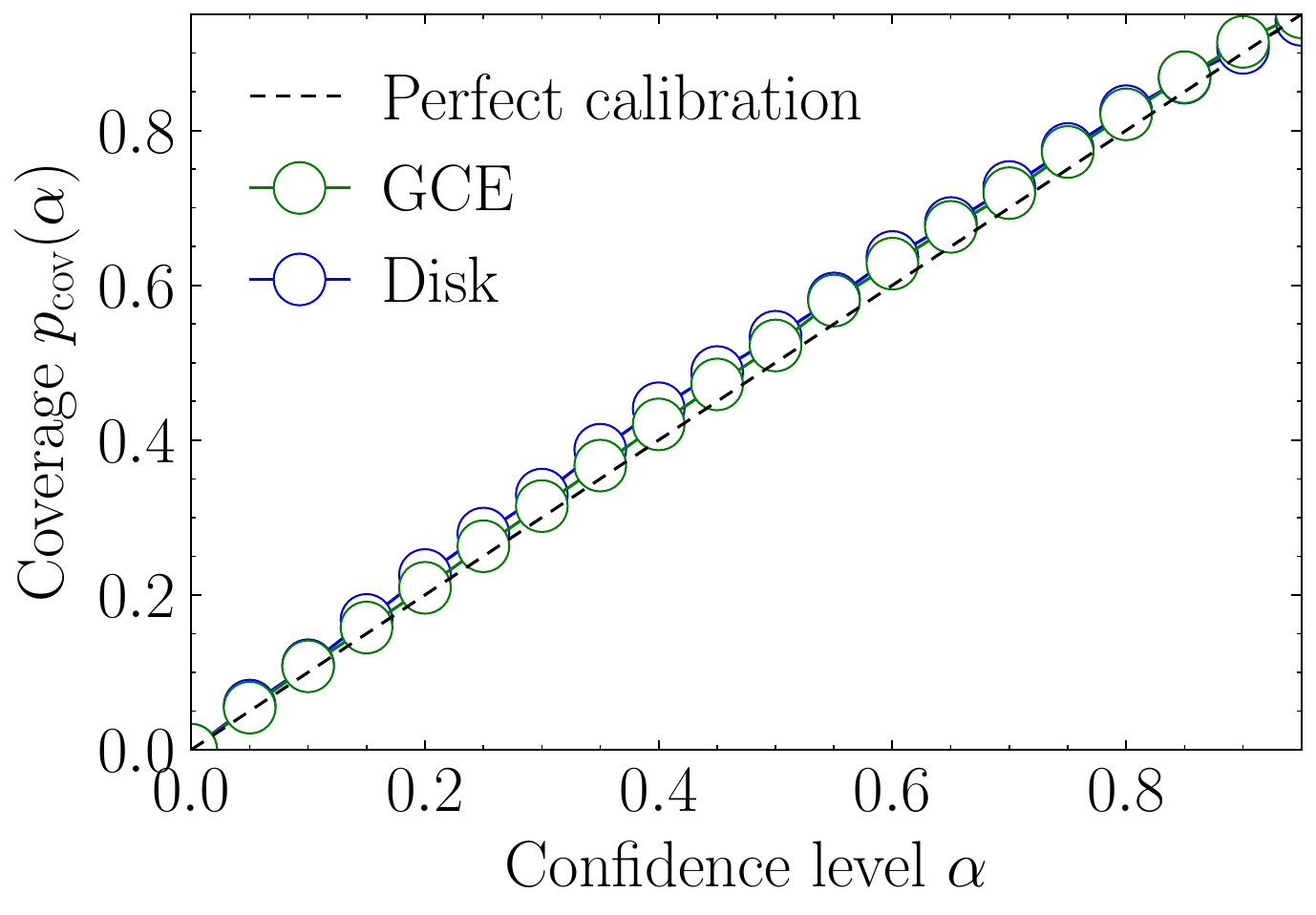}\hspace{0.2cm}
\includegraphics[width=.32\linewidth]{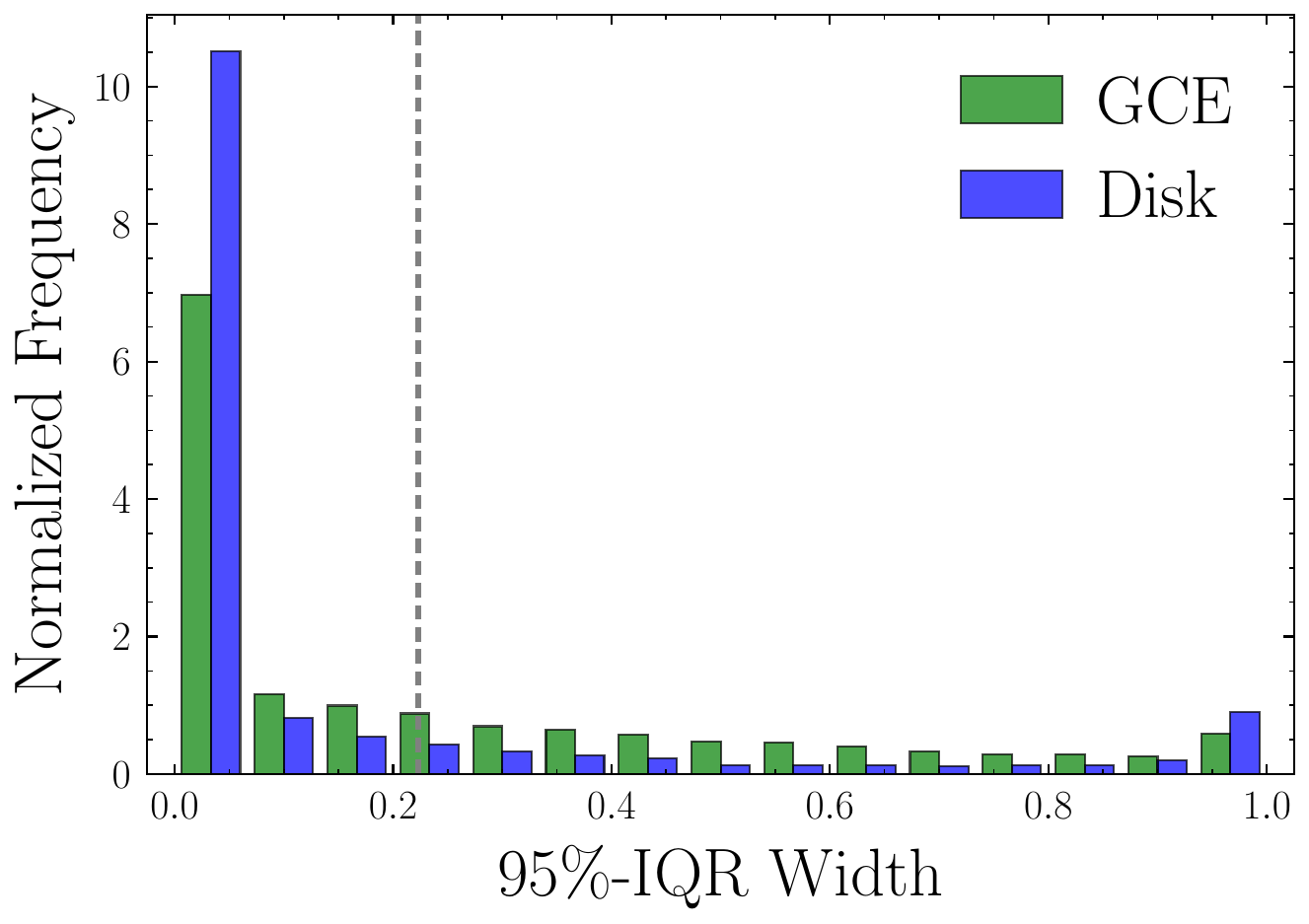}\hspace{0.2cm}
\includegraphics[width=.32\linewidth]{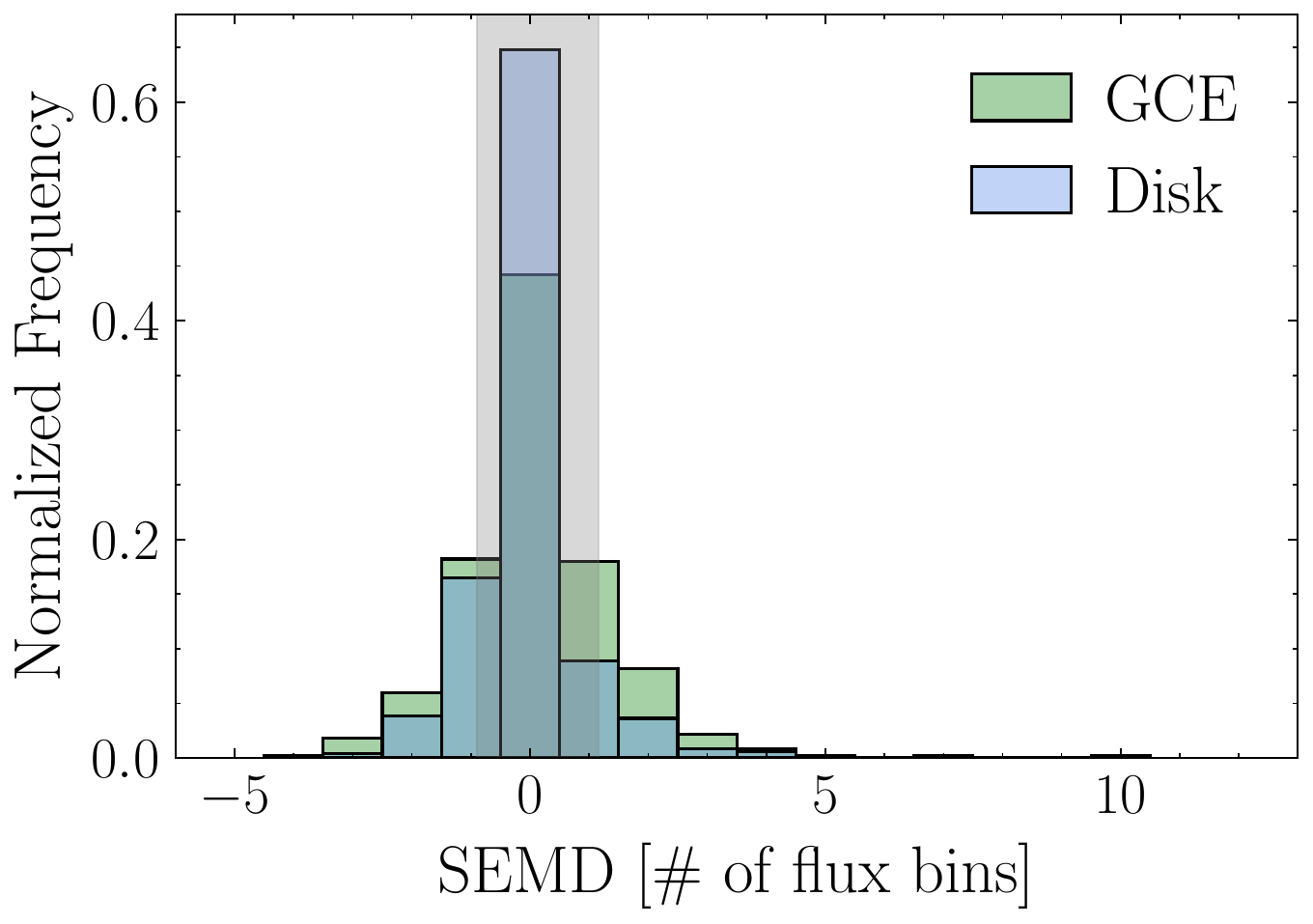}
\vspace{-0.2cm}
\caption{Performance of the SCD inference for our default CNN on 500 test maps.
On the left we show the error bars for both SCDs are well calibrated; for example, the truth value falls between the 25-75\% quantiles for roughly 50\% of the error bands produced over all test maps.
(The hollow circles are simply markers for presentation such that the underlying perfect calibration curve can be seen.)
The middle figure depicts the sharpness of our error bands, or in more detail the 95\% inter-quantile range; as long as they are well calibrated, we would like the method to return the smallest error band it can (indicating more confident predictions).
For the GCE, the median IQR is shown as a vertical dashed line.
On the right we show the signed Earth mover's distance (SEMD), measured as the number of log-flux bins one needs to move the true SCD into the median prediction.
A negative value corresponds to an overly dim prediction and similarly positive implies overly bright.
The gray band depicts the [16\%, 84\%] quantile range for the GCE.
The outliers at a large number of flux bins are a mix of both the GCE and thin disk.
Further details regarding these metrics are provided in the text.}
\label{fig:sharpcalibration_hist}
\end{figure*}

Beyond examples, it is also important to quantify the performance of the network across the ensemble of test maps.
Results studying this are shown in Fig.~\ref{fig:sharpcalibration_hist}.
In particular, we show three measures of performance.
The first of these is calibration.
For a given confidence level $\alpha \in [0,\,1]$, the calibration plot shows how often the true value falls within $[(1-\alpha)/2,\,(1+\alpha)/2]$ of the median CNN prediction.
Flux bins with truth value below the threshold value, $\epsilon =10^{-5}$ or above $1-\epsilon$ in CDF space are excluded from the analysis.
This prevents our calibration metric from being skewed by numerical inaccuracies well below the level of affecting physical results, and by biases introduced by the tight clustering of all the quantiles around 0 or 1. 
We have checked that the results are stable to variation of the threshold value.
If the results were perfectly calibrated, then the coverage -- fraction of the maps falling within the appropriate range -- would satisfy $p_{\rm cov}(\alpha) = \alpha$.
On the other hand, $p_{\rm cov}(\alpha) < \alpha$ would indicate a tendency towards overconfidence (the predicted error bands are too small) whilst $p_{\rm cov}(\alpha) > \alpha$ implies underconfidence.
The calibration is computed in aggregate over all 26 flux bins in the SCD and 500 test maps.
The results show that the errors appear to be well calibrated.
(Note that for the SCD we only estimate the quantiles between 2.5\% and 97.5\%, which is why we only plot the calibration up to $\alpha=0.95$.)

As we review in the following section, well calibrated error bands need not be useful.
To be useful, one also wants precise predictions and therefore error bands in the CDF space that are as small as possible, whilst remaining calibrated.
Sharpness quantifies this intuition, and in the middle of Fig.~\ref{fig:sharpcalibration_hist} we show the 95\% inter-quantile range (IQR), which is the distance between the 2.5\% and 97.5\% quantile.
Again this distribution is computed over the full set of bins and maps.
In general, we see that the range peaks towards smaller values, although in general predictions for the disk tend to be sharper than those for the GCE. 
We can attribute this to two factors.
Firstly, referring back to Tab.~\ref{tab:priors}, the test data for the GCE includes dimmer maps (where the exact distribution becomes harder to infer) than for the disk.
Secondly, the bimodality of the GCE distribution can make it harder to infer, and the CNN can smooth over the two peaks (see Fig.~\ref{fig:exCDFs} bottom center).

Lastly, beyond ensuring predictions are accompanied by reliable error bands, it is also useful to quantify how reliable the median SCD predictions are.
For this, in the right panel of Fig.~\ref{fig:sharpcalibration_hist} we consider a third performance criterion: the signed Earth mover's distance (SEMD), which we define as
\begin{equation}
\operatorname{SEMD}(\bm{y},\hat{\bm{y}}) = \sum_i (Y_i - \hat{Y}_i),
\end{equation}
where $\bm{Y} = \{Y_i\}$ and $\hat{\bm{Y}} = \{\hat{Y}_i\}$ are the true and predicted binned CDFs associated with the binned PDFs $\bm{y}$ and $\hat{\bm{y}}$, respectively.
This is a measure of how much one needs to adjust the predicted median histogram in order to adjust it to align with the truth.
The magnitude of the SEMD measures how much flux needs to be moved across the log-spaced flux bins in order for the distributions to agree, and the sign is determined such that a negative value indicates that the prediction was too dim, whereas a positive value indicates the prediction was too bright.
We note that alternative definitions of a ``signed'' EMD are also possible. In particular, one may instead retain the standard (unsigned) EMD -- given by the sum of \textit{absolute} CDF differences -- and subsequently assign a sign based on the direction of the median shift relative to the truth~\cite{stossi2024space}.
In contrast, our definition introduces the sign intrinsically at the level of the transport cost by omitting the absolute value inside the sum over CDF differences.
As a consequence, our signed EMD directly measures the net signed displacement of probability mass, allowing positive and negative contributions to cancel, for which reason a SEMD of zero does not imply the distributions are identical.
Further, our SEMD is a comparison of normalized distributions, it does not measure any difference in total flux.

\begin{figure*}[!t]
\centering
\includegraphics[width=.45\linewidth]{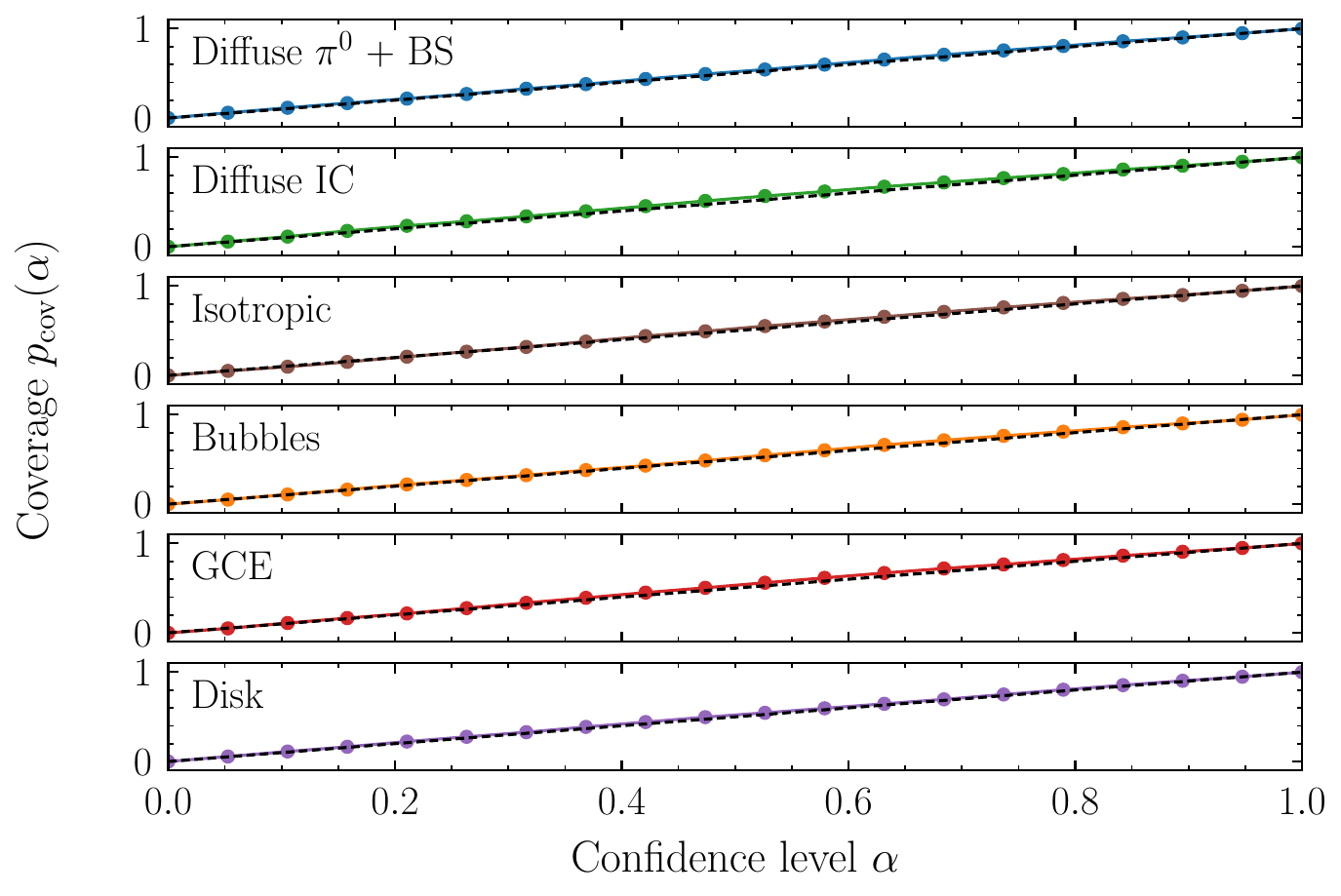}\hspace{0.5cm}
\includegraphics[width=.45\linewidth]{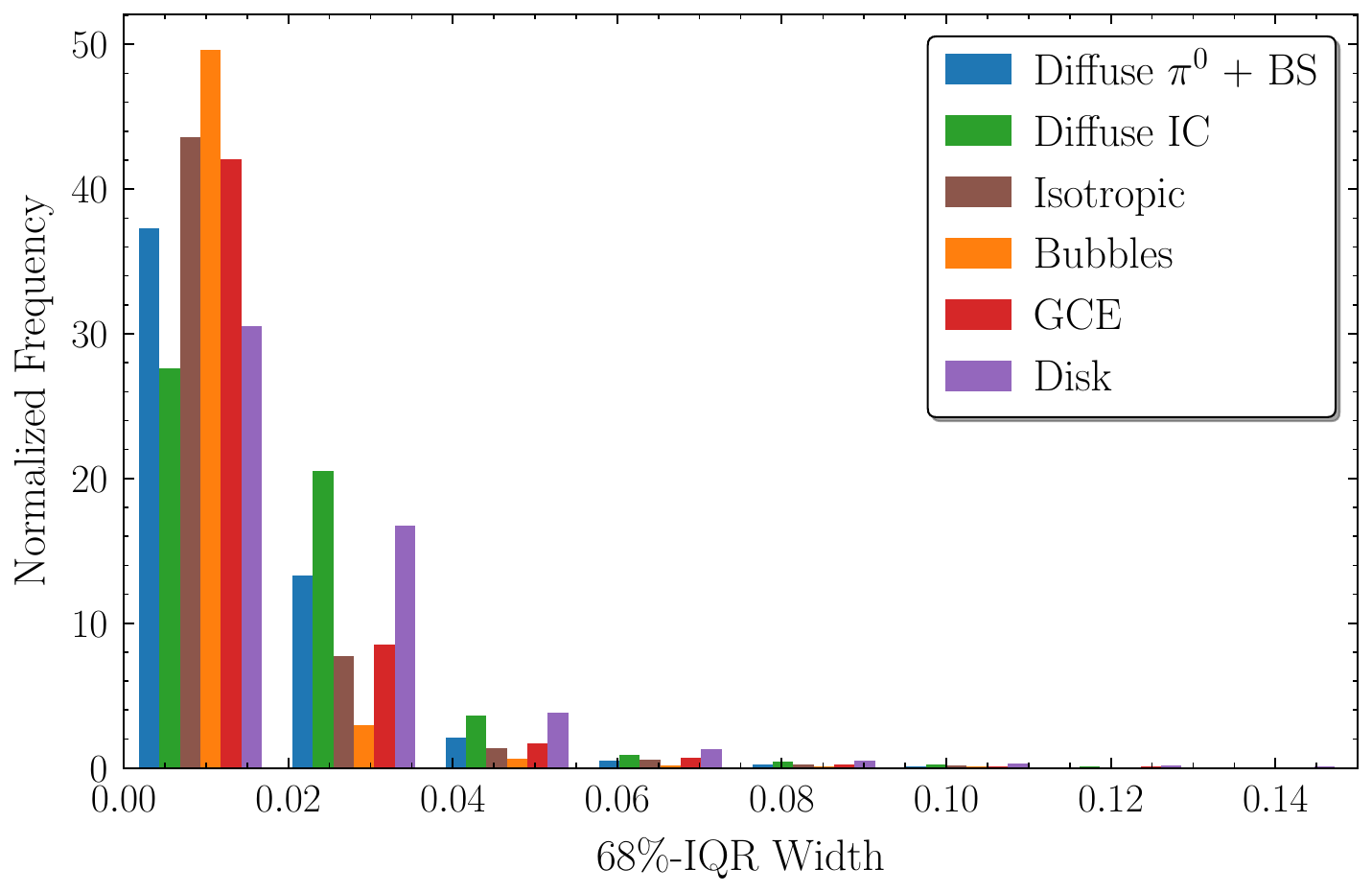}
\vspace{-0.2cm}
\caption{The calibration (left) and sharpness (right) of the spectral predictions made by the first CNN.
The values on the $x$-axis of the sharpness plot are in units of flux fraction.
Good performance is observed across the board: this is generally a far simpler task for the CNN than the inference of the SCD.}
\vspace{-0.5cm}
\label{fig:spectral-sharpcalibration}
\end{figure*}

The results in Fig.~\ref{fig:sharpcalibration_hist} show that for the disk and GCE the SEMD is peaked near zero, although there are isolated outliers in both cases.
To provide context for the numerical value of the SEMD, we note that the (unsigned) Earth mover's distance between the energy dependent and independent predictions for the GCE SCD in Fig.~\ref{fig:Results} is $\simeq 3.3$.

So far we have discussed exclusively the validation of SCD reconstruction, which is the task of our second CNN.
We can similarly study the performance of the first CNN whose task is to infer the bin-by-bin flux fractions in energy for each template.
The performance here is very similar to that of the energy independent CNN, which was studied extensively in Refs.~\cite{List:2020mzd,List:2021aer}, and we refer there for further details.
Nevertheless, in Fig.~\ref{fig:spectral-sharpcalibration} we show the calibration and sharpness for the flux fraction predictions (where each energy bin is treated as a separate data point in the calculation of calibration and sharpness).
For the spectra, although the network returns Gaussian errors, given a standard deviation we can compute the expected coverage values for each $\alpha$.
Given these are Gaussian errors, however, in this case we choose to show sharpness as the 68\% interquantile range.
We observe that all results are well calibrated with relatively small interquantile ranges.
One comment we make is that the IC and thin disk have the least sharp results.
We attribute this to the spatial similarity between their templates: a degree of misattribution of flux between these two templates should be expected (see the negative correlation between these two templates with a neural network trained to predict a full covariance matrix in \cite[Fig.~S20]{List:2020mzd}).

\subsection{On the Importance of Sharpness}

\begin{figure*}[!b]
\centering
\includegraphics[width=0.45\linewidth]{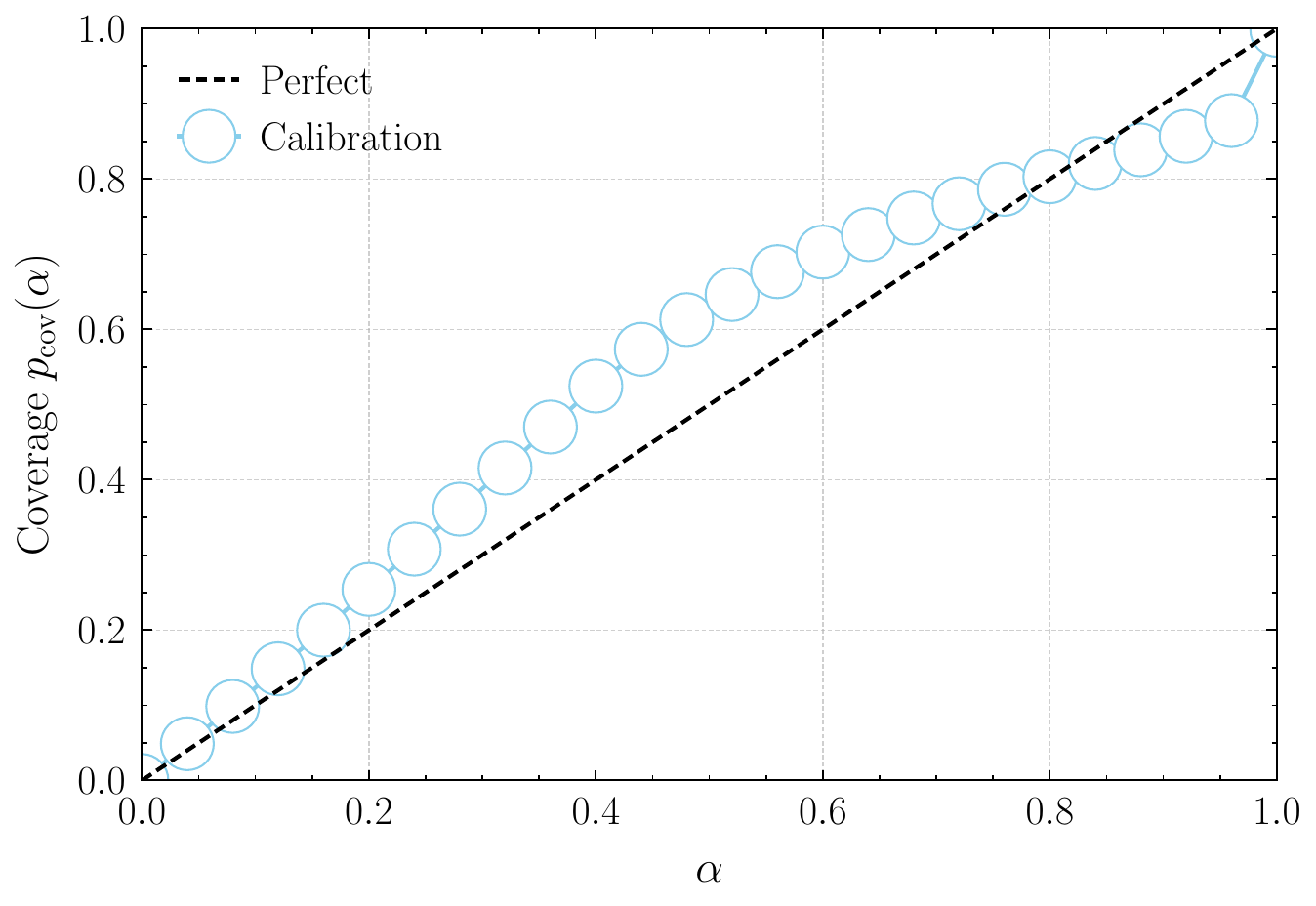}
\hspace{.5cm}
\includegraphics[width=0.45\linewidth]{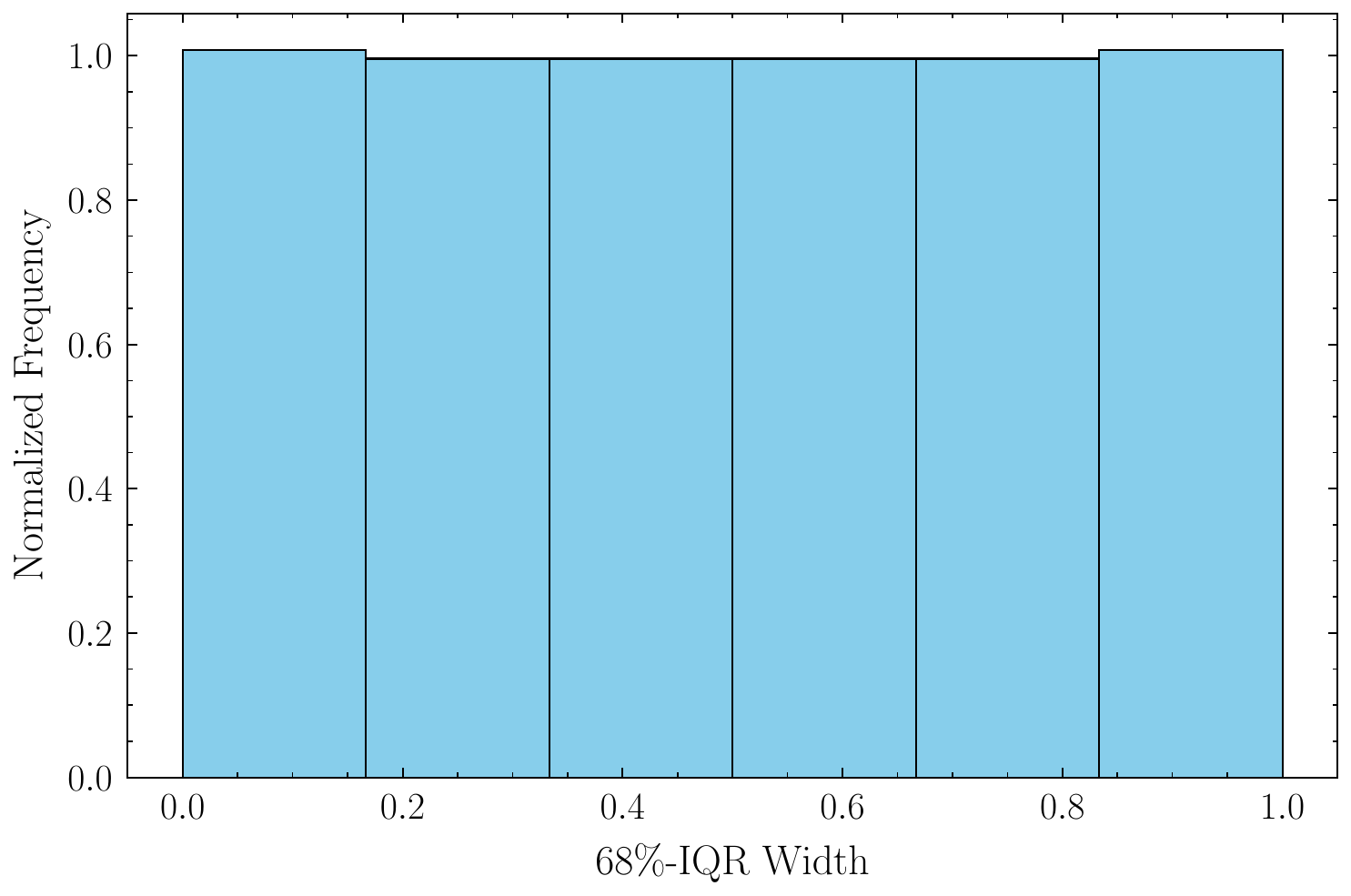}
\vspace{-0.2cm}
\caption{A simple toy example showing why sharpness is required in coordination with calibration to diagnose the reliability of error bands.
These results assess the performance of a strategy for guessing a random number drawn from $U([0,1])$ of guessing a mean value of 0.5 and then attributing an error band of width randomly drawn between 0 and 1.
The resulting predictions are reasonably well calibrated as shown on the left, but as can be seen on the right the uniform sharpness distribution reveals this strategy as not particularly informative.}
\vspace{-0.5cm}
\label{fig:sharpnessexample}
\end{figure*}

Sharpness and calibration are useful criteria for assessing the performance of machine learning methods which predict not only values but error bars.
A more detailed discussion of both can be found in, for example, Ref.~\cite{List:2021aer}.
For the moment we recall that calibration dictates how often the truth falls within the method's error bars, while sharpness is a measure of the width of the predictions.
As mentioned in the previous section, both requirements are necessary to assess whether the error bands one has are useful or not.
Specifically, one should not just rely on calibration to assess the network performance as the method could appear to be well calibrated by learning only global properties of the ensemble of datasets whilst ignoring any aspect of the particular dataset it is assessing.
In general, as sharper predictions have smaller error bars it is an overall measure of the specificity of predictions.

In this section we provide a simple example to drive this point home.
Calibration and sharpness are criteria we can apply to assess any set of inferred error bars, a neural network need not be involved.
In particular, consider the task of trying to guess a value that will be drawn from the uniform distribution $U([0,1])$.
In this example one can only rely on global properties of the dataset and one strategy would be to guess a value of 0.5 every time.
For the error bars, although it is not physically meaningful, to ensure calibration we can draw the width of the error band from a uniform distribution $U([0,1])$, and then as shown on the left of Fig.~\ref{fig:sharpnessexample} the errors would be reasonably well calibrated.
This does not mean they are useful, as the sharpness distribution emphasizes.
Its complete uniformity shows that the error bars that are totally uninformative---arbitrarily precise predictions are just as likely as complete uncertainty.
This simple example highlights that sharpness provides key information beyond calibration.

For completeness, let us remark that stronger notions of calibration than the one considered herein -- also known as ``average calibration'' -- exist, such as individual calibration or (adversarial) group calibration~\cite{zhao2020individual}.
These measures require a predictor to be calibrated when conditioned on groups of inputs or even every individual input (although verifying the latter is typically not feasible).
For instance, while we confirm the calibration of the predictions in Figs.~\ref{fig:sharpcalibration_hist} and \ref{fig:spectral-sharpcalibration} template by template, this does not guarantee that the neural network predictions are well calibrated individually for each flux and energy bin or when evaluating the neural network on selected subsets of the data distribution, as different miscalibrations could average out.

In our experiments, we observed that the predicted spectra for the diffuse backgrounds and the disk can become somewhat miscalibrated when restricting the evaluation to narrow regions in parameter space -- for instance, realizations for our best-fit Fermi parameters. 
This suggests that expanding the coverage of our training data across the relatively broad parameter space allowed by our priors could further enhance the performance of our method.
As an alternative to generating ever more training data, sequential inference techniques (for example, Refs.~\cite{papamakarios2019sequential, anau2024scalable}) offer a promising approach by focusing data generation on regions of high posterior density during training.
We plan to explore such extensions in future work.

\subsection{Hypothesis Testing in the Presence of Dim Sources}
\label{ssec:nn3}

\begin{figure*}[!b]
\centering
\includegraphics[width=0.45\textwidth]{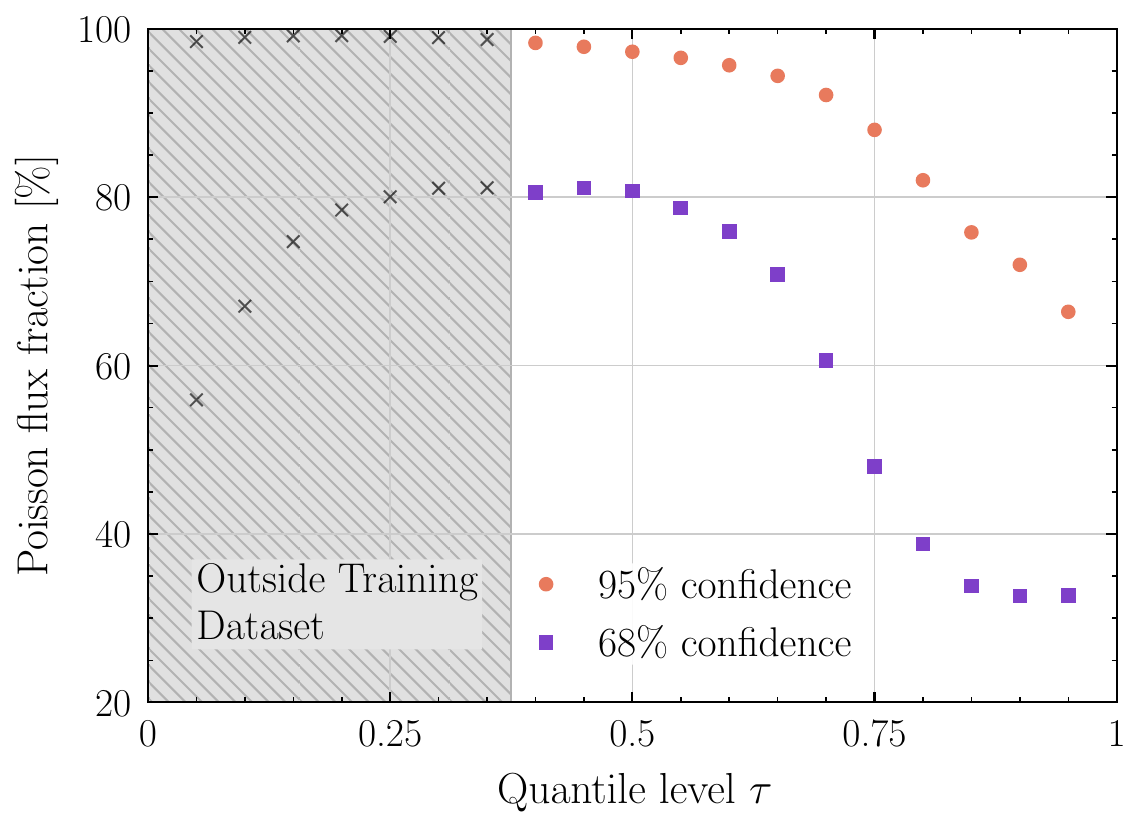}
\vspace{-0.2cm}
\caption{We plot predictions from NN3 for the fraction of the Fermi flux with a Poisson origin.
Specifically, we show two curves: the 95\% and 70\% confidence predictions of NN3.
These predictions are made by feeding NN3 the SCD quantiles predicted for the Fermi data, treating each quantile as an independent input.
The hatched region marks where the input SCD lies outside the 99th percentile of the training set, indicating extrapolation beyond the training domain.
The predictions for quantiles $\tau\le0.35$ are dimmer than 99\% of the training data.
If we focus on median SCD ($\tau=0.5$), we can see that at 95\% confidence the predicted Poisson flux fraction is $\simeq 97\%$---very little Poisson emission can be genuinely excluded.
}
\vspace{-0.5cm}
\label{fig:NN3_median}
\end{figure*}

As discussed in the main text, our inferred GCE SCD is peaked noticeably below the 1 photon threshold even when plotted as $F^2dN/dF$, which tends to shift the distribution to higher flux values. 
Given that this is the range where we expect purely Poisson emission to be reconstructed (cf. Fig.~\ref{fig:poissonGCE}), one can ask to what extent we can exclude a fraction of the observed flux as being associated with Poisson emission.
Note that this is a well posed question: sufficiently bright sources are inconsistent with the Poisson hypothesis, so we can always look to exclude some fraction of Poisson emission given an SCD.
(The alternative question of excluding the point-source hypothesis as a description for an observed flux is not well posed as we can always make the sources arbitrarily dim---in a Bayesian analysis, the results will then be entirely prior driven, and the unequal prior spaces for the Poisson vs. point-source parameters might lead to unintended bias \cite{Collin:2021ufc}.)

In order to address this question we follow the procedure outlined in Ref.~\cite{List:2021aer}: we train a third neutral network (NN3) for the task.
Specifically, we use a simple feed-forward neural network trained to predict the percentage of the flux associated with a given SCD that can be attributed to Poisson emission.
The SCDs are generated from simulated Fermi maps where the GCE emission corresponds to a fraction arising from exactly Poisson emission and the remainder being generated by point sources drawn from the same set of priors in Tab.~\ref{tab:priors}.
In particular, across the maps the GCE total flux remains roughly constant, but the amount arising from Poisson or point sources varies.
Importantly, we ensure that there is an almost uniform distribution in Poisson versus point sources across the training datasets for NN3, as this implicitly sets the prior for the network predictions.
Further, there are training maps where the point source emission is so dim that it becomes indistinguishable from Poisson: this degeneracy is a reality the neural network has to account for in its predictions.
Once these maps are generated, NN3 is trained by passing these maps through our default CNN framework to obtain an SCD, the median prediction is passed as an input to NN3 along with the true Poisson flux fraction value.
NN3 is then trained to predict the fraction of the emission which is Poisson.
Given the procedure, it is important that the earlier networks are trained on bimodal SCDs, which can account for a simultaneous contribution from Poisson emission and point sources.
Similarly as for the $dN/dS$, we do not simply infer a point estimate for the Poisson fraction; rather, we recover the full posterior in terms of quantiles.
This allows us to obtain upper bounds on the Poisson fraction associated with an input $dN/dS$ conditional on any given confidence level. Regarding the technical implementation of this approach, we train NN3 using the pinball loss~\cite{Steinwart2011}, with quantile (i.e.\ confidence) levels uniformly drawn during training.

In Fig.~\ref{fig:NN3_median} we show the results of NN3 when passed the SCD we infer for the GCE on the actual Fermi data for confidence levels 70\% and 95\%.
We pass NN3 all SCD-quantiles $\tau$ in order to analyze the degree of consistency across a range of SCDs deemed plausible by the CNN, however, the hatched region corresponds to distributions that are dimmer than 99\% of the training data and therefore should be disregarded as it is in a regime where NN3 is forced to extrapolate.
This is because NN3 was trained on the median predictions of the SCD network and, although we can of course evaluate NN3 at test time on any SCD, the predictions for the Fermi map for small $\tau$ become even fainter than the typical median SCD predictions for a genuinely Poisson GCE, thus rendering the NN3 constraints unreliable.
For each quantile of the SCD, NN3 can predict what is the associated maximum fraction of the flux that has a Poisson origin at a given confidence value $\alpha$.
Focusing on the median prediction, $\tau=0.5$, at $\alpha=$95\% confidence the network can only exclude 3\% of the emission as having a Poisson origin.
For comparison, in Ref.~\cite{List:2021aer}, where an SCD consistent with that of the energy independent CNN shown in Fig.~\ref{fig:Results} was inferred, the equivalent value was 34\%, indicating a preference for a far greater contribution from point sources bright enough to be distinguished from Poisson emission.

We emphasize that although our inferred SCD for the GCE would be consistent with Poisson emission, scenarios relying on a large number of faint astrophysical sources have been proposed in literature, which might equally be compatible with our results.
In particular, very faint SCDs would not be in tension with the non-observation of a large number of point sources in the Galactic bulge to date (see Ref.~\cite{Dinsmore:2021nip} for a comprehensive study of the allowed parameter space for luminosity functions).
For instance, millisecond pulsar formation through accretion-induced collapse might give rise to $O(10^5)$ sources that could explain the GCE~\cite{Gautam2021}.
Reference~\cite{Ploeg2020} finds a preference for 20,000 - 50,000 millisecond pulsars in the bulge at 68\% confidence and argues that a common evolutionary trajectory for the disk and bulge millisecond pulsars is physically viable.
This range of bulge point sources would be consistent with our uncertainty regions.

\section{Template Variations}
\label{sec:templatevariations}

The existence of the GCE is a visceral reminder that there remain significant gaps in our ability to understand and model the $\gamma$-ray sky.
As it pertains to the results presented in the main text, this motivates the question of how our conclusions would vary with changes in the assumed spatial templates.
We study this in the present section, by considering four variations to our analysis.
Firstly, we show results obtained if we entirely retrain our network using a different diffuse emission model.
Secondly, in a purely simulated environment, we study the stability of our CNN when it is shown data from a background model it was not trained on and in particular compare the performance of the energy dependent and independent networks.
Third, we demonstrate that our results are stable under variations to the GCE template that are consistent with what is observed in the data.
Fourth, we study the extent to which including a more accurate and energy dependent data simulation plays a role in the results we show in the main text.
Finally, we demonstrate that our results are not especially sensitive to the explicit Fermi data set we use by showing that on a subset of the Fermi data the results vary although are broadly consistent.

Before presenting our studies, let us state clearly the two ways energy can be included in our analysis.
As mentioned, energy is treated as an additional input channel to our neural networks and we study networks that use either one or ten input channels.
However, energy can also be used differently in terms of how we generate the simulated datasets.
For our energy dependent (10 channel) CNN, we generate the data in ten energy bins.
For our energy independent (1 channel) CNN, we either generate the data in ten bins and then sum them together or else generate the data in one bin.
Of course, the single bin generation will be less accurate as it does not account for the energy dependent variation in the Fermi PSF, amongst other effects.
In Figs.~\ref{fig:Results} and \ref{fig:SCD-Disk} the results labeled ``with energy'' correspond to 10 input channels and 10 energy bins, whereas those denoted ``no energy'' had a single input channel and bin, as this is the approach that had been used in earlier energy independent works (for example, Refs.~\cite{Mishra-Sharma:2016gis,List:2021aer}).
Studies including the case of 1 channel and 10 bins are considered in this section.

\subsection{Results for a Network Trained on Model A}
\label{sec:modA_NN}

\begin{figure*}[!b]
\centering
\includegraphics[width=0.45\textwidth]{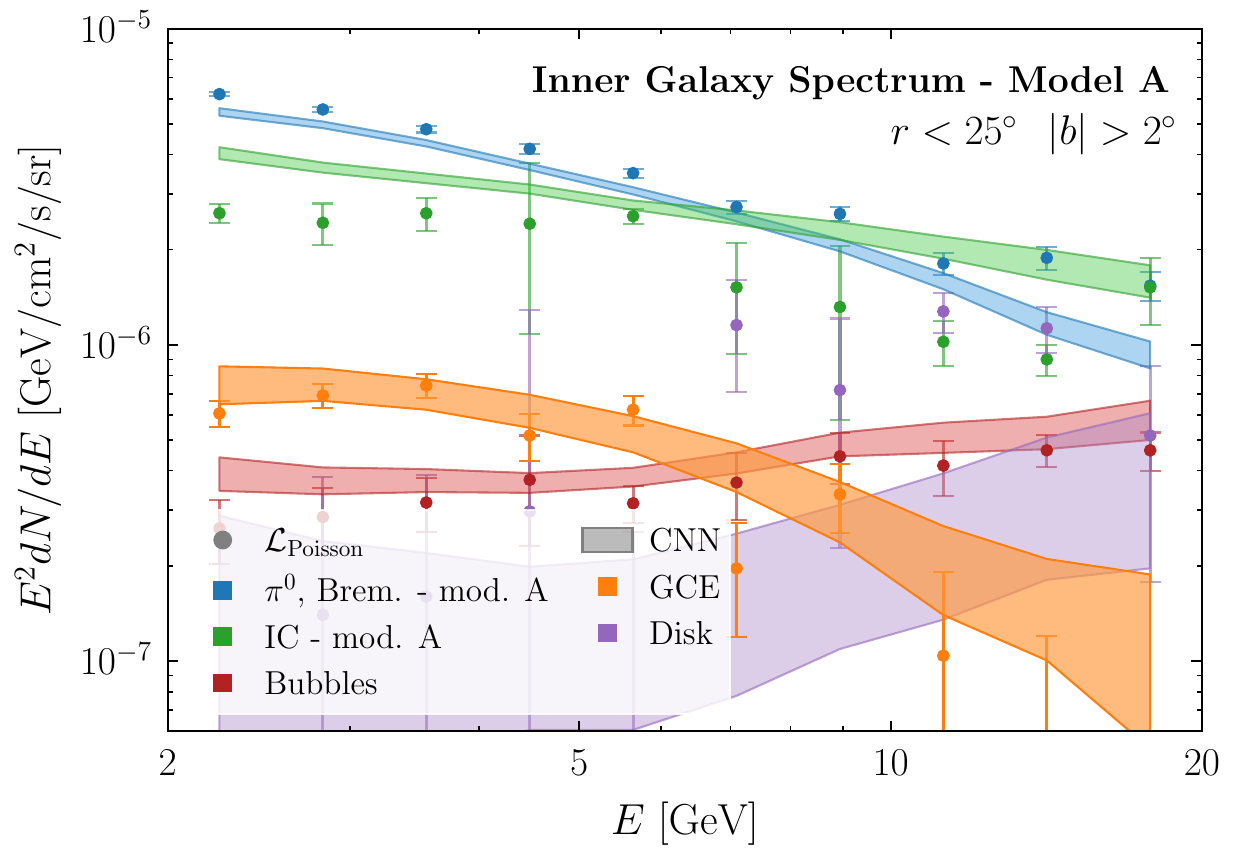}\\
\includegraphics[width=0.45\textwidth]{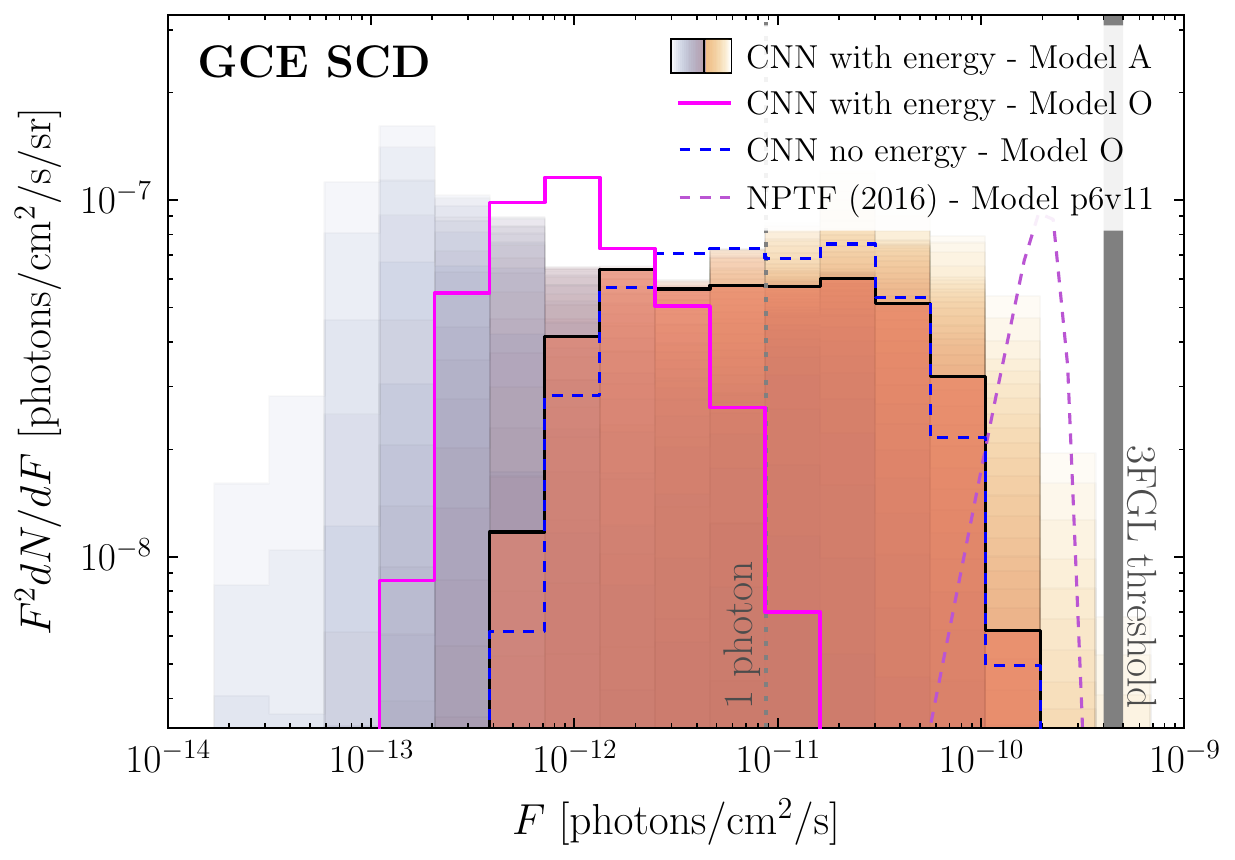}
\hspace{0.5cm}
\includegraphics[width=0.45\textwidth]{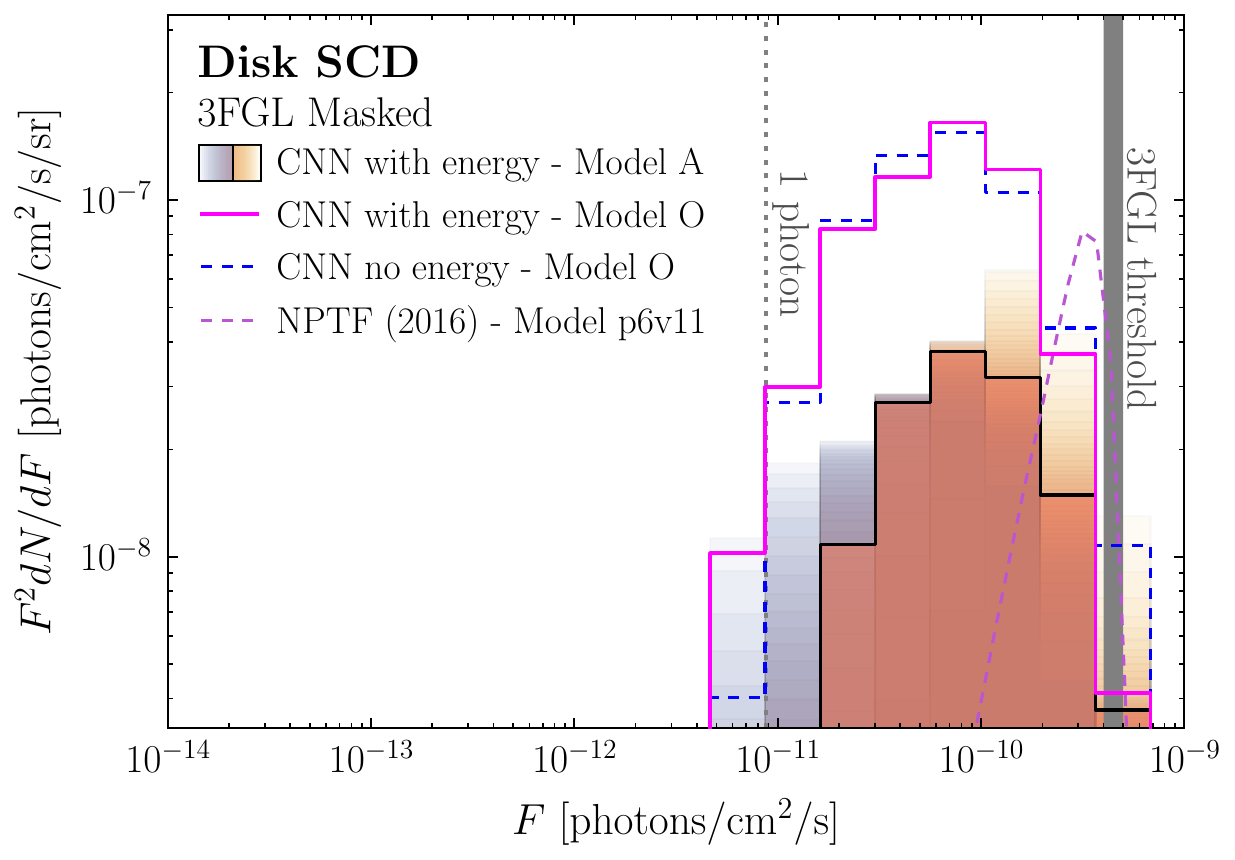}
\vspace{-0.2cm}
\caption{Equivalent results to those shown in Figs.~\ref{fig:Results} and \ref{fig:SCD-Disk} but for a network trained using the Model A diffuse emission template.
The top figure shows the energy spectrum inferred in a Poisson likelihood (points) or CNN (bands) approach, whereas on the bottom we show the GCE (left) and disk (right) SCD.
Note that the energy independent results we compare to are generated with Model O, whereas the NPTF used \texttt{p6v11} (see Sec.~\ref{ssec:NPTFCompare}).}
\vspace{-0.5cm}
\label{fig:ModAResults}
\end{figure*}

The primary systematic concern in any template fitting analysis of the Fermi sky is mismodeling of the diffuse emission.
As one method for studying robustness, we fully retrained our default CNN using a different set of templates for the $\pi^0$ and Bremsstrahlung as well as for the inverse Compton emission.
The alternative model we use is called Model A, and it was originally presented in Ref.~\cite{Calore:2014xka}, where it was shown that above 1\,GeV this provides a better fit to the diffuse emission than the official Fermi background model \texttt{p6v11}.
Nevertheless, as shown in Ref.~\cite{Buschmann:2020adf}, it is a worse fit to the Fermi data over our chosen energy range than our default Model O.
Beyond changing the diffuse model used to generate the training data, all other aspects of the CNN are left unchanged.

Equivalent results to those shown in the main text are given in Fig.~\ref{fig:ModAResults}.
For the spectrum, we note that the Poisson likelihood fit is also recomputed using the Model A background templates.
At the level of the spectrum it is clear that many components are less well constrained than for Model O, as indicated by the generally larger error bars for the Poisson likelihood fit.
There is also reduced agreement between the CNN and likelihood approaches, especially at high energies, where the methods disagree on how to divide the emission between the inverse Compton and disk components.
(There is also a notable disagreement between $\pi^0$ and IC at low energies.)
Nevertheless, the inferred GCE spectrum remains consistent between methods and also to the results determined with Model O, although the peak brightness is somewhat reduced.
We note that except in the case of the disk the error bands inferred by the CNN tend not to be larger than those for Model O.
This suggests that the network is overconfident in its predictions and we explicitly confirm that in the next section: no attempt has been made to have the CNN account for systematics in the error band it predicts; this is a point that would be an interesting extension to consider for future work.

Fig.~\ref{fig:ModAResults} further shows the GCE and disk SCDs inferred when Model A is used.
Note that in the plots we also show the comparison to the Model O energy independent results (1 input channel to the neutral network and 1 bin for the simulations) and the 2016 NPTF values.
For the disk, the shape of the SCD is consistent to that found with Model O, except for the fact the normalization is noticeably reduced, as was apparent already from the spectral plots.
We note that large variations in the disk flux fraction have previously been observed as one alters the background templates even across different methods, see for instance Ref.~\cite[Tab. S1]{List:2020mzd}.
For the GCE, the overall flux remains roughly consistent, whereas the shape changes considerably: the energy dependent Model A results are roughly consistent with the energy independent Model O findings.
Specifically, the move to Model A leads to the results CNN being almost an order of magnitude brighter, being centered around the 1 photon line.
We emphasize this remains a considerably dimmer SCD than found with earlier likelihood analyses discussed in the main text; however, such an SCD would be suggestive of an ${\cal O}(1)$ contribution to the flux from point sources.

\subsection{Performance in the Presence of a Mismodeled Background}

\begin{figure*}[!t]
\centering
\includegraphics[width=0.45\linewidth]{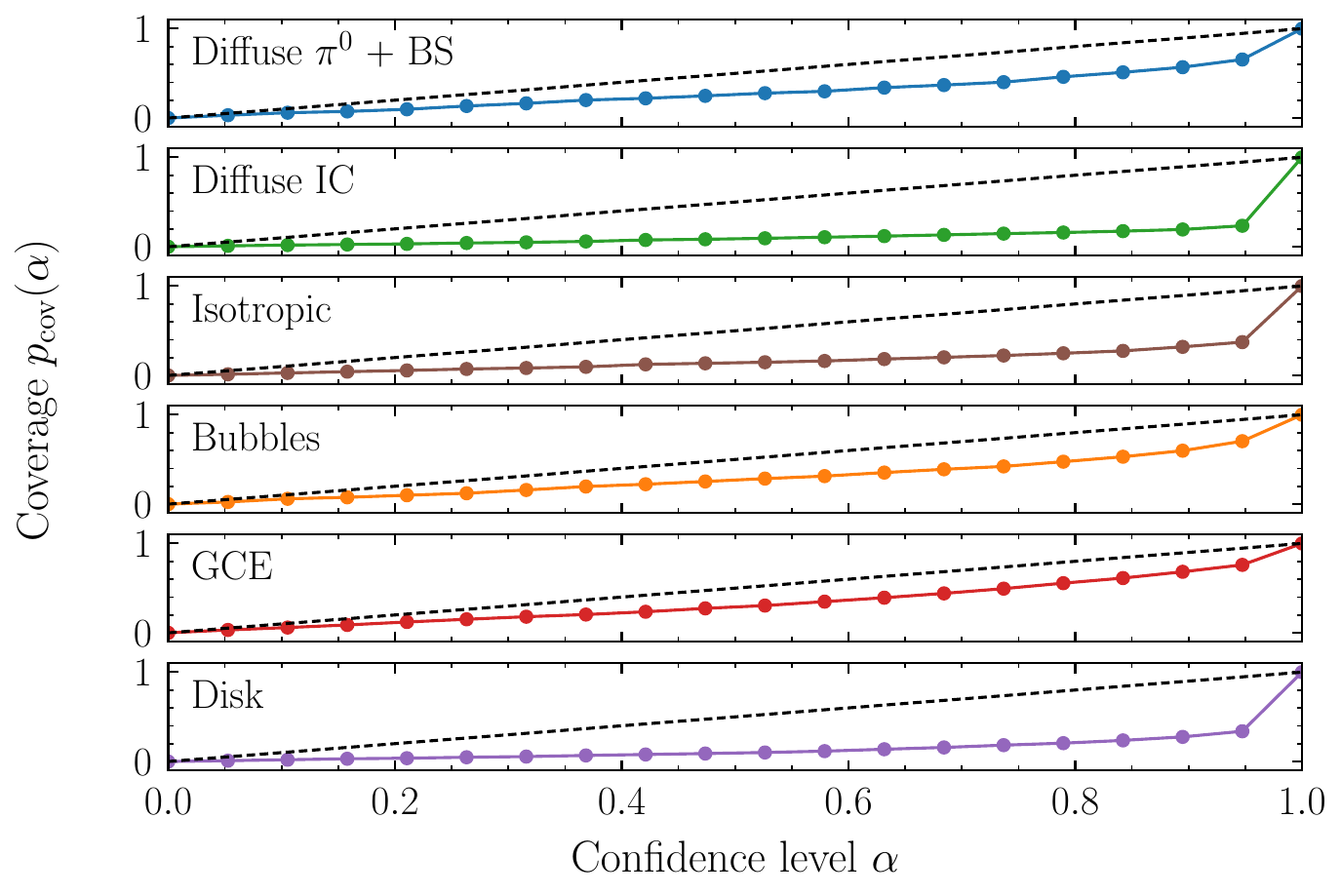}
\hspace{0.5cm}
\includegraphics[width=0.46\linewidth]{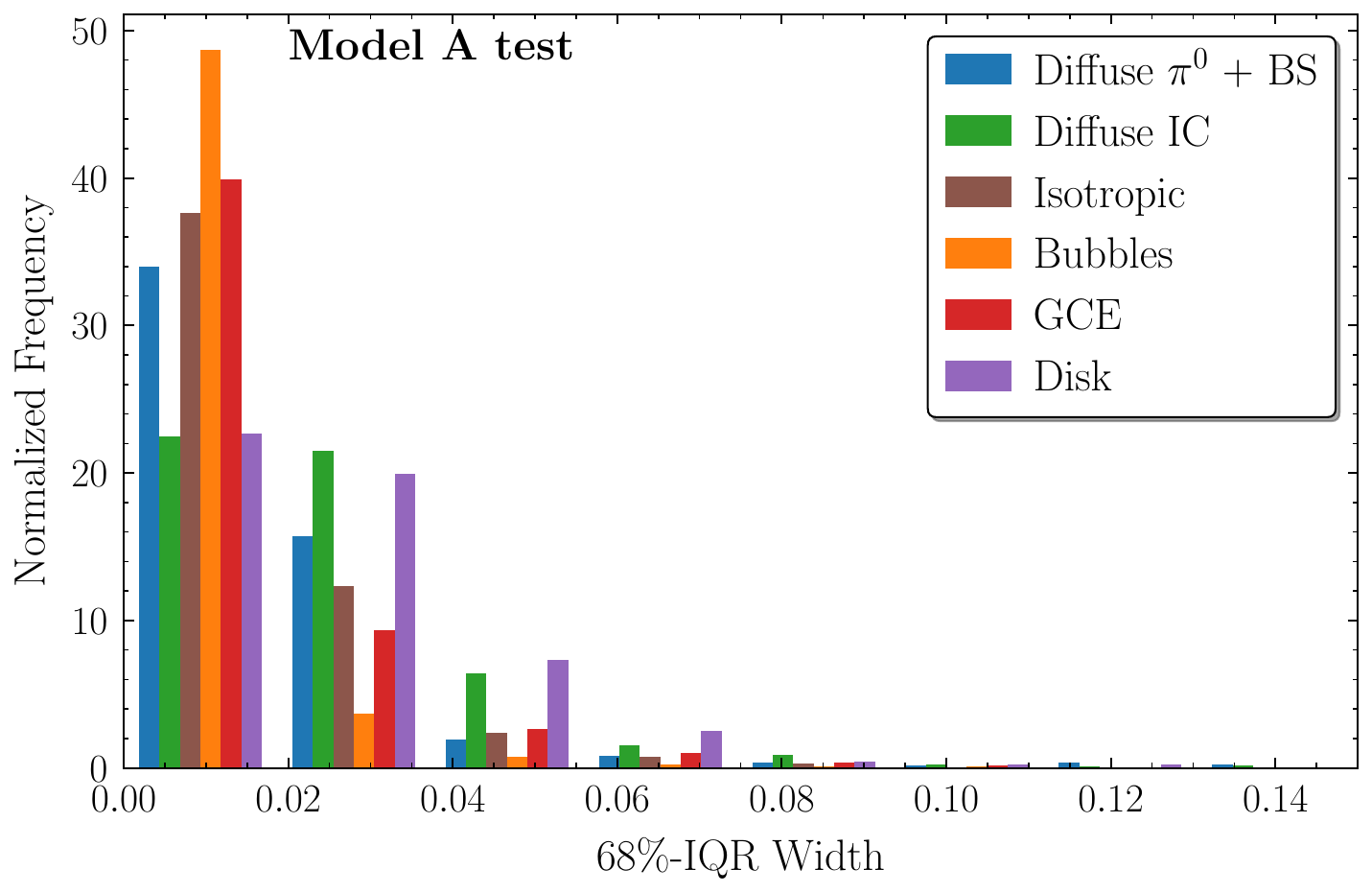}
\\ \vspace{0.05cm}
\includegraphics[width=0.45\linewidth]{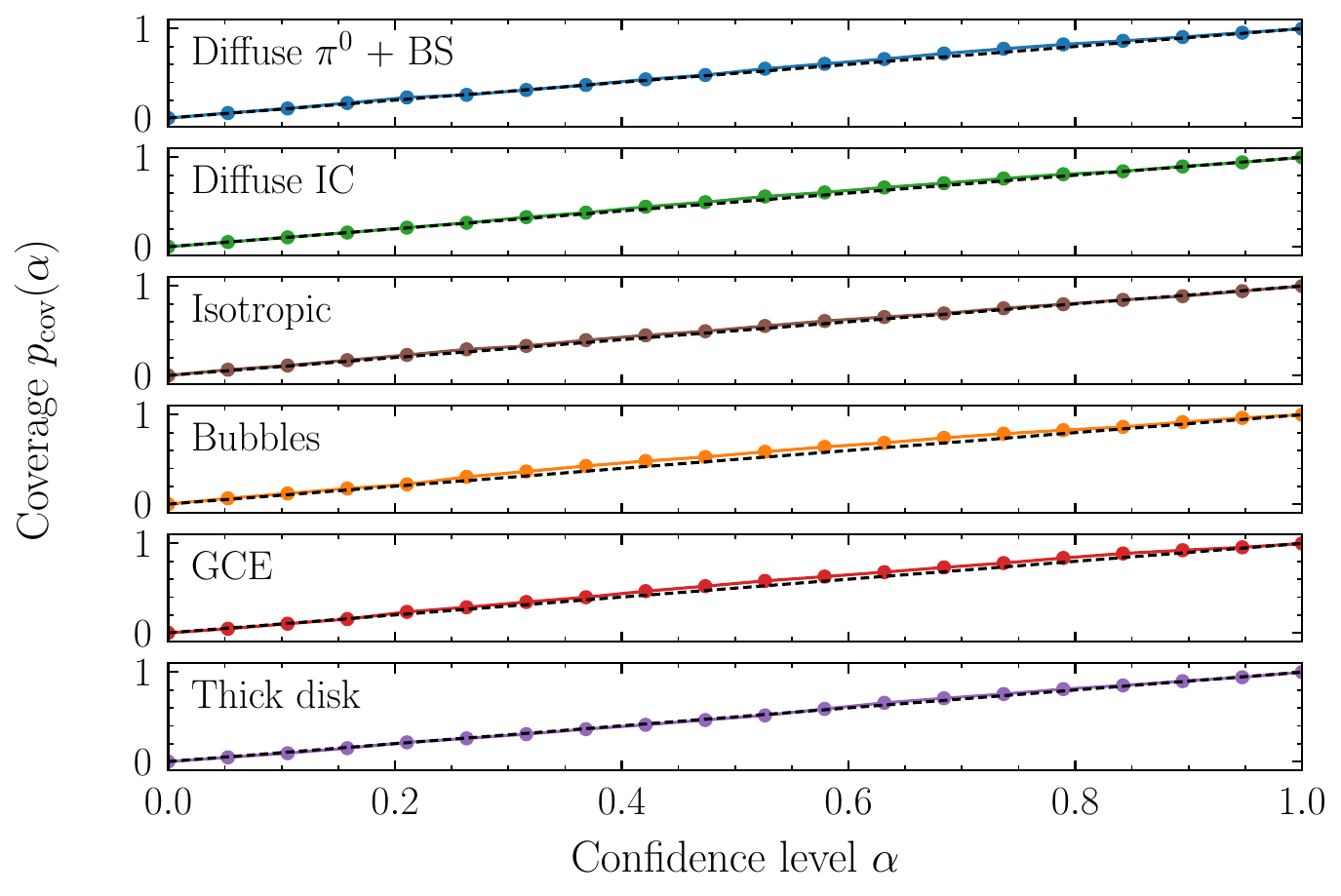}
\hspace{0.5cm}
\includegraphics[width=0.46\linewidth]{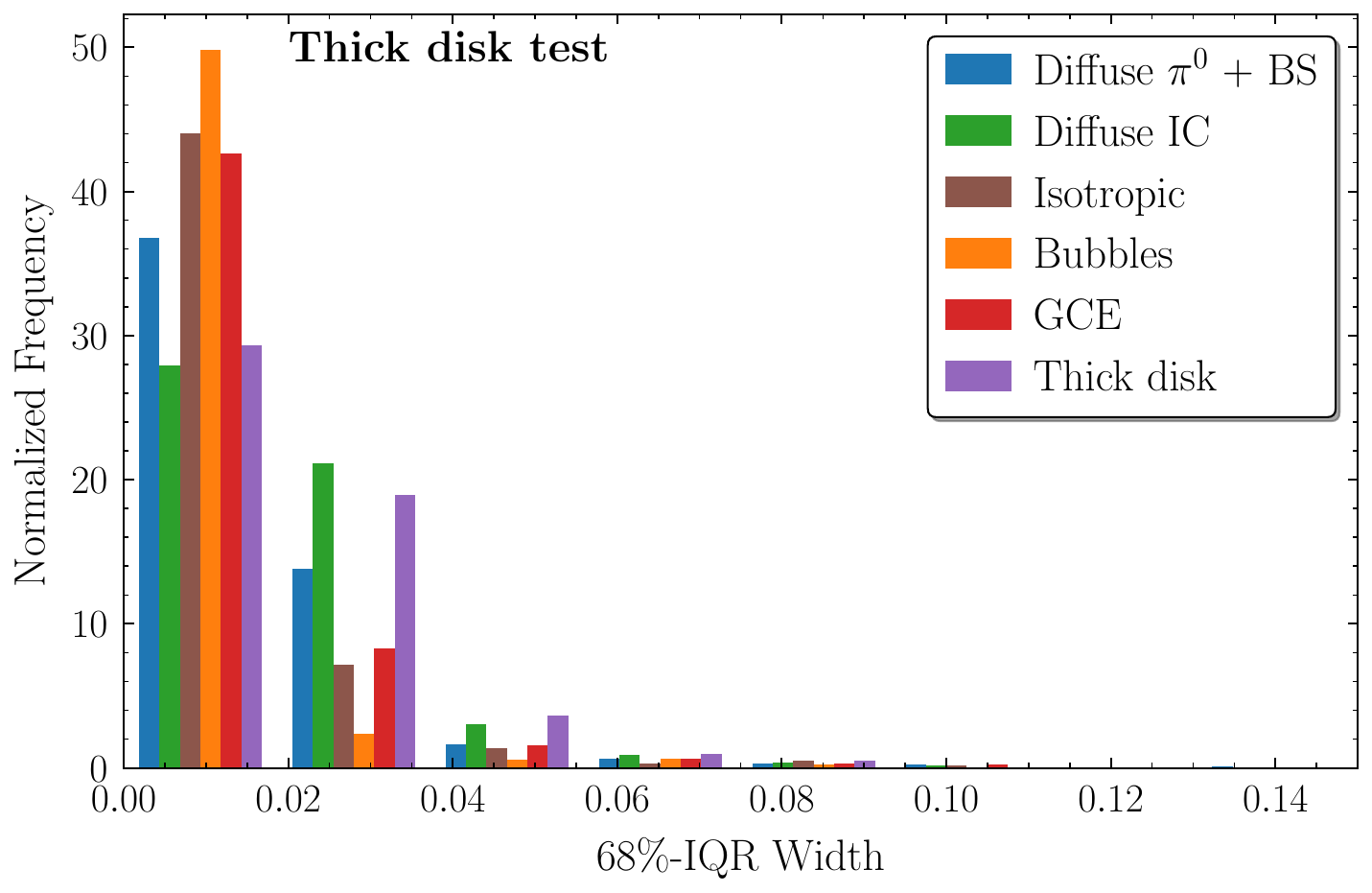}
\\ \vspace{0.05cm}
\includegraphics[width=0.45\linewidth]{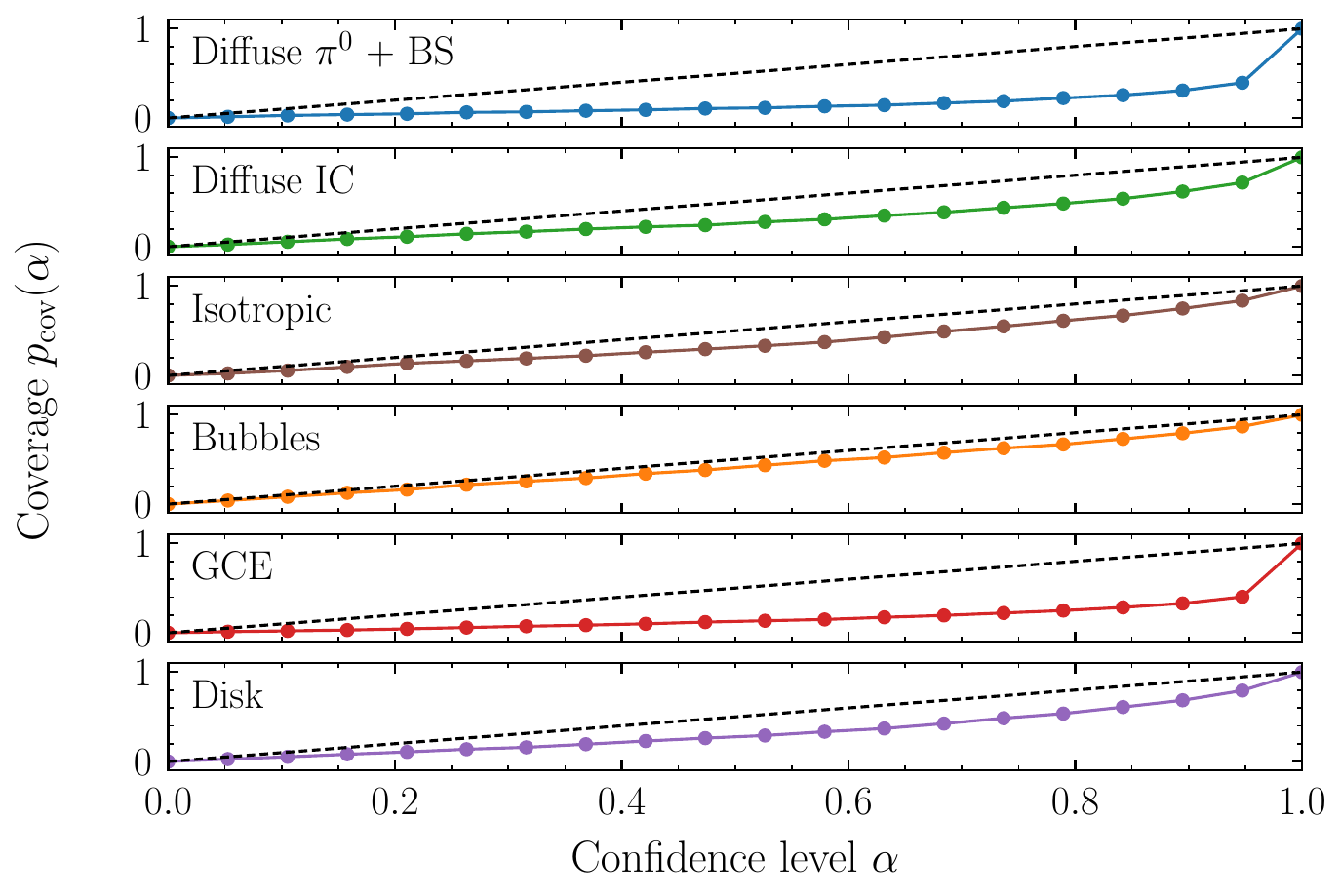}
\hspace{0.5cm}
\includegraphics[width=0.46\linewidth]{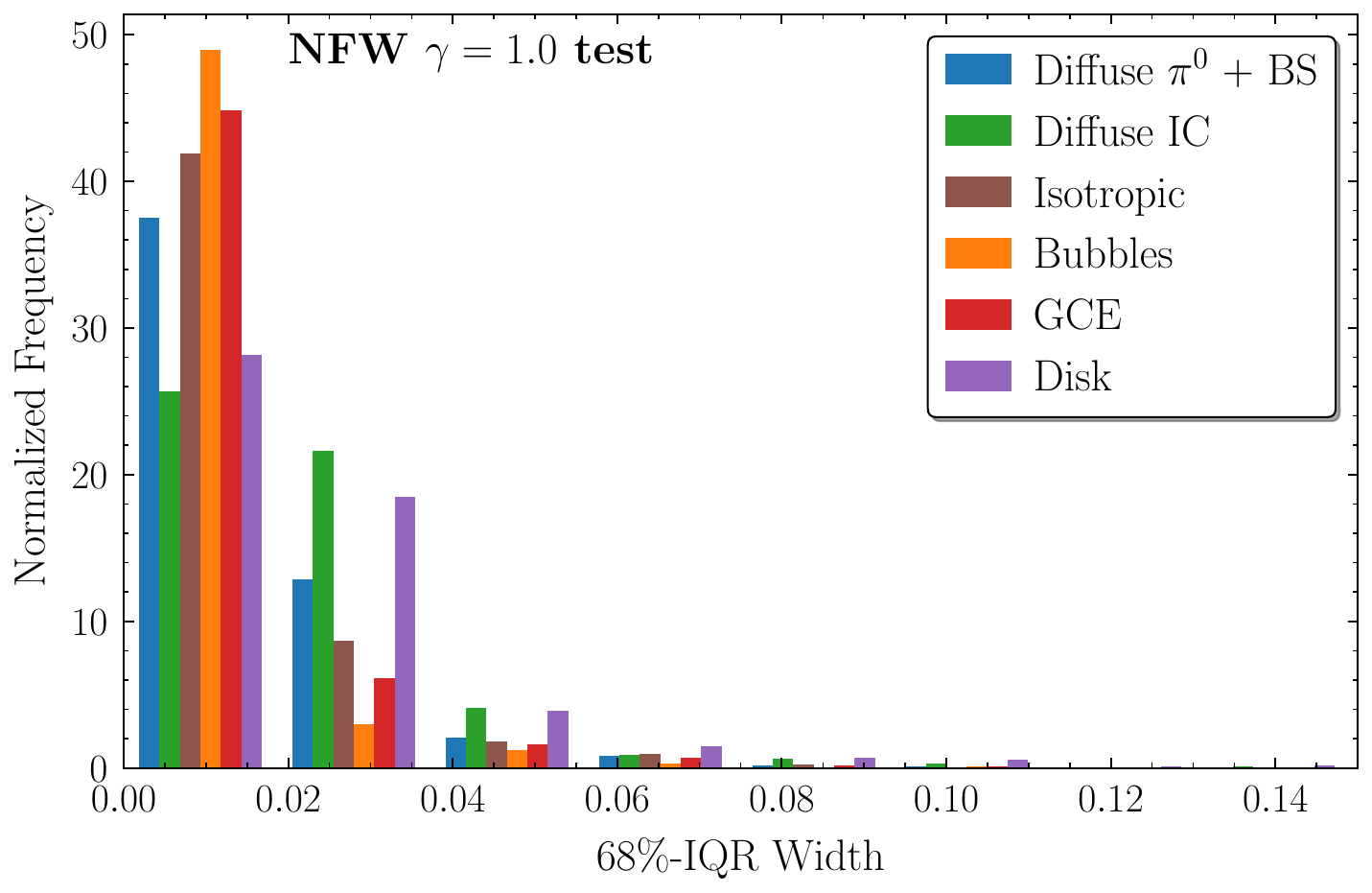}
\vspace{-0.3cm}
\caption{Performance of the energy spectrum recovery in the presence of mismodeling.
We show the calibration (left) and sharpness (right) for data generated with the Model A (top row), a thick disk (middle row), an NFW with $\gamma=1.0$ (bottom row).
The performance of Model F, not shown, is similar but slightly worse than that of Model A.
One hundred simulated maps are generated in each case and analyzed with our default Model O trained CNN.}
\vspace{-0.5cm}
\label{fig:mismodeling}
\end{figure*}

In the previous section we showed that changing the diffuse emission model impacts the Fermi conclusion.
One can point to the fact that the spectral results for Model A look less consistent or that the model is overall a worse fit to the Fermi data, but neither fact is truly definitive in weighing on which inference is closer to the true GCE SCD.
In this section we consider a second method of testing the impact of background mismodeling where a reference to the truth is genuinely possible.
In particular, we take our default CNN trained on Model O that was used to generate the predictions shown in the main text, and ask it to evaluate maps generated using different templates to those on which it was trained.
In particular, we study four variations: (1) data where the diffuse emission is specified with Model A; (2) models specified with another unique diffuse background denoted Model F; and simulations where the diffuse emission remains the same, but where our model for the (3) disk and (4) GCE are systematically varied.
To provide more context, Model F is another diffuse emission model that was introduced in Refs.~\cite{Fermi-LAT:2012edv,Calore:2014xka} and was shown in Ref.~\cite{Buschmann:2020adf} to provide a comparable fit to the Fermi data over our energy range to Model A, although both were noticeably worse than Model O.
We modify the disk by adopting a ``thick disk'' template with the scale height increased from our default 0.3\,kpc to 1\,kpc, and adjust the GCE profile from an NFW with $\gamma=1.2$ to one with $\gamma=1.0$.
(A far broader set of GCE spatial morphologies have been considered in the literature and would be interesting to study in the context of our method, see e.g. Refs.~\cite{Macias:2016nev,Macias:2019omb,Song:2024iup,Bartels:2017vsx,Cholis:2021rpp,McDermott:2022zmq,Zhong:2024vyi,Ramirez:2024oiw}.)

\begin{figure*}[!t]
\centering
{\includegraphics[width=0.3\linewidth]{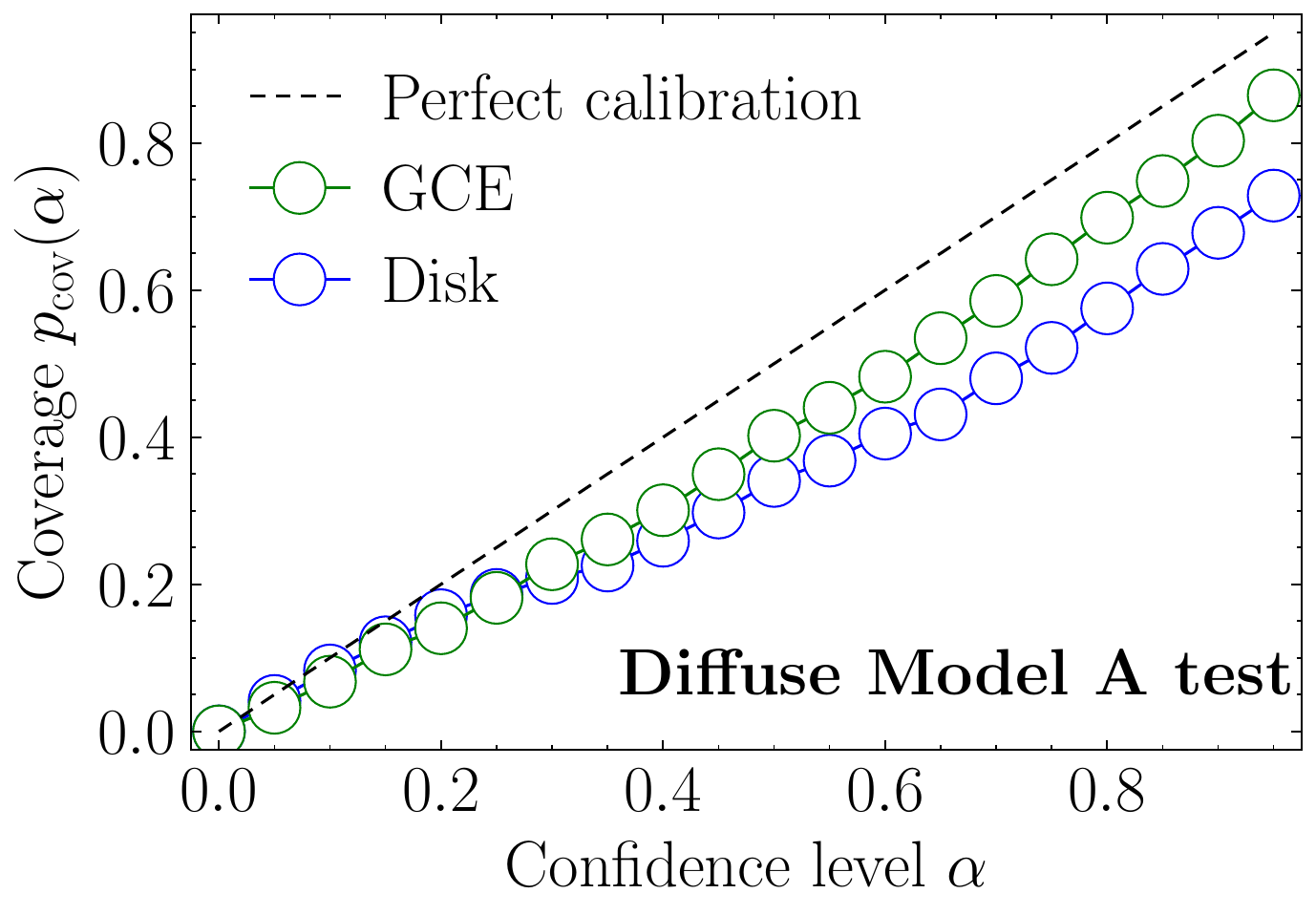}}
\hspace{0.1cm}
\includegraphics[width=0.3\linewidth]{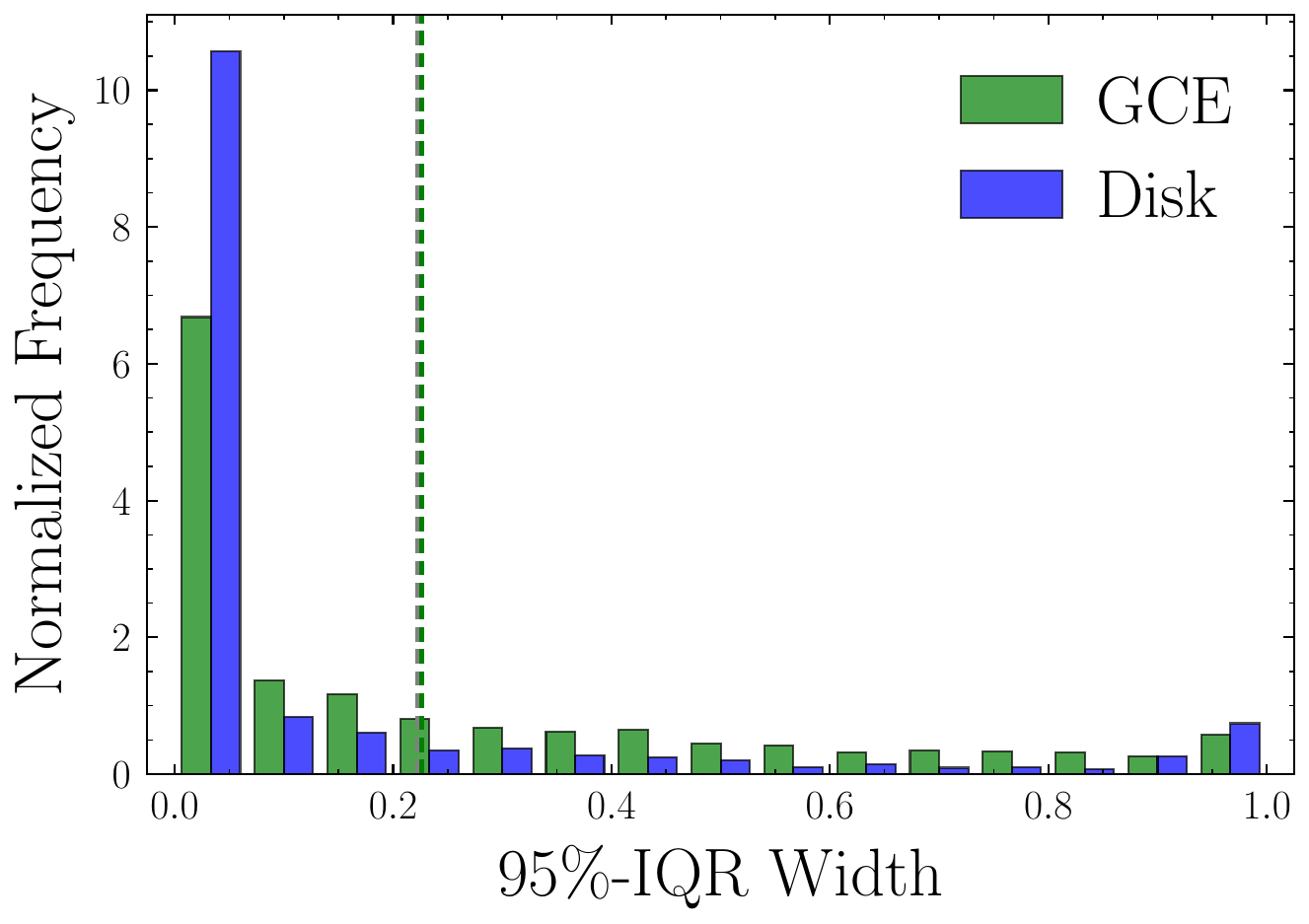} 
\hspace{0.1cm}
\includegraphics[width=0.3\linewidth]{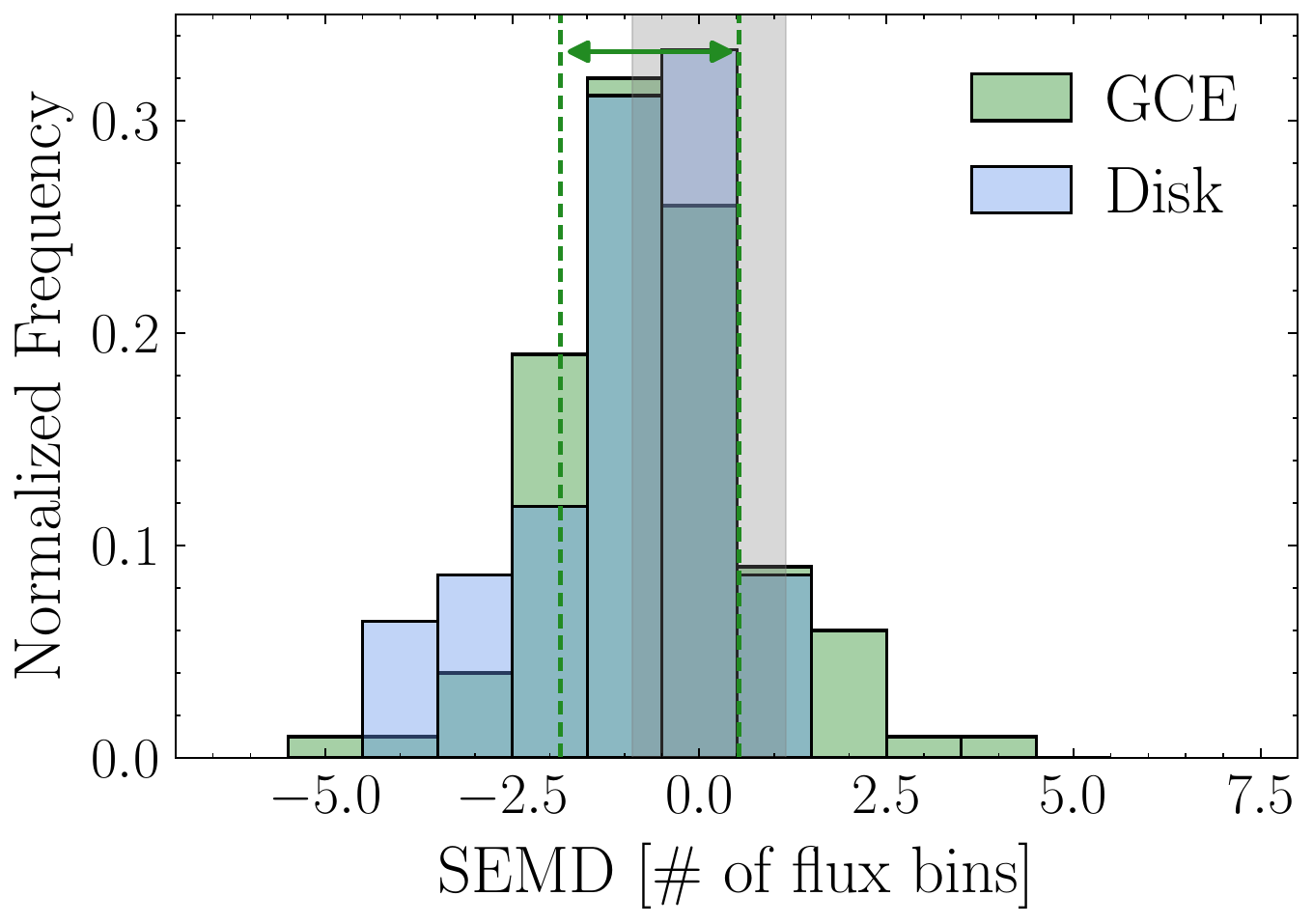}\\ \vspace{0.1cm}
\includegraphics[width=0.3\linewidth]{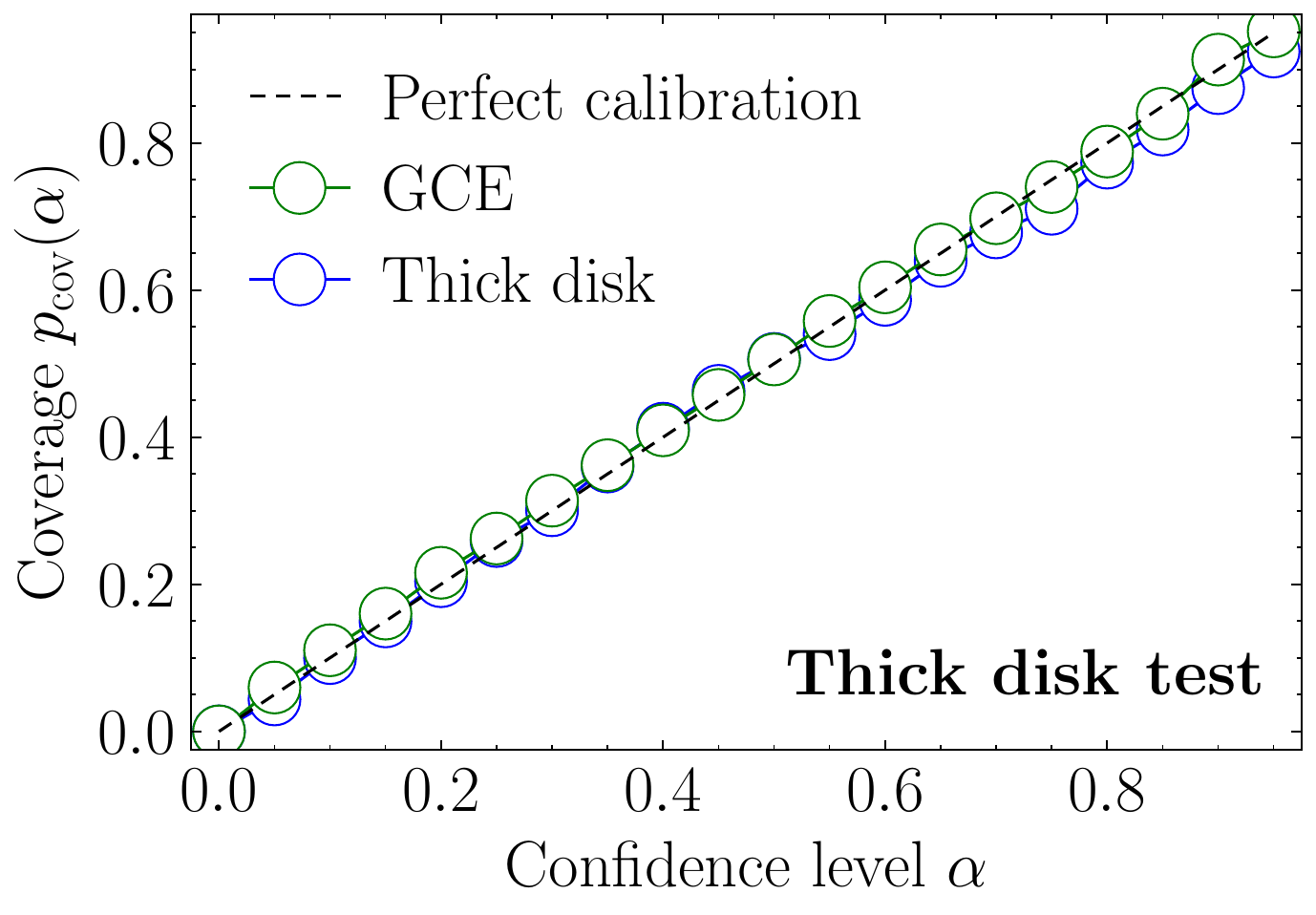}
\hspace{0.1cm}
\includegraphics[width=0.3\linewidth]{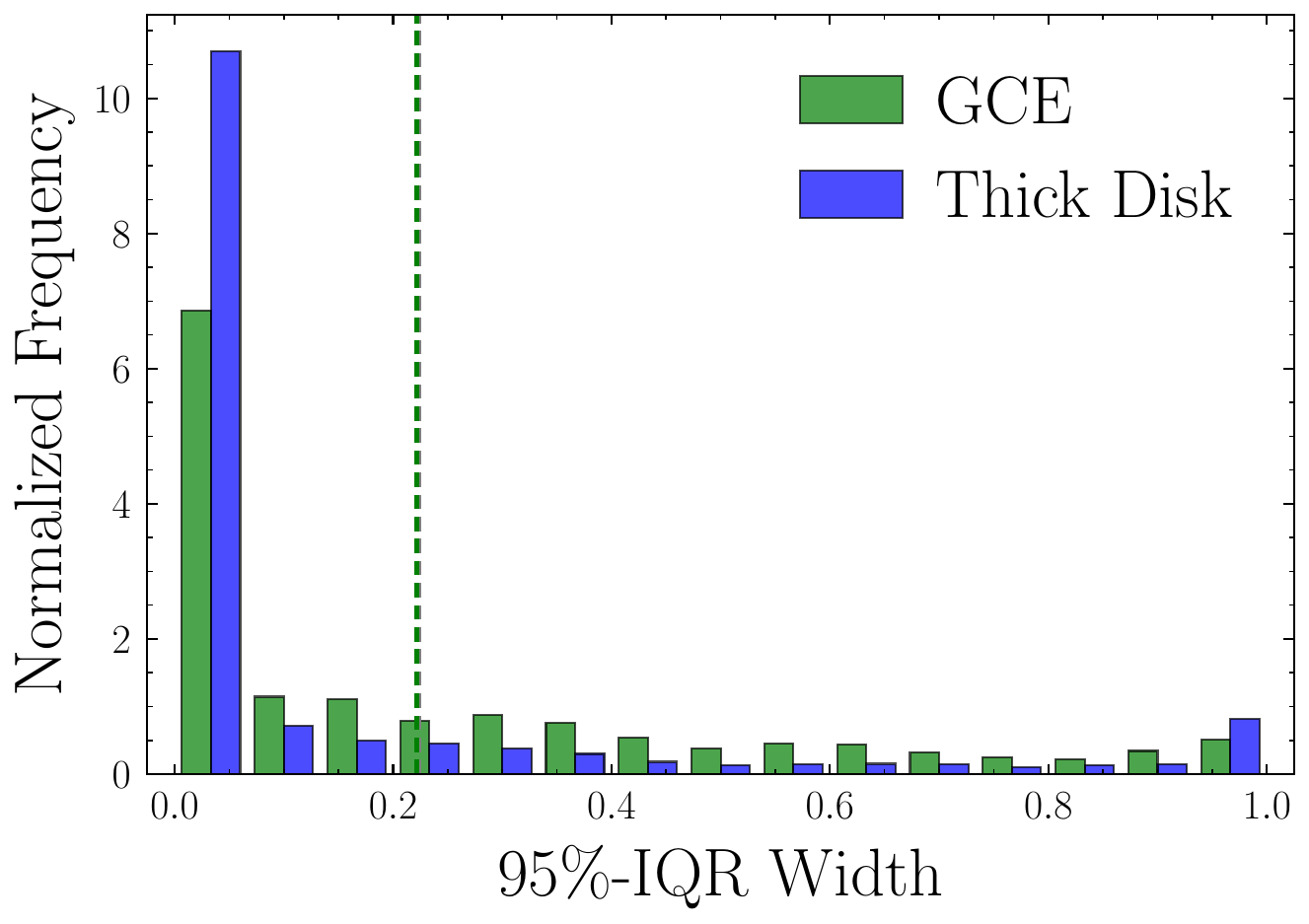} 
\hspace{0.1cm}
\includegraphics[width=0.3\linewidth]{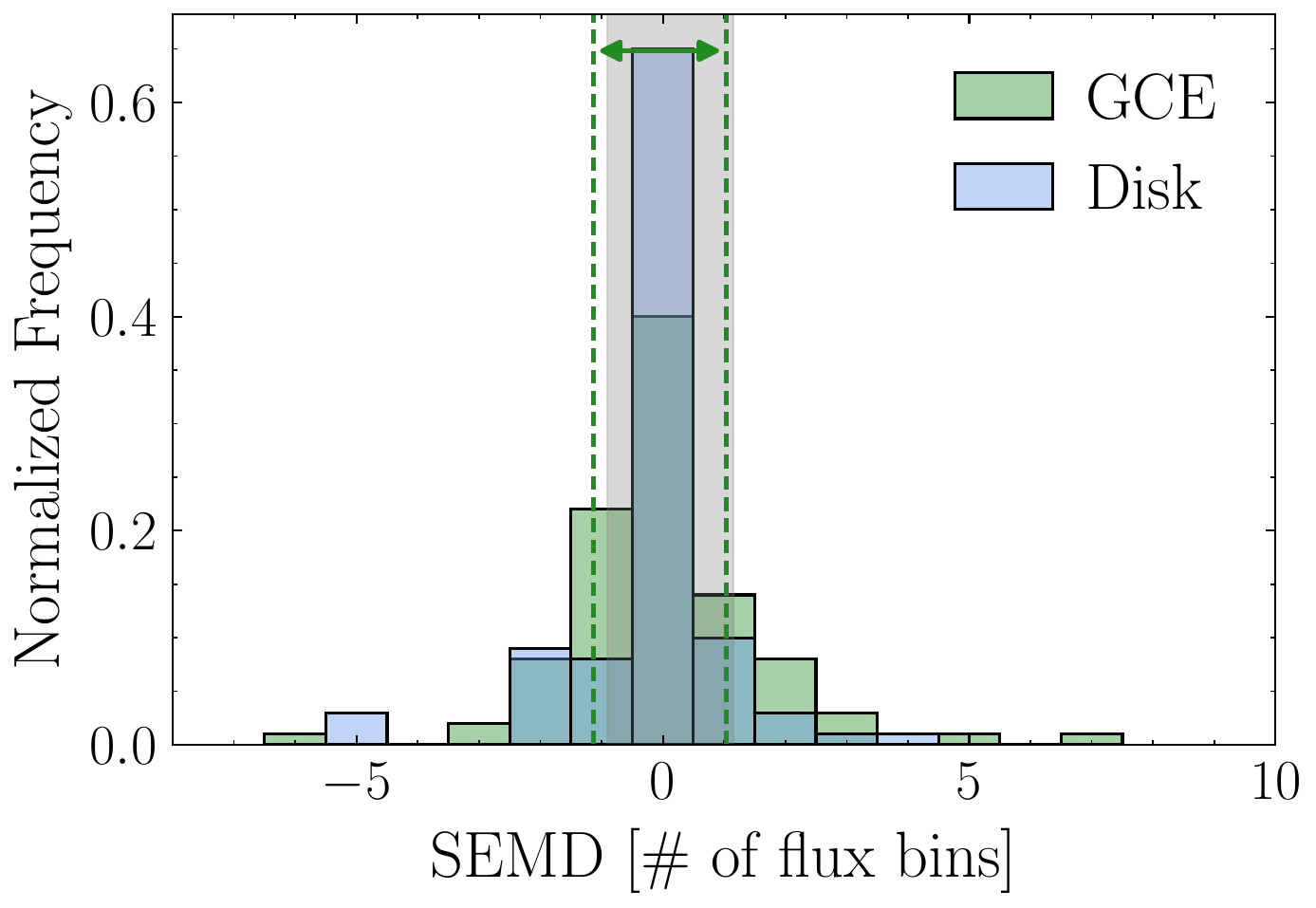}
\vspace{0.1cm}
\includegraphics[width=0.3\linewidth]{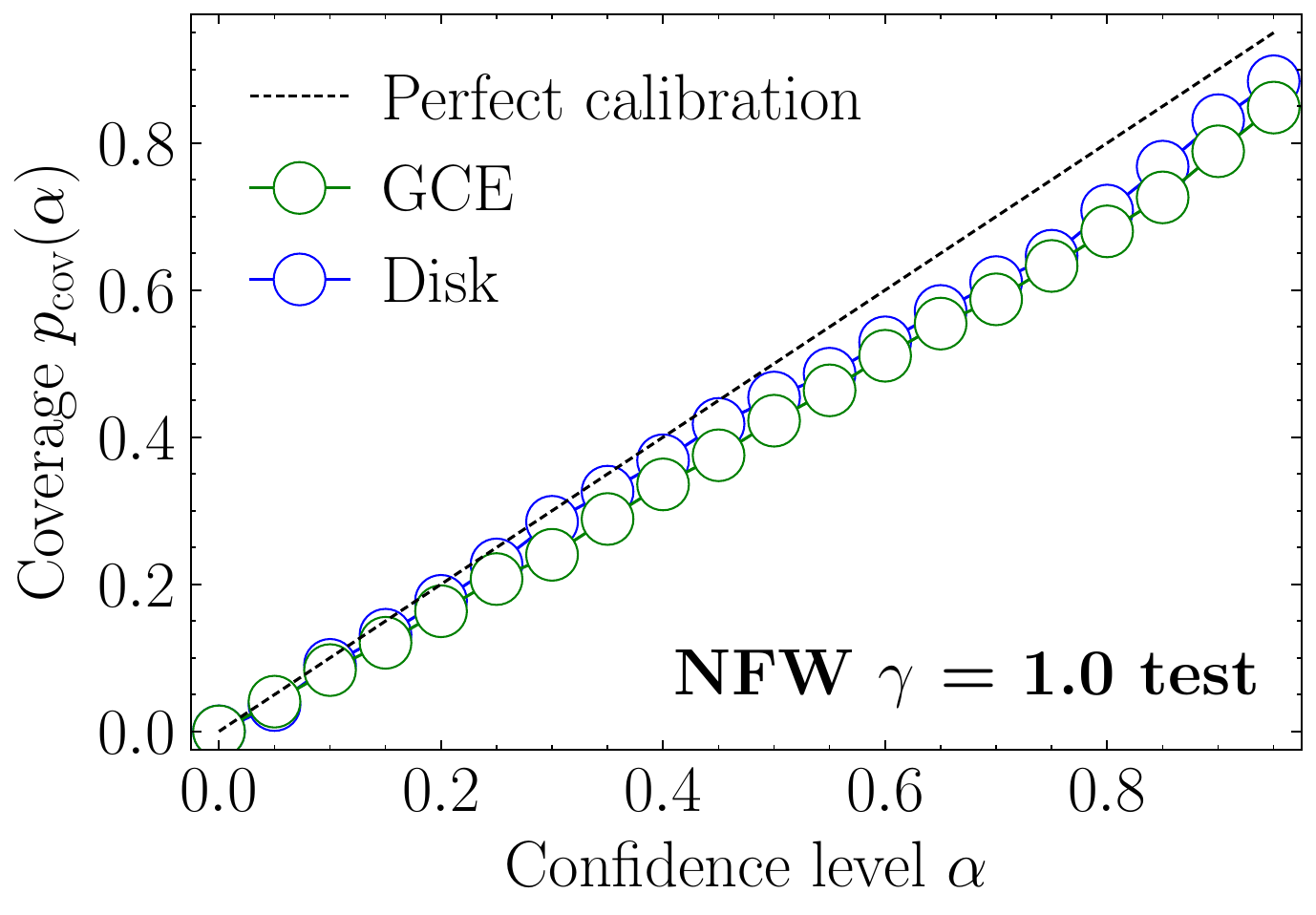}
\hspace{0.1cm}
\includegraphics[width=0.3\linewidth]{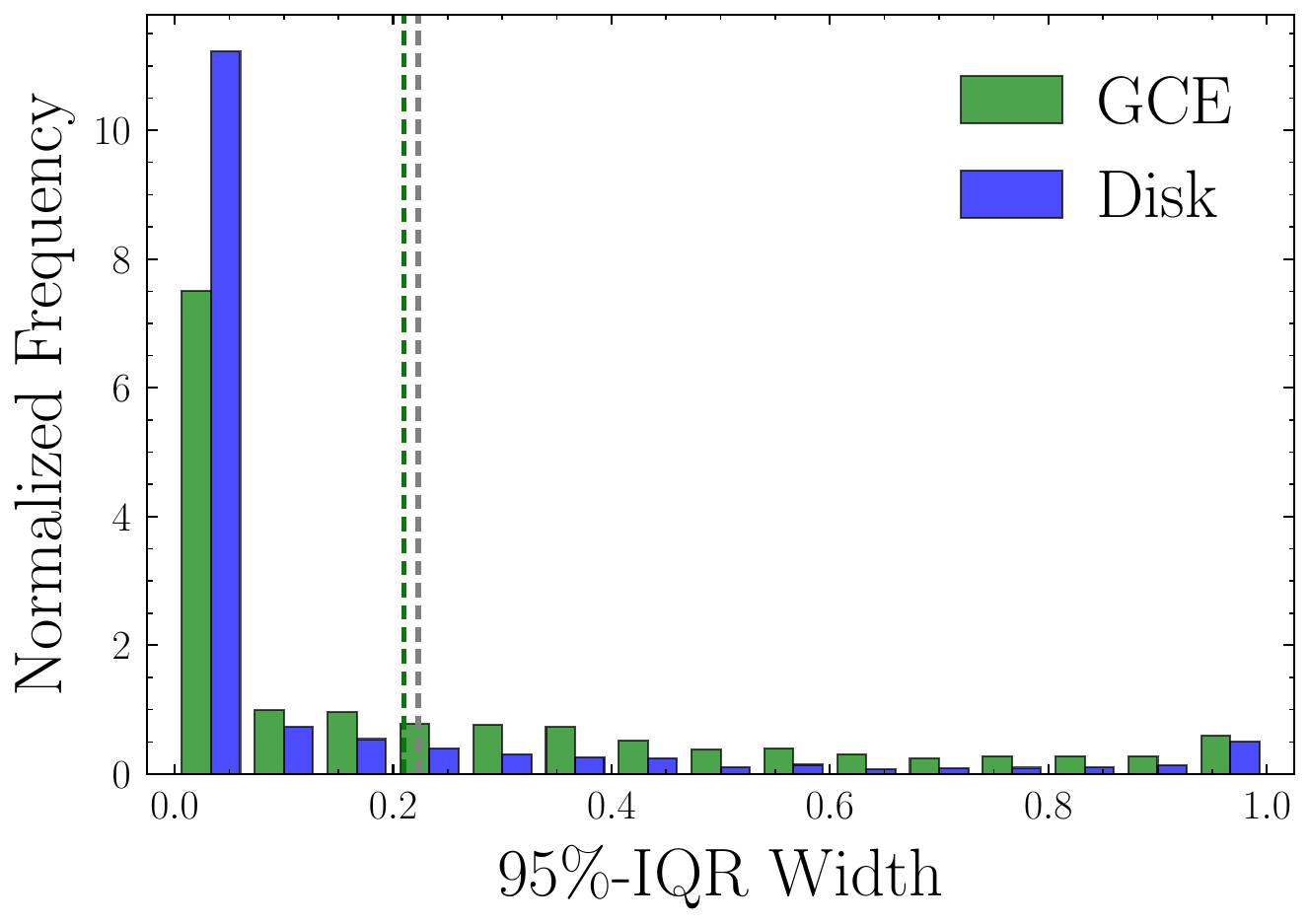} 
\hspace{0.1cm}
\includegraphics[width=0.3\linewidth]{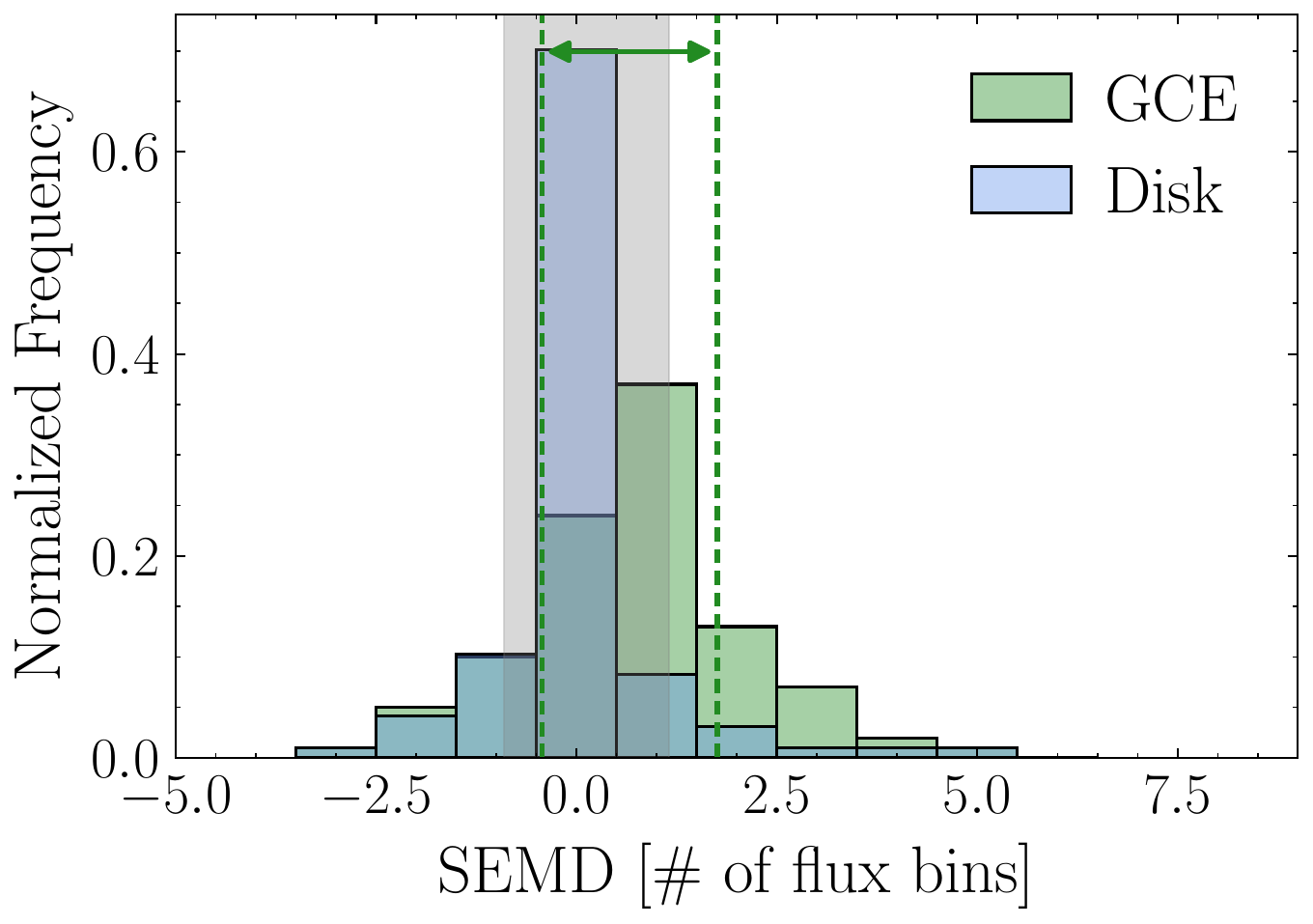}
\vspace{-0.3cm}
\caption{Similar to Fig.~\ref{fig:mismodeling} but for SCD recovery.
We show the calibration (left), sharpness (middle), and SEMD (right).
Again, the Model F performance was found to be similar but slightly worse than Model A.
The sharpness plots also depict the median GCE value as a green vertical line, which can be compared to the default value (gray) obtained from Fig.~\ref{fig:sharpcalibration_hist}.
In the SEMD figures, the gray band indicates the [16\%, 84\%] quantile range of the GCE SEMD from our default CNN on 500 test maps (see Fig.~\ref{fig:sharpcalibration_hist}), while the dashed green lines indicate the same range for the respective GCE SEMD.}
\vspace{-0.5cm}
\label{fig:mismodeling_hist}
\end{figure*}

For each variation, we demonstrate the impact on the calibration and sharpness of the spectral predictions in Fig.~\ref{fig:mismodeling}.
These results should be compared to those for our default validation in Fig.~\ref{fig:spectral-sharpcalibration}.
(Note that our validation analysis used 500 maps whereas for the comparison here we only show 100 maps in each case.)
Broadly, changing the disk template leaves the performance comparable.
Changing the diffuse emission templates has a more discernible impact: the predictions have become less sharp, and what is especially noticeable is that the predictions become generically overconfident, exactly as one would expect by injecting a source of error that the network was not trained to anticipate.
Unsurprisingly there is large overconfidence for the diffuse templates, but also the disk template which has a large spatial degeneracy with the inverse Compton map.
Interestingly, the overconfidence for the GCE is generally less than for other templates with the exception of the extremely spatially unique Fermi bubbles.
Performance for a variation of the GCE template is in between: the method becomes overconfident for the GCE and $\pi^0$ component, but the remaining templates are relatively unaffected.

How the SCD predictions are impacted by mismodeling is shown in Fig.~\ref{fig:mismodeling_hist}, results which should be compared to Fig.~\ref{fig:sharpcalibration_hist}.
(Note when comparing these distributions to Fig.~\ref{fig:sharpcalibration_hist}, recall that these have 100 maps whereas the default validation had 500---this helps explain the additional SEMD outliers in the default results.)
As for the spectra, there is generic overconfidence in the predictions when the incorrect diffuse model is used, with results again far less impacted when the disk or GCE templates are changed.
Sharpness on the other hand remains largely unchanged.
This is not surprising, as sharpness is the only quantity shown that is not computed relative to the truth but purely based on the CNN output.
Therefore, a change in sharpness in the presence of mismodeling would require the CNN to recognize the presence of mismodeling in the first place---a task it has not been trained on. 
It would be interesting to study extensions in the direction of making the CNN account for such a covariate shift (i.e. an offset between data distributions at training and evaluation time) in future work.
In the presence of diffuse mismodeling the method is biased towards reconstructing the SCD as a few bins too dim.

\begin{figure*}[!t]
\centering
\includegraphics[width=0.3\linewidth]{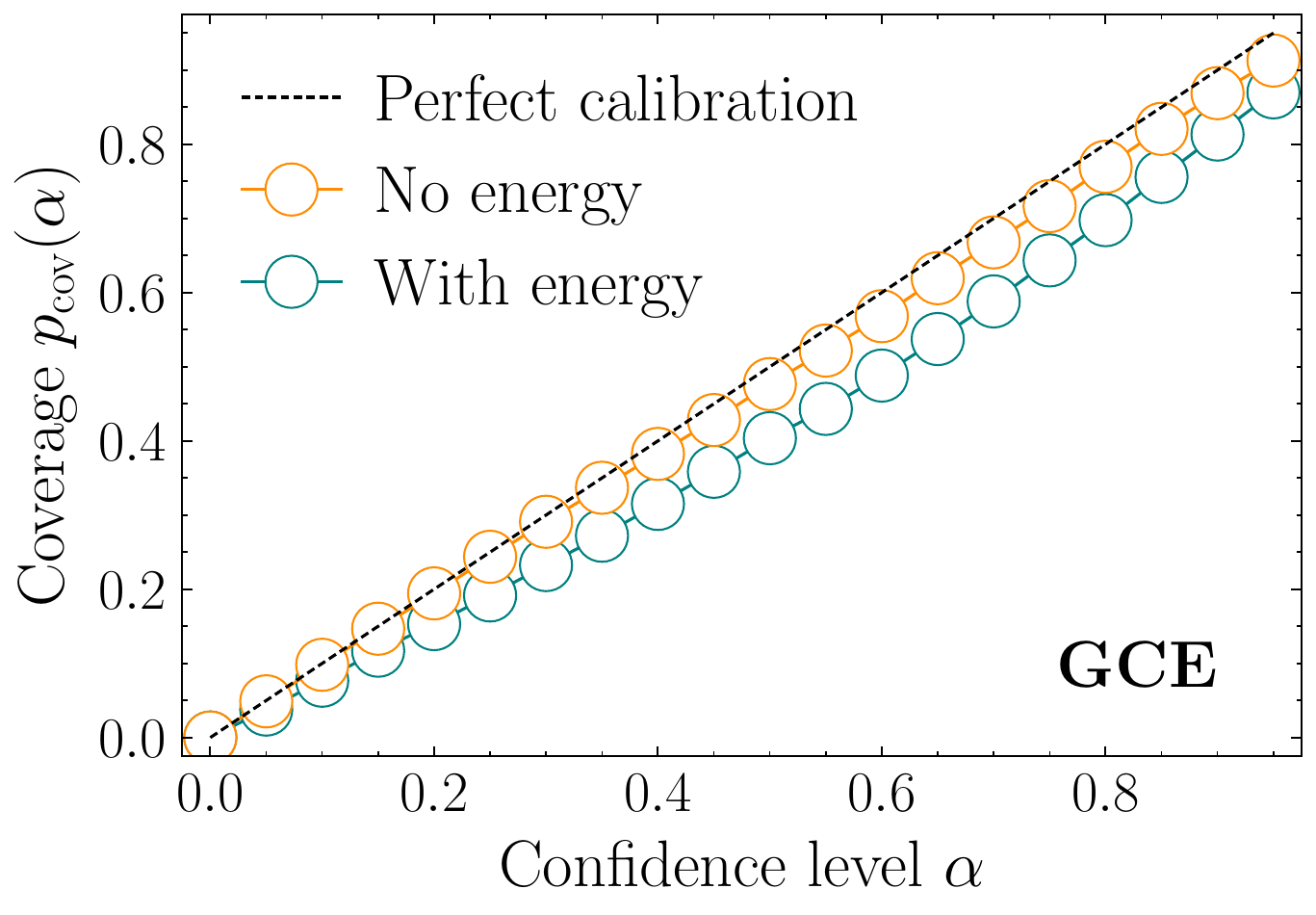} \hspace{0.1cm}
\includegraphics[width=0.3\linewidth]{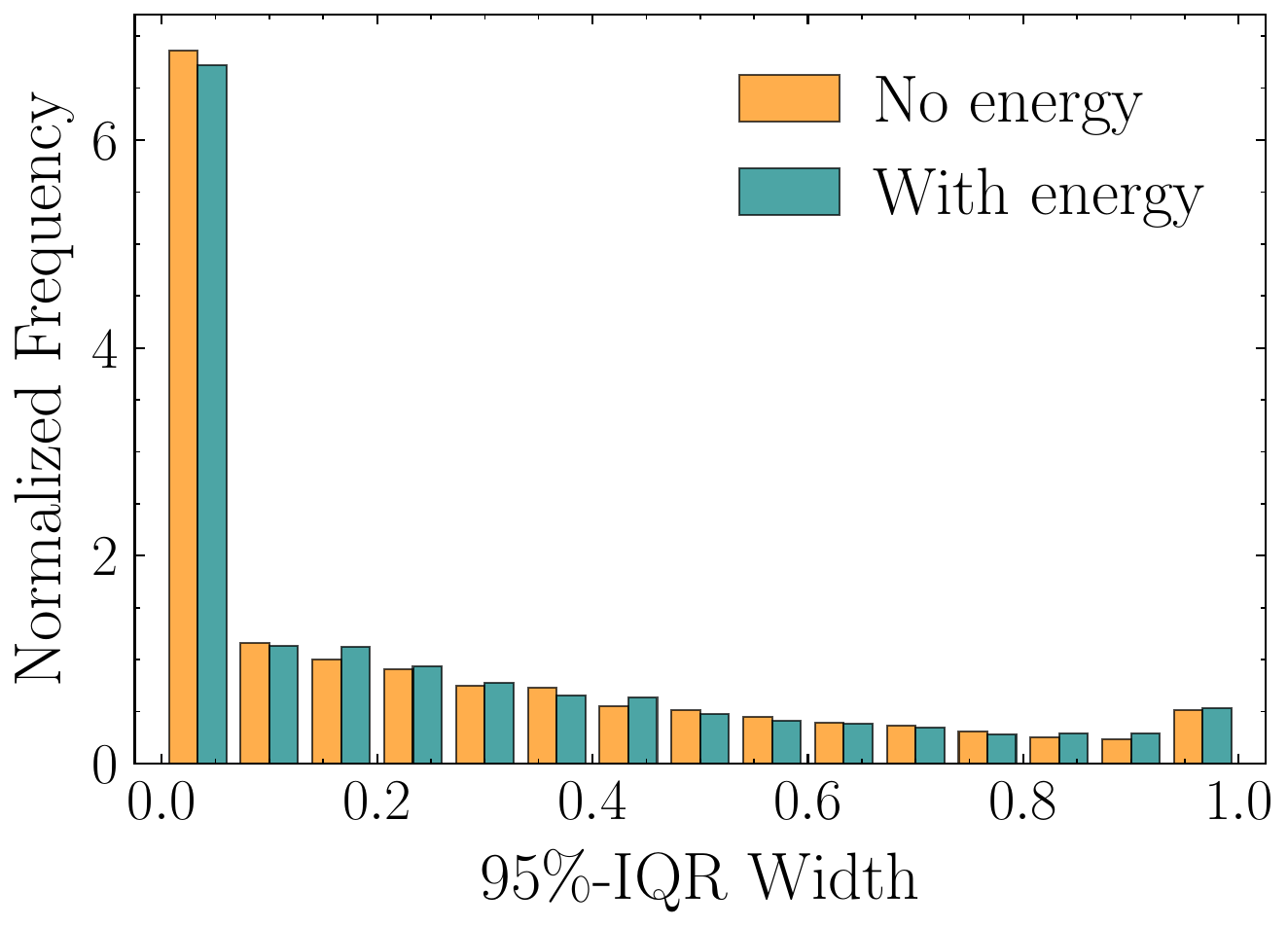} \hspace{0.1cm}
\includegraphics[width=0.3\linewidth]{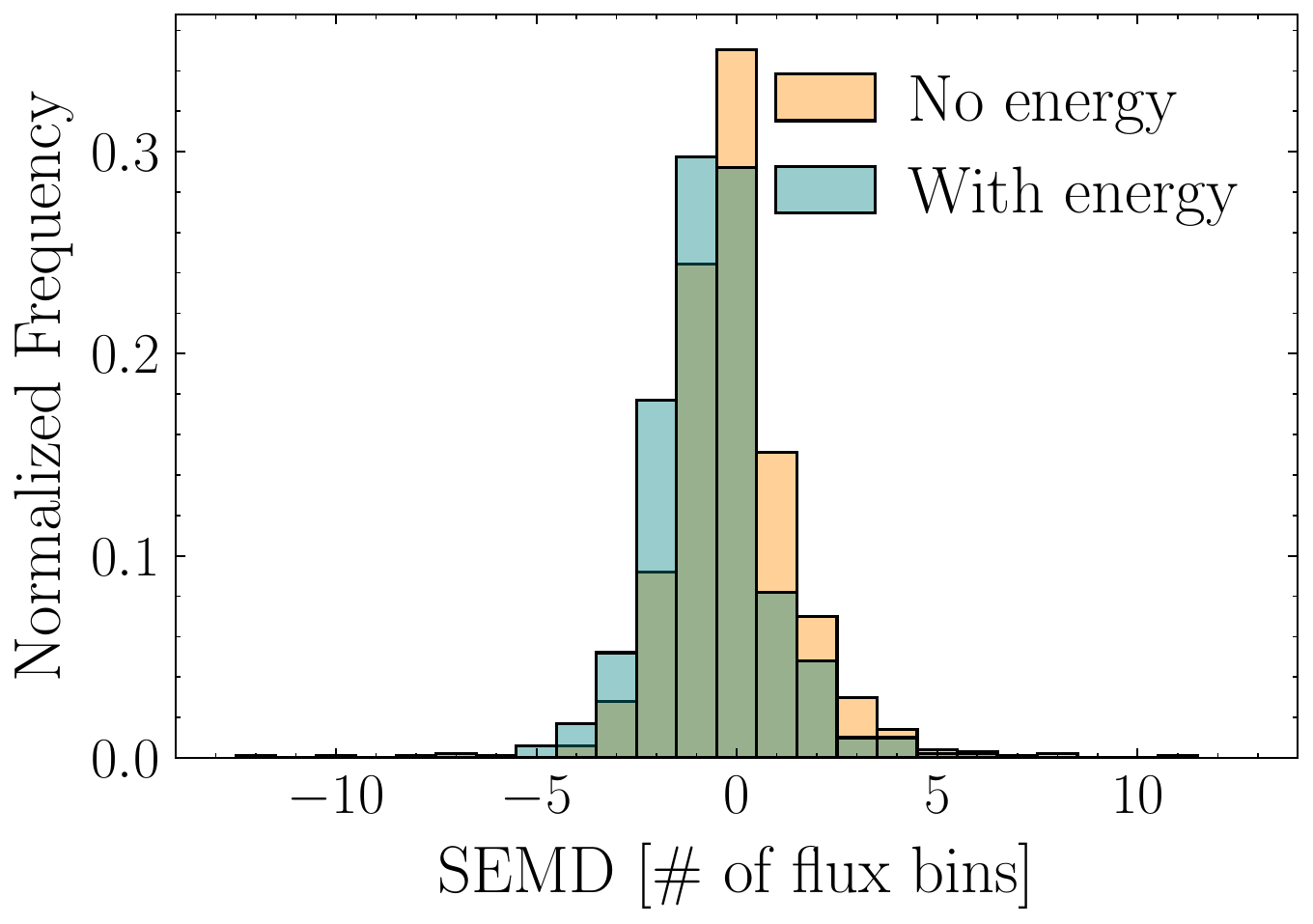}\\ \vspace{0.1cm}
\includegraphics[width=0.3\linewidth]{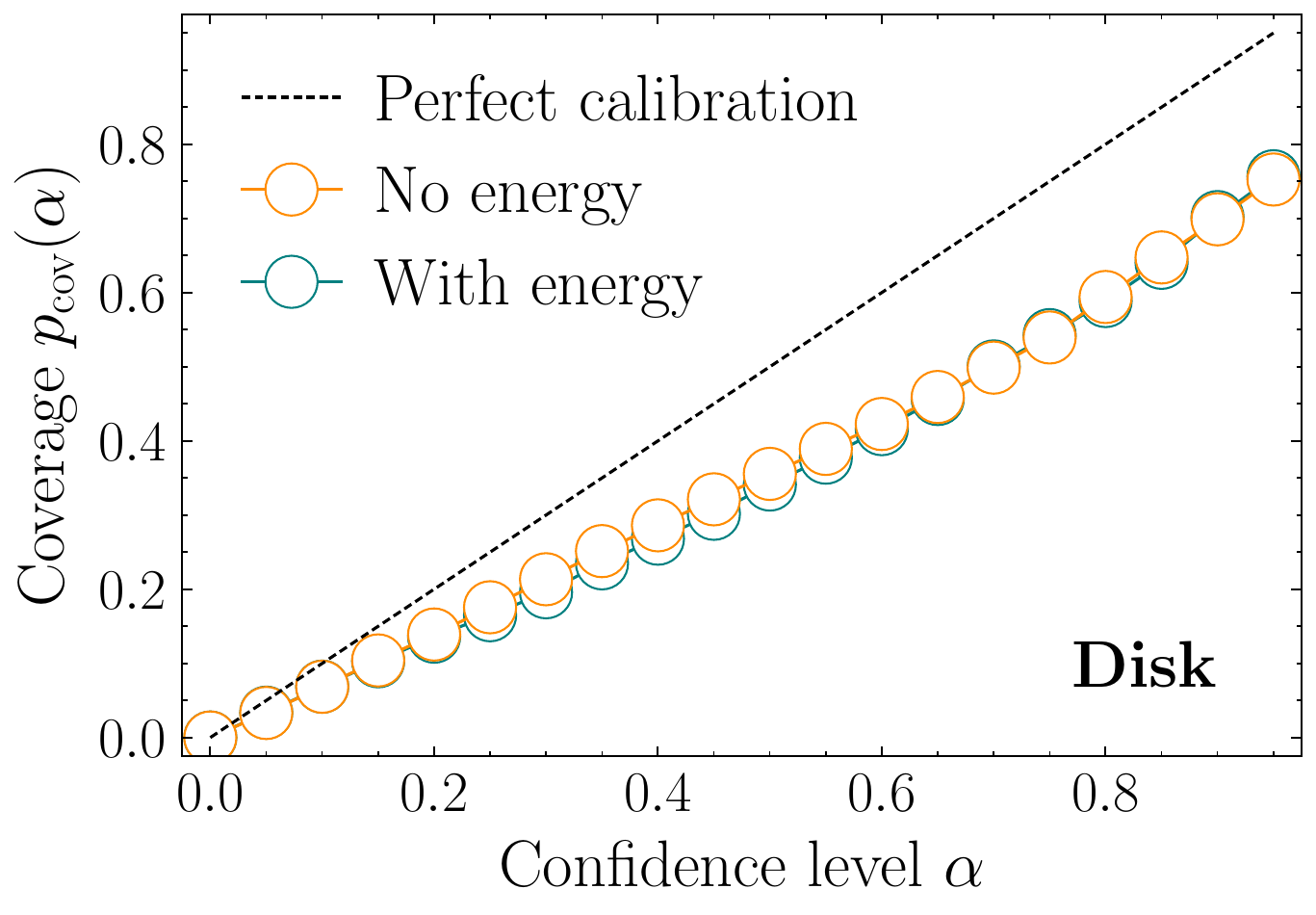} \hspace{0.1cm}
\includegraphics[width=0.3\linewidth]{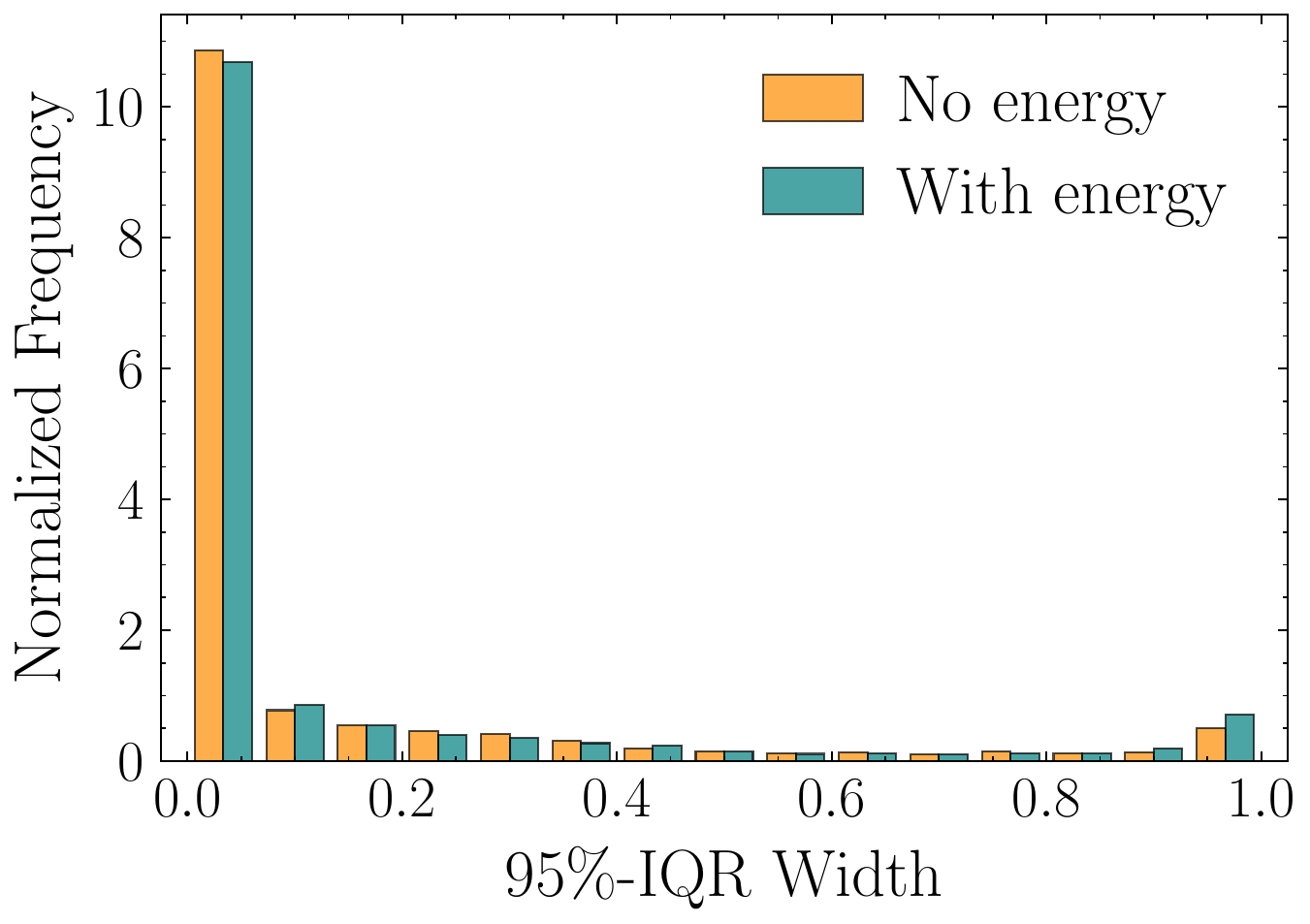} \hspace{0.1cm}
\includegraphics[width=0.3\linewidth]{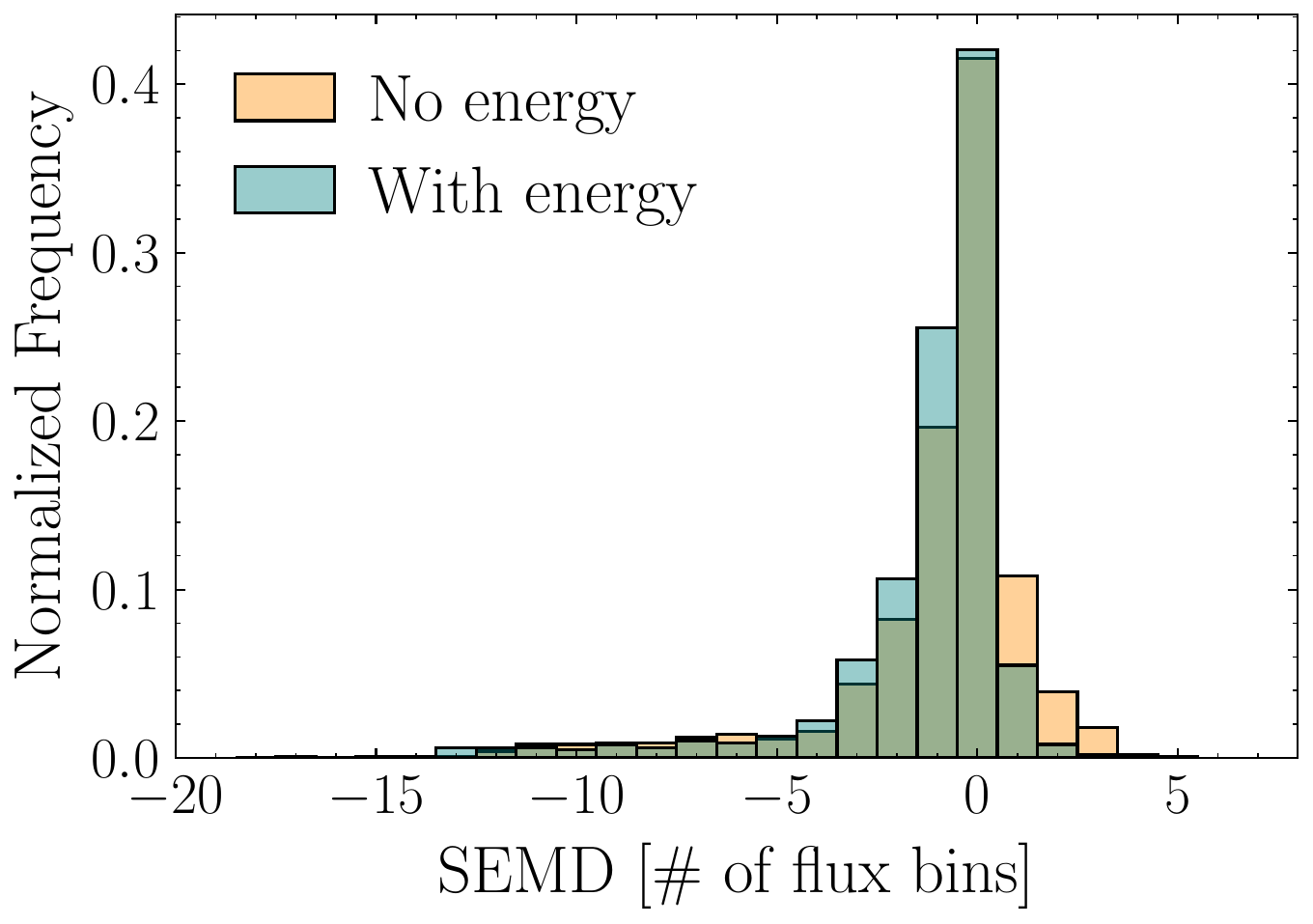}
\vspace{-0.3cm}
\caption{A direct comparison of the performance of the energy dependent and independent networks in the presence of mismodeling as simulated by using data from Model A.
The three columns are as in Fig.~\ref{fig:mismodeling_hist}, except now we compare the two networks directly for the GCE performance in the top row and the Disk on the bottom.
For this test we increase the number of test maps to 1,000 and the same maps are shown to both networks, simply summed together before being passed to the energy independent network.
In order to be able to show both network the same maps, here the network labeled ``no energy'' has one input channel but was trained on data generated in ten bins and then summed together.}
\vspace{-0.5cm}
\label{fig:mismodelA-indep-dep}
\end{figure*}

\begin{figure*}[!t]
\centering
\includegraphics[width=0.3\linewidth]{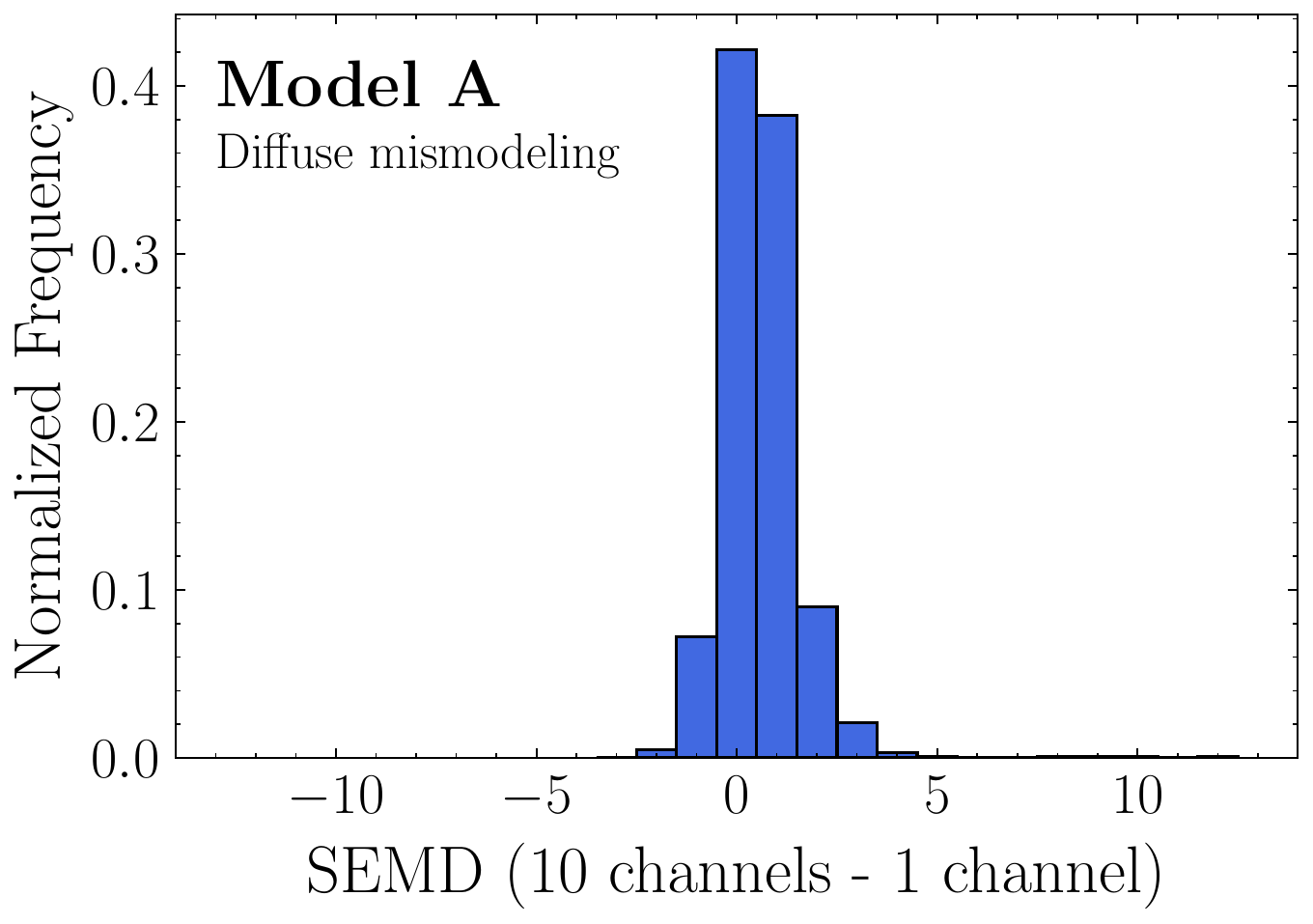} \hspace{0.1cm}
\includegraphics[width=0.3\linewidth]{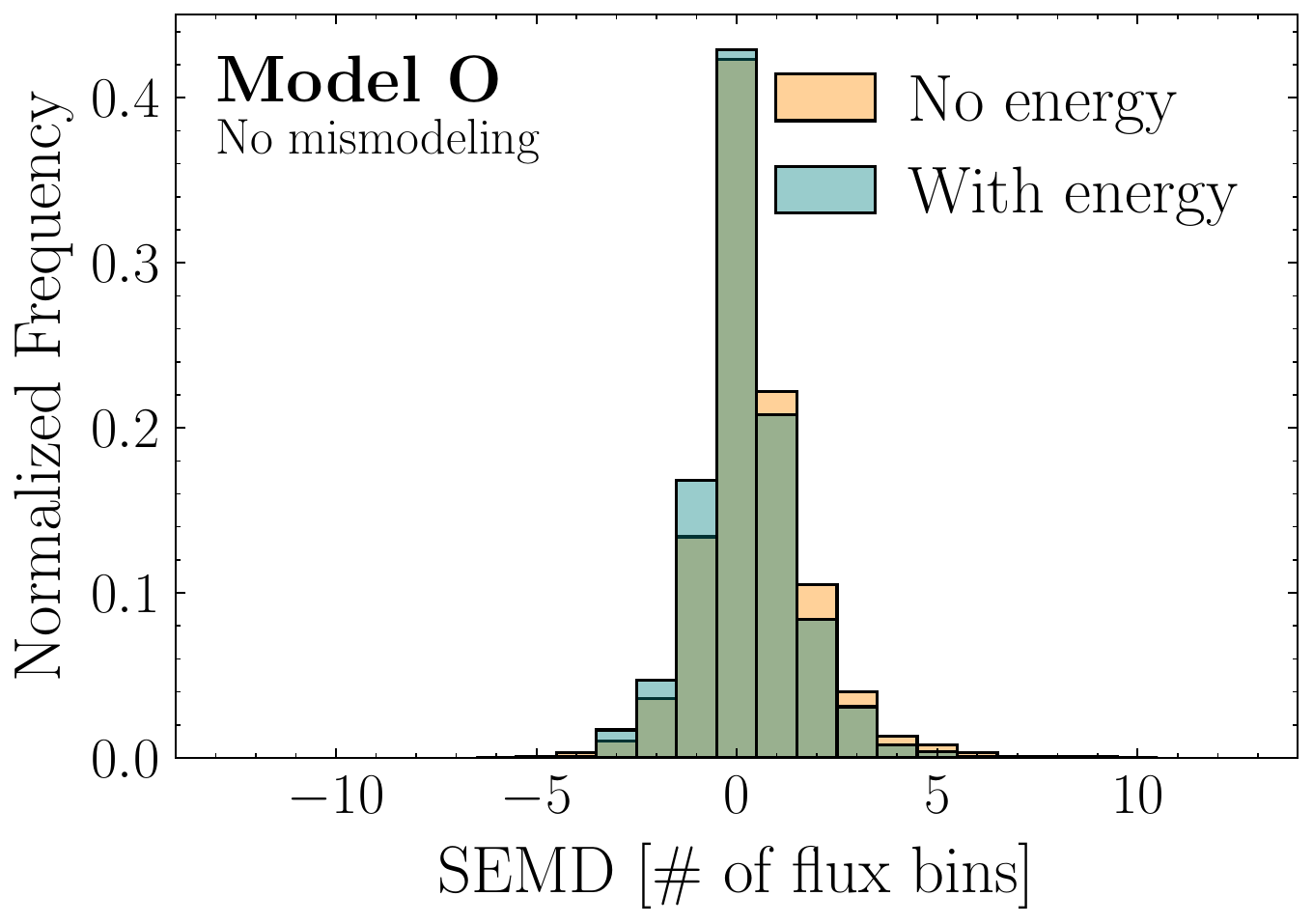} \hspace{0.1cm}
\includegraphics[width=0.3\linewidth]{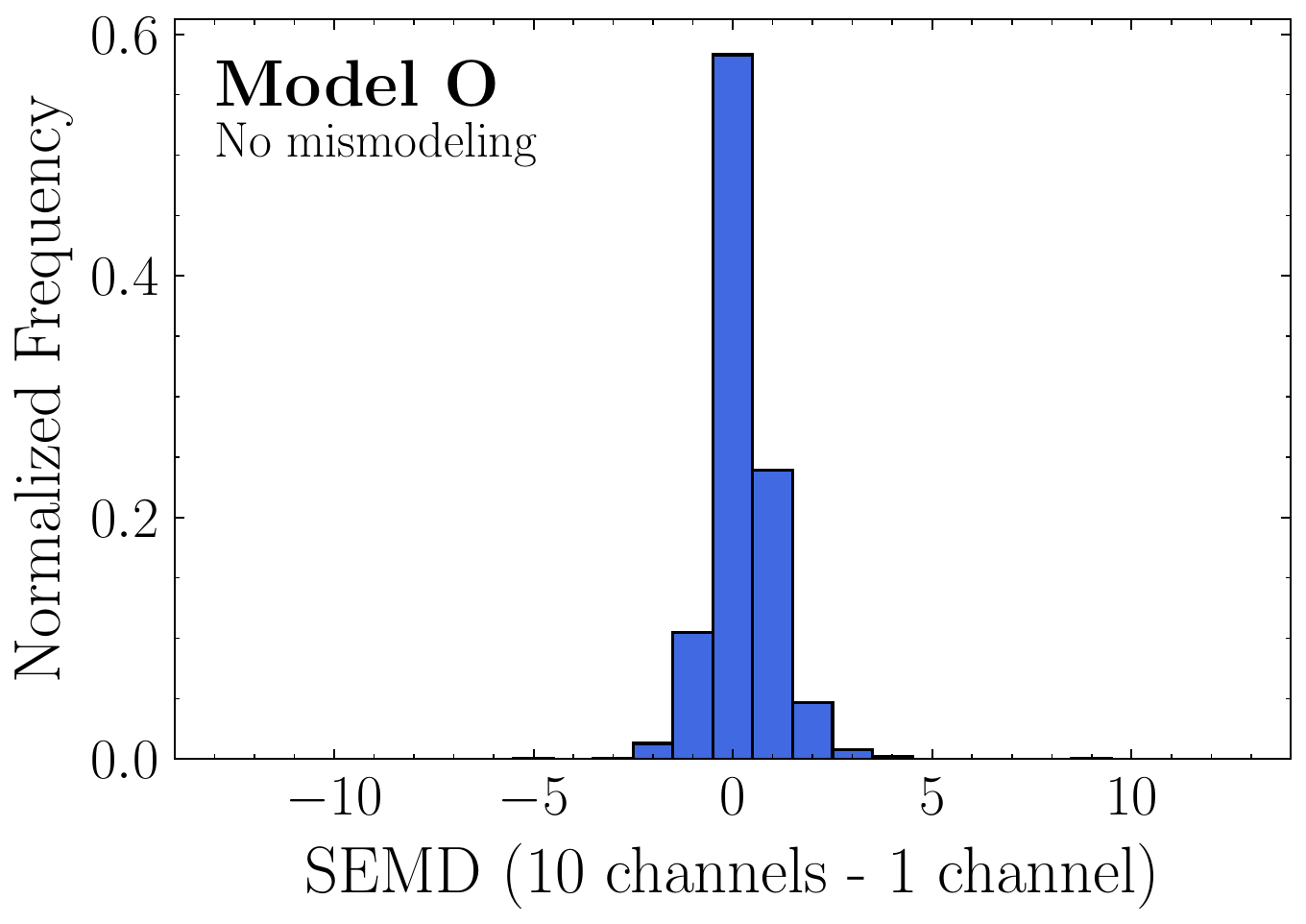}
\vspace{-0.3cm}
\caption{(Left) The SEMD between the energy dependent and independent CNN results as given in the top right of Fig.~\ref{fig:mismodelA-indep-dep}.
This is a measure of whether errors in the reconstruction are correlated between the two networks.
We show the SEMD computed by subtracting the energy independent network from the energy dependent one, so that a positive value implies that energy dependent CNN has returned a dimmer SCD.
In order to help calibrate these findings, we show analogous results in the absence of mismodeling -- maps are generated with Model O which both CNNs were trained on -- both for the individual networks as compared to the truth (middle) and the difference between networks (right).
}
\vspace{-0.5cm}
\label{fig:EMD-Diff}
\end{figure*}

As a final test of the impact of mismodeling, we cover the case discussed in the main text: how the performance in recovering the SCD of the energy dependent and independent networks compares when shown maps outside the space they were trained on.
In this case, for the independent network we use a CNN that is trained with 1 input channel, but on data generated in 10 energy bins.
The rationale is that we can then run the methods on an identical simulated dataset: for the energy independent case we simply sum the energy bins to provide the data passed to the input channel.
We show these results in Fig.~\ref{fig:mismodelA-indep-dep}, where now we have increased the number of simulated maps shown to 1,000 and the same sets of maps were shown to both networks, simply being summed together in the energy independent case.
We consider exclusively the case where the mismodeling is generated by simulating with Model A, whereas both networks were trained on Model O.
Broadly the performance of the two networks is very comparable, with the energy independent network slightly more calibrated in the case of the GCE.
The most noticeable differences are in the case of the SEMD: whilst the performance of both networks is peaked around zero, both networks demonstrate a slight bias towards recovering the SCD as dimmer than it actually is for both the GCE and the disk, with the effect modestly more pronounced for the energy dependent network.
We attribute the long tail to lower values in the case of the disk to confusion between the disk model and the inverse Compton emission.
For the GCE, we can also check whether the errors in reconstruction are correlated between the network with and without energy channels.
As shown in Fig.~\ref{fig:EMD-Diff}, they are.
To aid the interpretation of these results, we also show in that figure results generated without mismodeling (so using the Model O diffuse maps).
The SEMD values are smaller although comparable to those when using Model A.

\subsection{GCE Templates Beyond NFW}

\begin{figure*}[!t]
\centering
\includegraphics[width=0.45\textwidth]{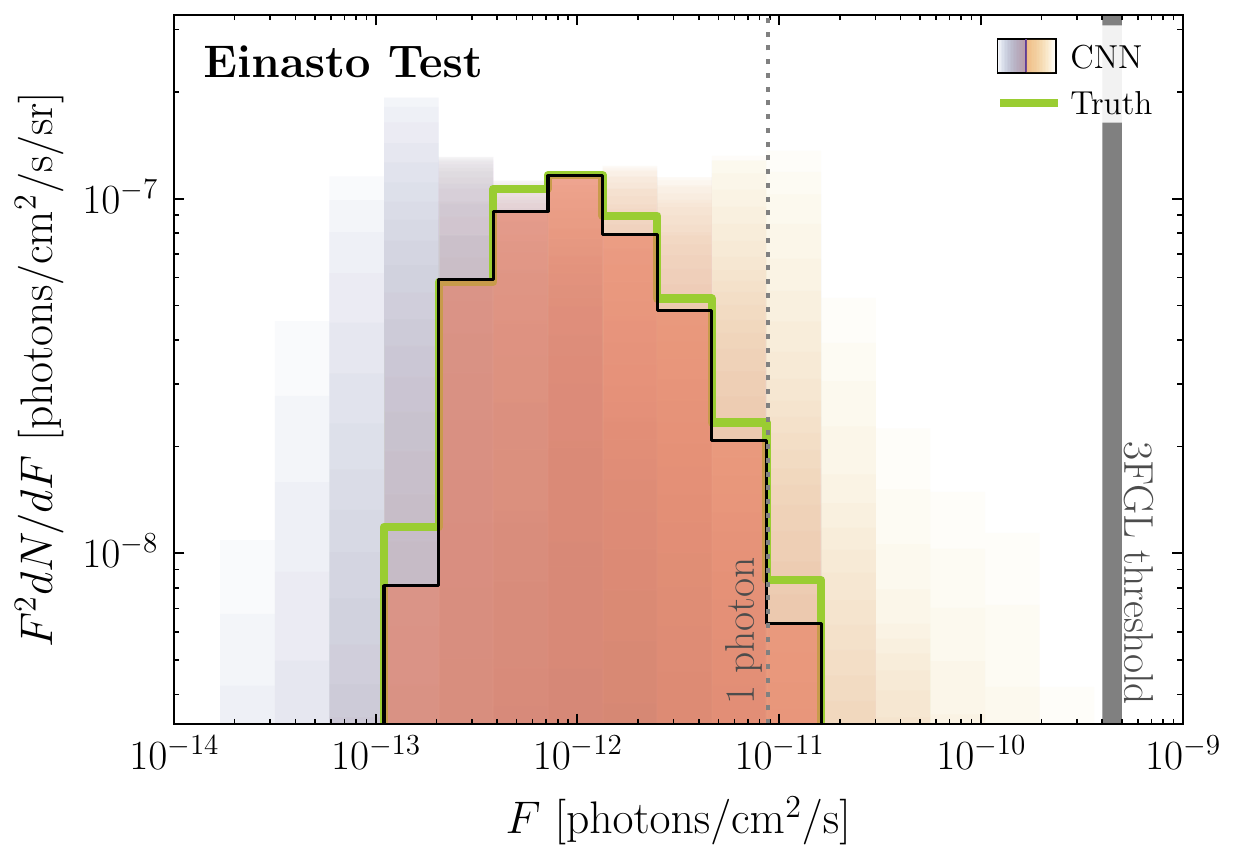} \hspace{0.5cm}
\includegraphics[width=0.45\textwidth]{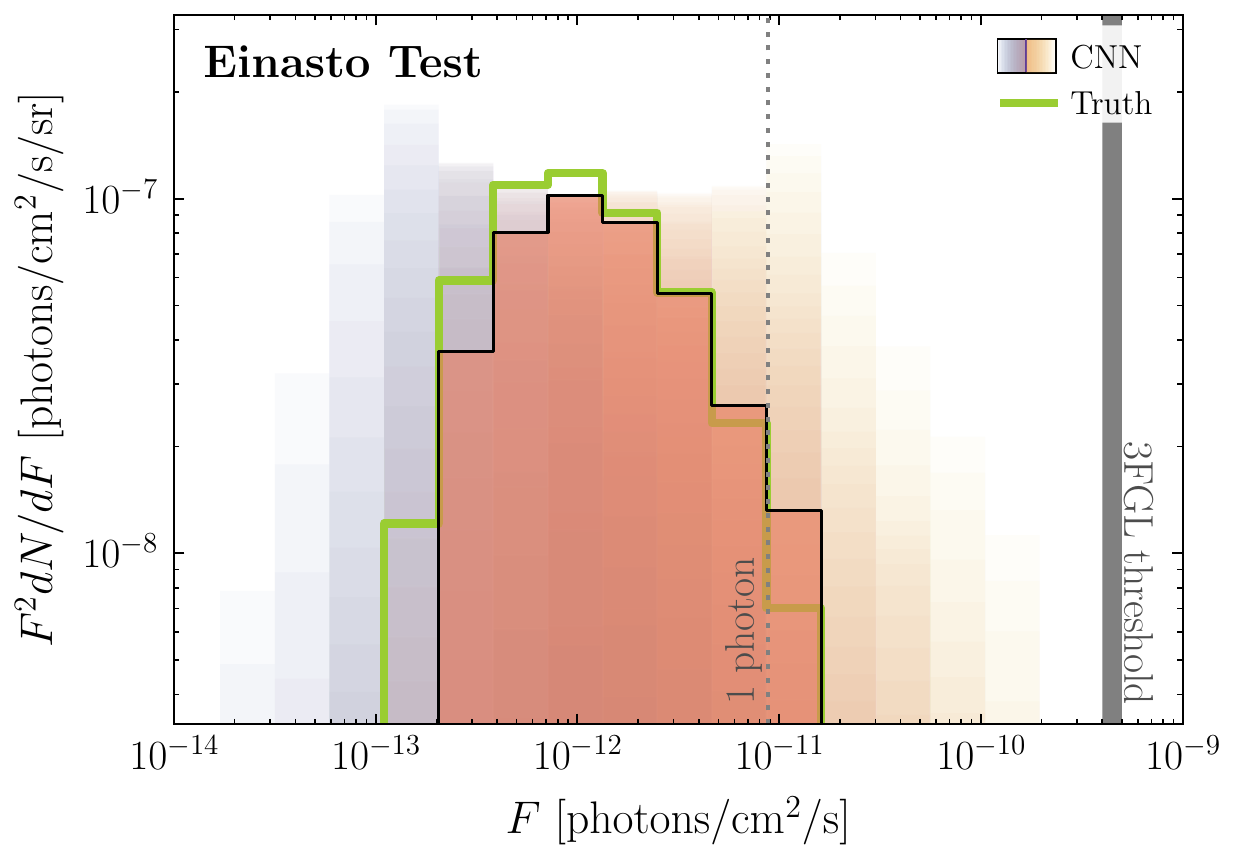} \hspace{0.5cm}\\
\includegraphics[width=0.45\textwidth]{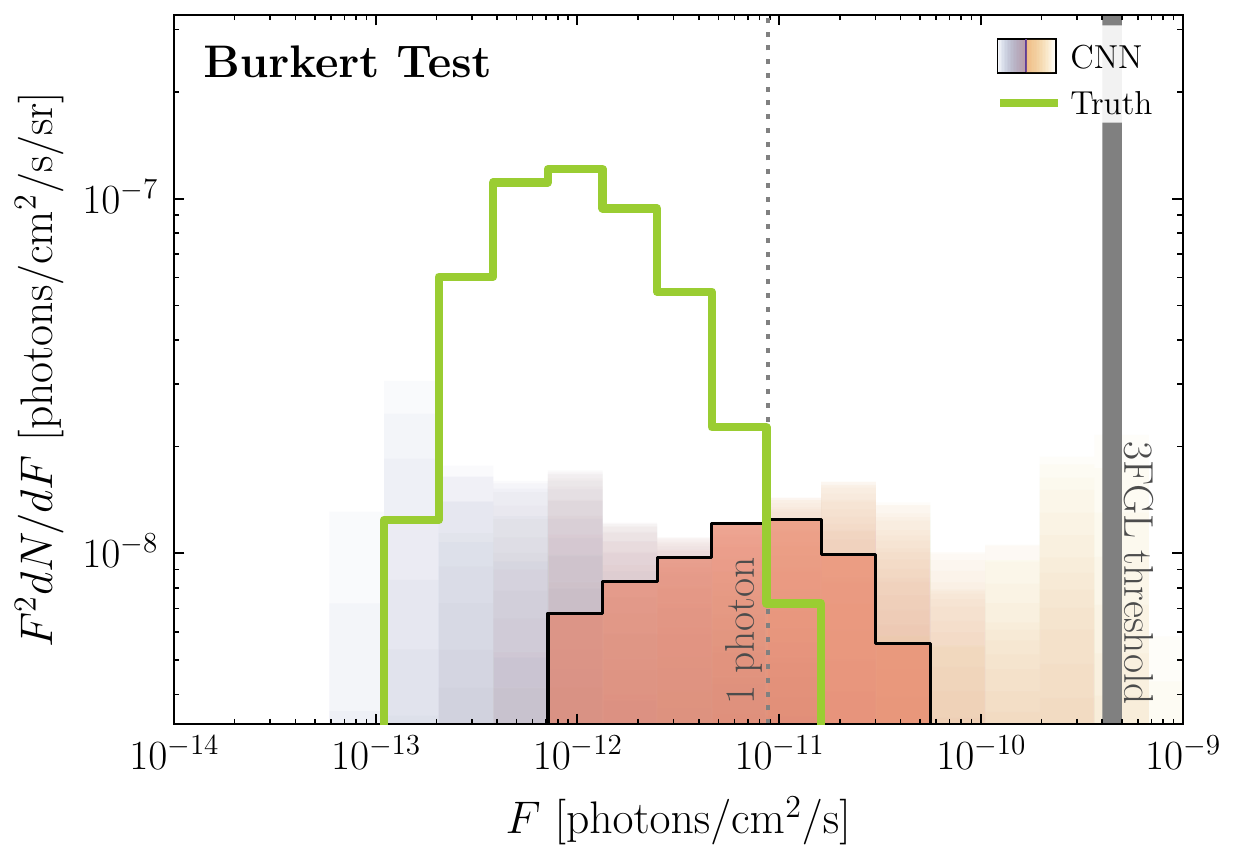}
\hspace{0.5cm}
\includegraphics[width=0.45\textwidth]{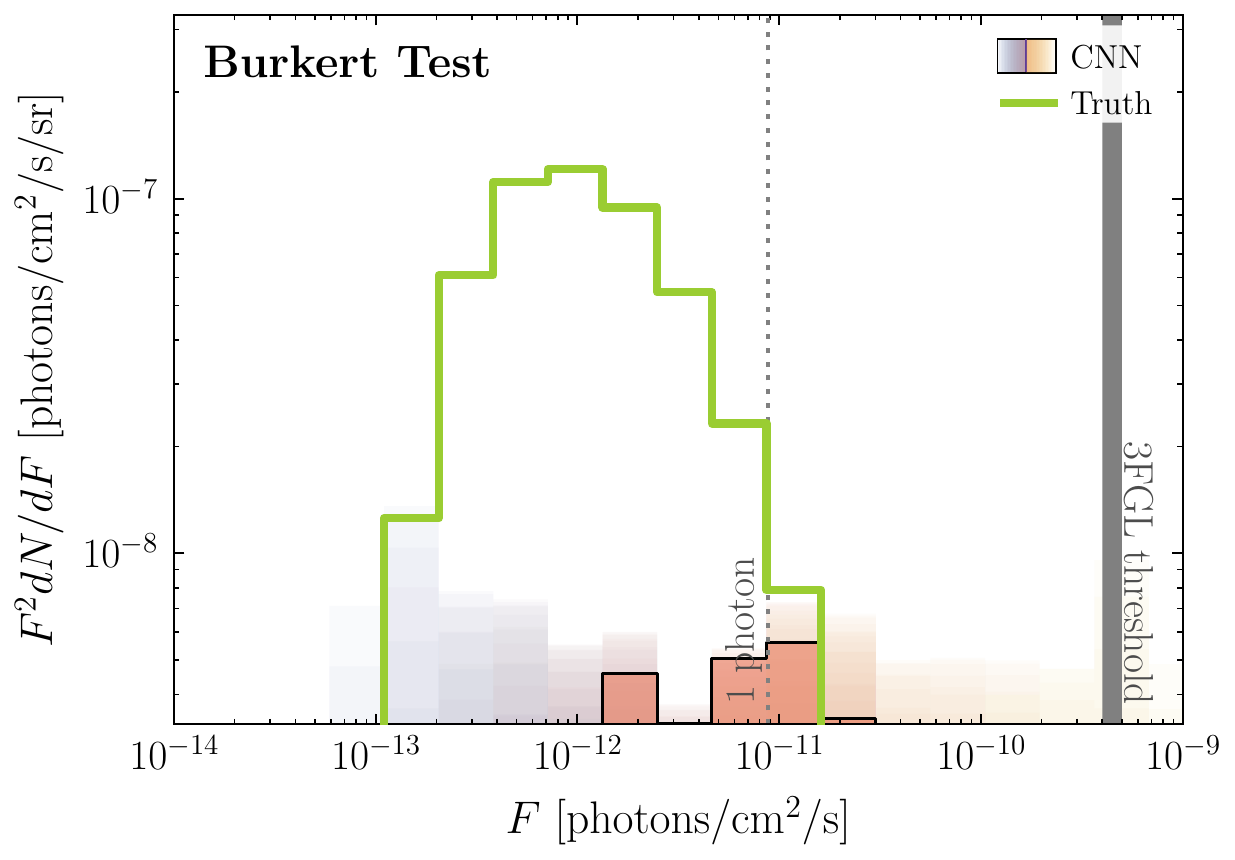}
\vspace{-0.2cm}
\caption{The impact of changing the spatial morphology of the GCE away from a generalized NFW profile to other commonly used DM distributions, the sharply peaked Einasto (top) or cored Burkert (bottom) distributions.
In both cases, we show results where mock Fermi datasets are generated with the GCE following either the Einasto or Burkert spatial profile and with the true $dN/dF$ fixed to the best fit found by our default analysis.
We then see how well our CNN trained on the NFW can reconstruct the truth.
In both cases, we show two realizations of the maps (left and right), although these are representative: the Einasto, which is close in shape to the generalized NFW, exhibits good performance, whereas the Burkert performance is poor, with the missing flux mostly being attributed to our isotropic template.
}
\vspace{-0.5cm}
\label{fig:GCEVariations}
\end{figure*}

The main statements in this work relate to properties of the GCE.
It is therefore reasonable to ask how our conclusions vary if we were to change the underlying morphology of the excess.
Of course, before doing so we emphasize that extensive studies of the GCE have found that a generalized NFW with $\gamma \simeq 1.2$ provides the best fit to the excess, see e.g. Ref.~\cite{Linden:2016rcf}.
Especially as our findings suggest that the best fit $dN/dF$ is largely consistent with Poisson emission, varying away from this would worsen the overall quality of the fit.
Nonetheless, we can still ask the question of how such a mismodeling could bias our findings in simulated data.

This was already studied in Figs.~\ref{fig:mismodeling} and \ref{fig:mismodeling_hist}, where we showed that softening the inner peak of the GCE template to an NFW with $\gamma=1.0$ could lead to a degree of overconfidence and on average a reconstructed $dN/dF$ that was modestly brighter than what was injected.
We extend that study here by considering two considerably different spatial templates beyond the NFW, the Einasto~\cite{1965TrAlm...5...87E} and Burkert~\cite{Burkert:1995yz} profiles.
The functional form of the profiles is given by
\begin{equation}\begin{aligned}
\textrm{NFW:}\hspace{0.5cm}& \rho \propto \frac{1}{(r/r_s)^{\gamma}(1+r/r_s)^{3-\gamma}}, \\
\textrm{Einasto:}\hspace{0.5cm}& \rho \propto \exp \! \left[ - \frac{2}{\alpha} \left( \left( \frac{r}{r_s} \right)^{\alpha} -1 \right) \right]\!, \\
\textrm{Burkert:}\hspace{0.5cm}& \rho \propto \frac{1}{(1+r/r_B)(1+(r/r_B)^2)},
\end{aligned}\end{equation}
where we take $r_s = 20\,$kpc, $\alpha = 0.17$, and $r_B = 0.7 r_s$.

In Fig.~\ref{fig:GCEVariations} we show two realizations (each) for a scenario where the simulated GCE follows an Einasto or Burkert profile, but is reconstructed using our default CNN which only saw an NFW profile during training.
The examples highlight that from the perspective of the inferred SCD, the use of an Einasto has only very little impact on the inference procedure.
This is certainly not true when the truth is a Burkert however: the performance is poor.
Most notably the CNN fails to reconstruct the flux in the Burkert and we find that much of the flux is instead attributed to the isotropic template.
This is not overly surprising however.
The sharp rise in the flux towards the Galactic Center is a hallmark of the GCE that is not exhibited by the Burkert profile, which is instead characterized by its flat central core.
Thus, we take from this result that the use of a significantly incorrect GCE template can undoubtedly influence the results, but as the Burkert is a poor fit to the actual data this specific systematic is unlikely to be influencing our primary findings.

\subsection{The Impact of Energy in the Simulated Data}

The key technical result of the present work is the development of an approach that facilitates a simultaneous inference of $dN/dE$ and $dN/dF$ in $\gamma$-ray data analysis.
However, as emphasized at the outset of this section, energy was added to our approach in two ways.
Most significantly, we developed a CNN that allows ten input channels corresponding to energy bins, which allow for the network to make inferences in energy.
To train the network, however, we also needed to generate simulated data in energy bins, whereas previous studies of the point-source nature of the GCE have generally simulated data in a single energy bin.
The data generated in multiple energy bins is more accurate: the instrument response, in particular the PSF, varies across energies and this is accounted for in the energy dependent data.
We note that even in the energy independent cases, the underlying diffuse models are developed in multiple energy bins before being summed back together, e.g. to account for the energy dependent cosmic-ray propagation parameters; for a discussion related to Model O see Refs.~\cite{Buschmann:2020adf,Macias:2016nev,Macias:2019omb}.
Still, the fully energy independent CNN only sees a single spatial structure for all templates, whereas when the spectra are drawn randomly the combined map will exhibit spatial variations corresponding to the different weightings between bins.

\begin{figure*}[!t]
\centering
\includegraphics[width=0.45\linewidth]{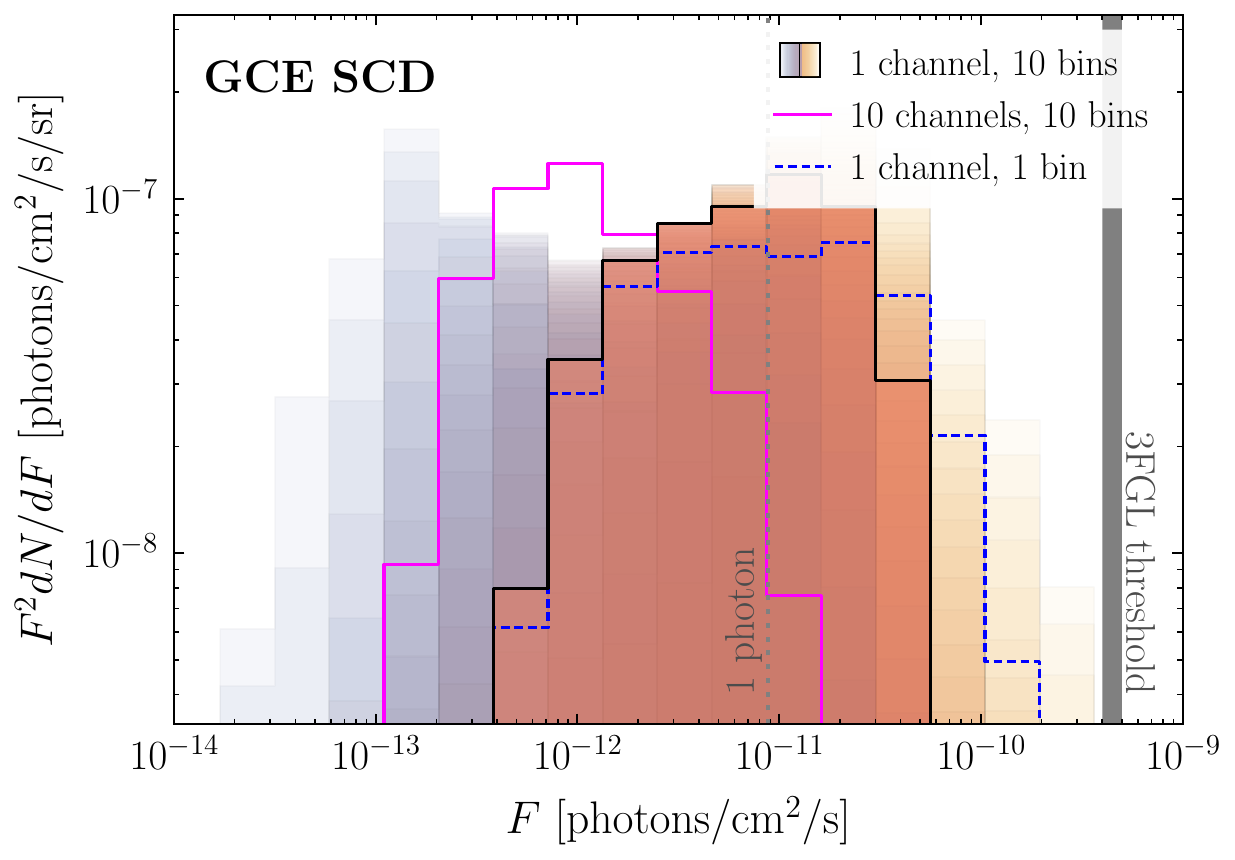}
\vspace{-0.2cm}
\caption{The inferred GCE SCD from three different networks.
The result labeled 10 channels, 10 bins corresponds to the median prediction from our default energy dependent CNN that has ten input channels and was trained on data generated in ten energy bins.
In dashed blue we show the median result for our default energy independent network, which had a single input channel and energy bin for data generation.
Using a more accurate data generation does not have a large impact; a CNN with a single input channel trained on data generated in ten energy bins gives broadly consistent results as shown.}
\vspace{-0.5cm}
\label{fig:eindep-fermi-prediction}
\end{figure*}

As already mentioned, the comparison performed in Fig.~\ref{fig:Results} was between our energy dependent CNN that used 10 input channels and data generated in 10 energy bins and an energy independent CNN with 1 input channel and data generated in 1 energy bin.
In Fig.~\ref{fig:eindep-fermi-prediction} we show the difference between those two cases and a case where the CNN has a single input channel and data generated in ten energy bins, which are then summed together.
As can be seen, the change in the number of input channels has a far larger impact than the change in the number of simulated energy bins; in other words, using the more accurate simulated data does not have a significant impact on the results.

\subsection{Stability In Time}

\begin{figure*}[!t]
\centering
\includegraphics[width=0.45\linewidth]{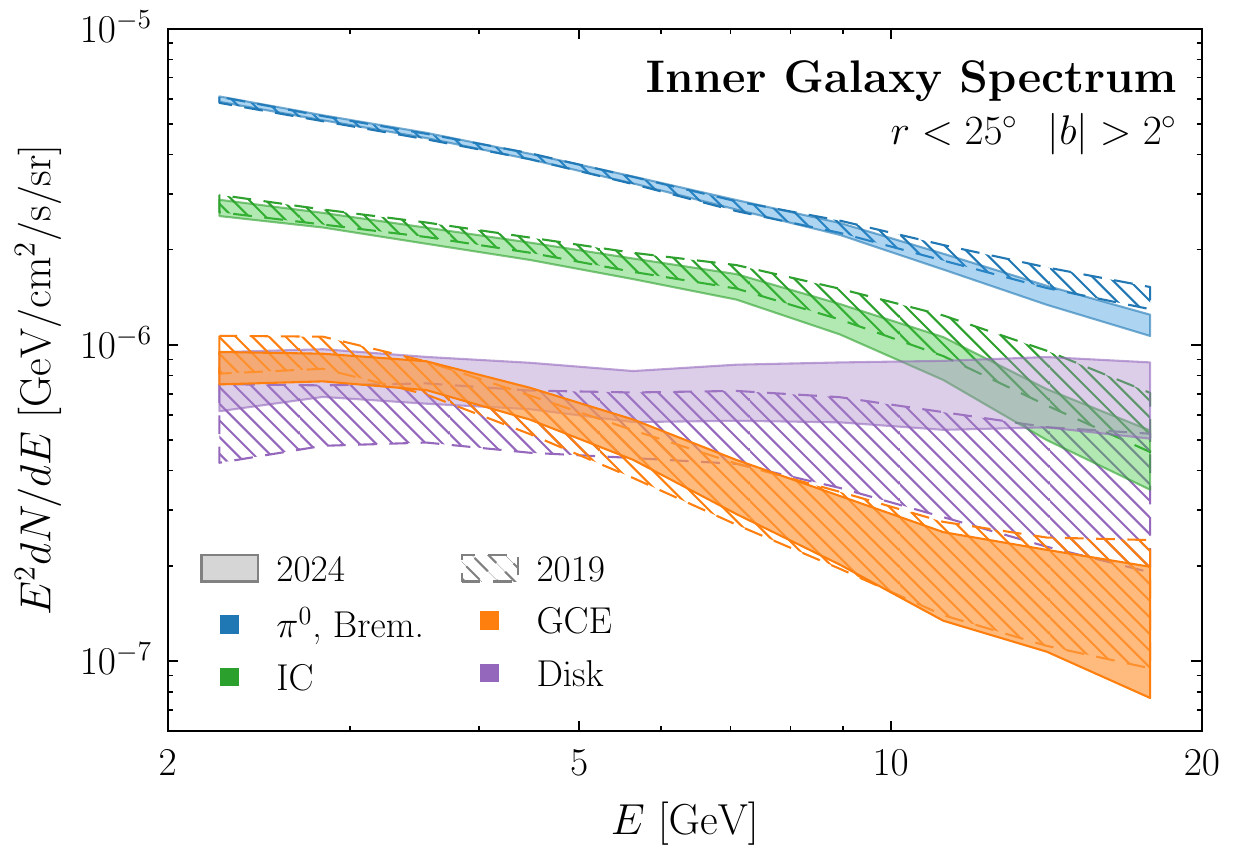}
\includegraphics[width=0.45\linewidth]{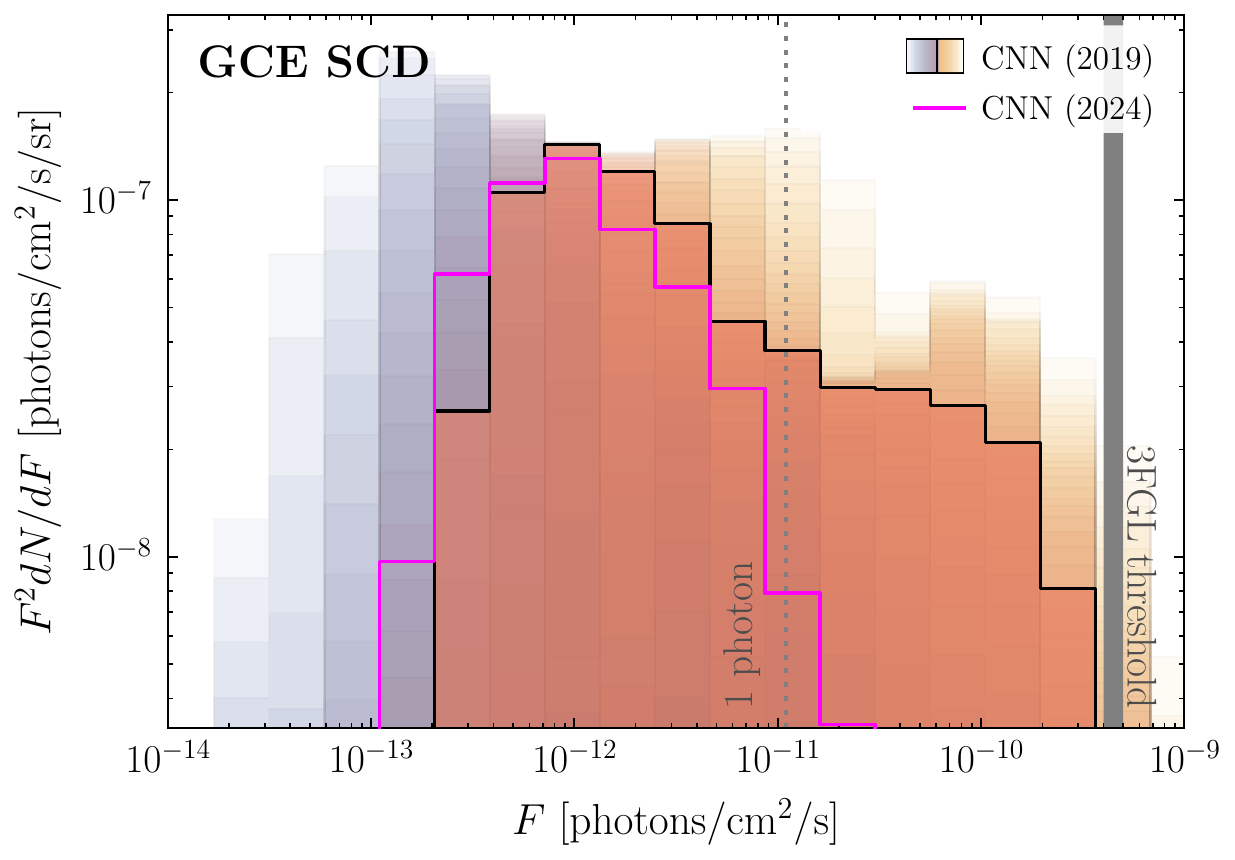}
\vspace{-0.2cm}
\caption{A comparison of our default CNN results to those obtained for a network trained on a subset of the data collected up to 2019 rather than 2024.
(Left) The primary spectra obtained from the Fermi data; $\pi^0$ and GCE spectra are highly consistent, whereas the disk and IC show noticeable deviations at high energies.
(Right) The 2019 GCE SCD as compared to the 2024 result.
The peak of the distribution remains well below the 1 photon line, although in the smaller dataset the CNN recovers a brighter second mode.
The SEMD between the two distributions is $\simeq 1.6$. }
\vspace{-0.5cm}
\label{fig:old-edep-results}
\end{figure*}

As a result of the systematic studies considered thus far we have observed several general trends.
Firstly, the returned spectra appear robust up to variations between the inverse Compton and disk templates.
Secondly, the GCE SCD is consistently returned as being peaked around or below the 1 photon threshold, indicating that the CNN prefers the excess to be largely consistent with Poisson emission, although a contribution from resolvable point-source cannot be robustly excluded.
As a final test of how our results vary with the underlying data and templates used to train the CNN, here we explore the stability of our conclusions with time.
In particular, we retrain our CNN on the same data used in Ref.~\cite{List:2021aer}, where 567 weeks of Fermi data was used, as collected by the instrument up to 19 June 2019.
All other selection criteria match those for our default dataset, so this simply represents a 30\% reduction in the available Fermi dataset.

A comparison of our key results between the 2019 (subset of the Fermi data) and 2024 (our full Fermi data) is provided in Fig.~\ref{fig:old-edep-results}.
(Results for the disk SCD are suppressed as there is strong consistency between CNN prediction in each dataset.)
Regarding the spectra, broad consistency is seen, especially for the GCE.
The disk exhibits the largest change, although the results are roughly consistent within the error bands.
Turning to the GCE SCD, the peak remains well below the 1 photon threshold.
In the subset of data, however, the CNN has a preference for a second mode with reduced amplitude of the GCE distribution above the 1 photon threshold; recall again that we are showing $F^2 dN/dF$ so that the distribution is skewed towards larger fluxes.
The SEMD between the two distributions is $\simeq 1.6$, suggesting the two are quite consistent given the level of SEMD calibration displayed by the default network, see Fig.~\ref{fig:sharpcalibration_hist}.
Nevertheless, the results do indicate that it would be premature to exclude any contribution from resolvable sources to the GCE.

\section{CNN Variations}
\label{sec:CNNvariations}

The last set of studies we perform relate to the variation of our results under changes in how we use and train the CNN.
We show the impact of three changes, all testing the impact of variation to the priors used to generate the simulated data, detailed in Sec.~\ref{ssec:priors}.
We consider: 1. adding jitters to the simulated spectra; 2. varying the brightness of the simulated sources; and 3. varying the modality of the simulated sources.
To briefly mention, we further investigated the use of instance~\cite{Ulyanov2016} rather than batch~\cite{Ioffe2015} normalization for the convolutional layers and whether there was a difference in passing the network counts per energy bin rather than total counts, which we pass by default.
The impact of these variations were indiscernible.
As in Ref. \cite{List:2021aer}, we use batch normalization, but our approach differs by providing counts per energy bin as input to the fully connected layers for all our energy dependent CNNs.

\subsection{Adding Jitters to the Simulated Spectra}

\begin{figure*}[!t]
\centering
\includegraphics[width=0.45\linewidth]{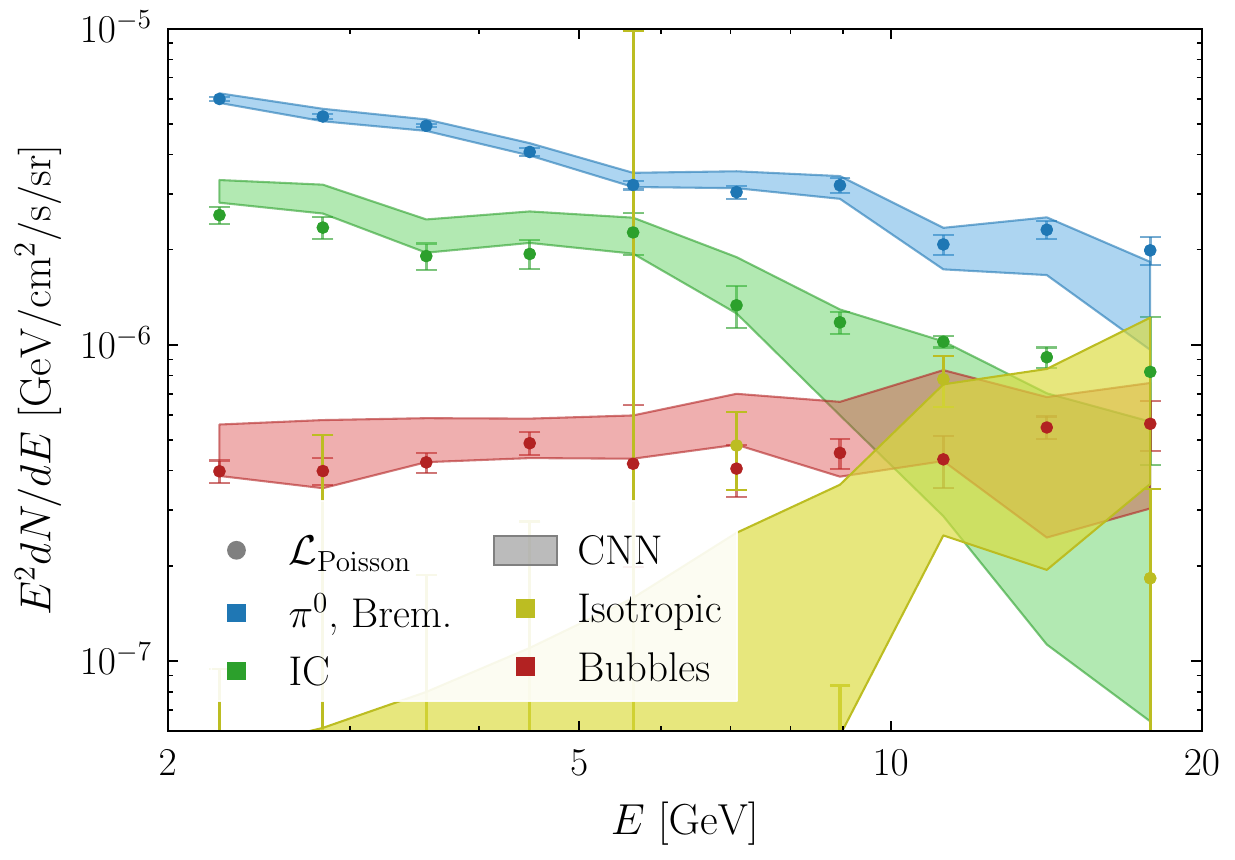} \hspace{0.5cm}
\includegraphics[width=0.45\linewidth]{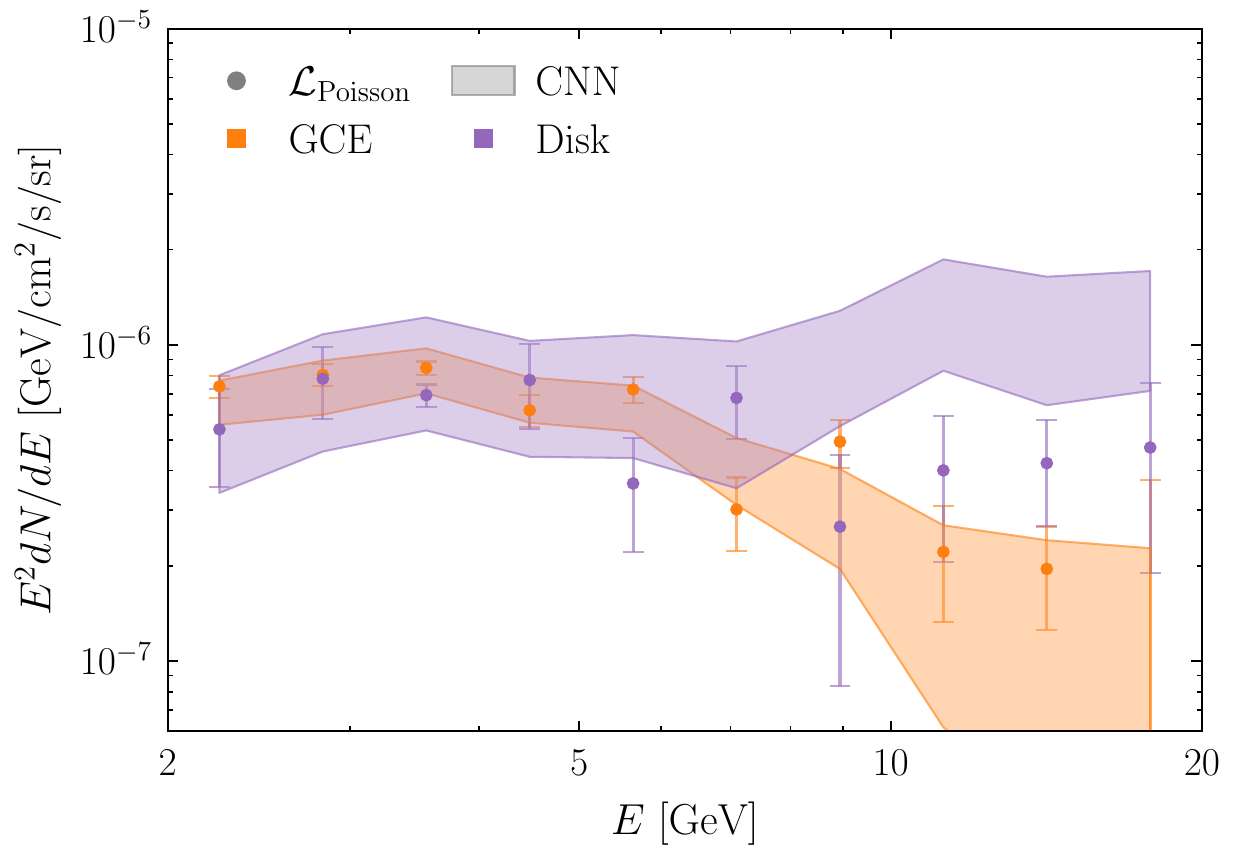}
\vspace{-0.2cm}
\caption{Spectral predictions as in Figs.~\ref{fig:Results} and \ref{fig:IsoBub}, but for a network that was trained with random jitters added into the spectrum.
The addition of the jitters leads to predictions that are far less smooth between bins than the default results, and in certain cases leads to improved agreement with the Poisson likelihood analyses.}
\vspace{-0.5cm}
\label{fig:spectral-jitters}
\end{figure*}

A noticeable aspect of the CNN inferred spectra shown throughout this work is their absence of bin-to-bin fluctuations.
In particular, it is clear in Fig.~\ref{fig:Results} that the Poisson likelihood results fluctuate between energy bins far more than the CNN, suggesting a degree of correlation across energies in the machine learning predictions.
This is largely unsurprising.
In the Poisson likelihood method each energy bin is analyzed independently implying the results are statistically independent.
The CNN however analyzes the data from all energy bins simultaneously in order to make its prediction and as it was trained on true spectra that are smoothed, that it returns a smooth prediction is to be expected.
(Note that the Fermi instrument response generates correlations across energy bins, although we neglect that effect in this work as it has been shown to be relatively small for the smooth spectra we study here~\cite{Linden:2016rcf}.)

To demonstrate that the smoothness originates largely from the training data, in this section we show that if we add jitters to the true spectra that the CNN is trained on it will return more jagged results.
In particular, the data generation procedure we use is almost identical, except once a random spectrum has been specified, we then add a perturbation on top of this: in each bin, if there is a flux prediction $\mu$, we redraw a value from ${\cal N}(\mu,\mu/2)$ to provide a new random value.
For the point-source templates, the total flux is distributed amongst weights in each bin and we instead fluctuate the weights.
(In either event, if this procedure leads to a negative value we simply repeat the process until a positive one is achieved.)
This procedure, albeit crude, allows the CNN to see spectra that are on average consistent with our default approach but with large bin-to-bin variations.

Once trained as above, the network applied to the real Fermi data returns the spectra shown in Fig.~\ref{fig:spectral-jitters}.
Compared to Figs.~\ref{fig:Results} and \ref{fig:IsoBub} the results are broadly consistent whilst far less smooth.
Focusing on the dominant emission component, the $\pi^0$ diffuse contribution, we see that there is now far greater agreement between the Poisson and CNN predictions; above $\simeq 6$\,GeV the CNN and Poisson methods both track each other's fluctuations, whereas in our default approach there was more apparent disagreement.

These results suggest that the correlations in the default CNN prediction do at least in part arise from the training data choice and suggest this as an interesting point for future studies.
For example, rather than a skew normal, one could in principle train the network on completely random values drawn in each bin and this would likely lead to results more consistent with the Poisson likelihood approach.
A drawback, however, is that a larger training sample may be required as in the real Fermi data the spectra, although displaying inter-bin variations, are broadly comparable across the energy range we study.

\subsection{Changing the Brightness of the Simulated Sources}

\begin{figure*}[!t]
\centering
\includegraphics[width=.45\linewidth]{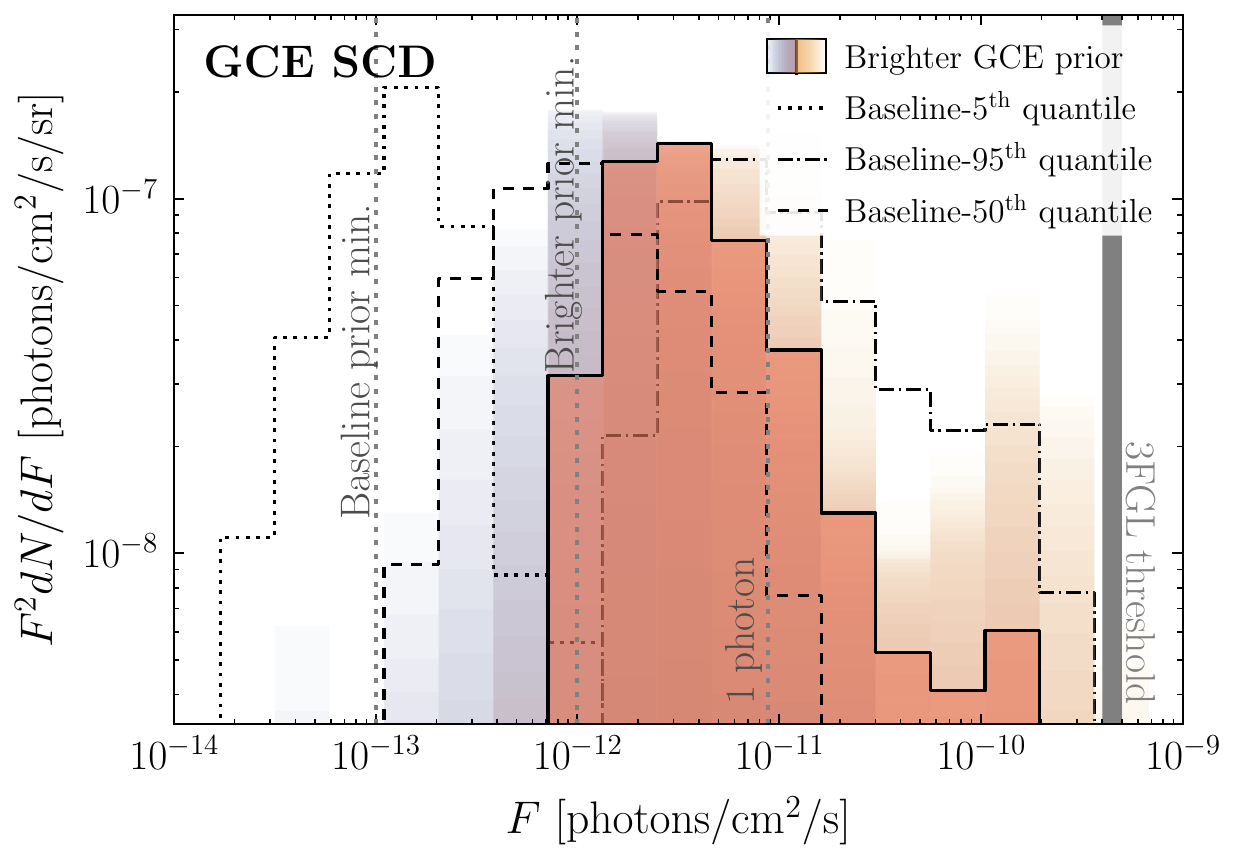}
\hspace{0.5cm}
\includegraphics[width=.45\linewidth]{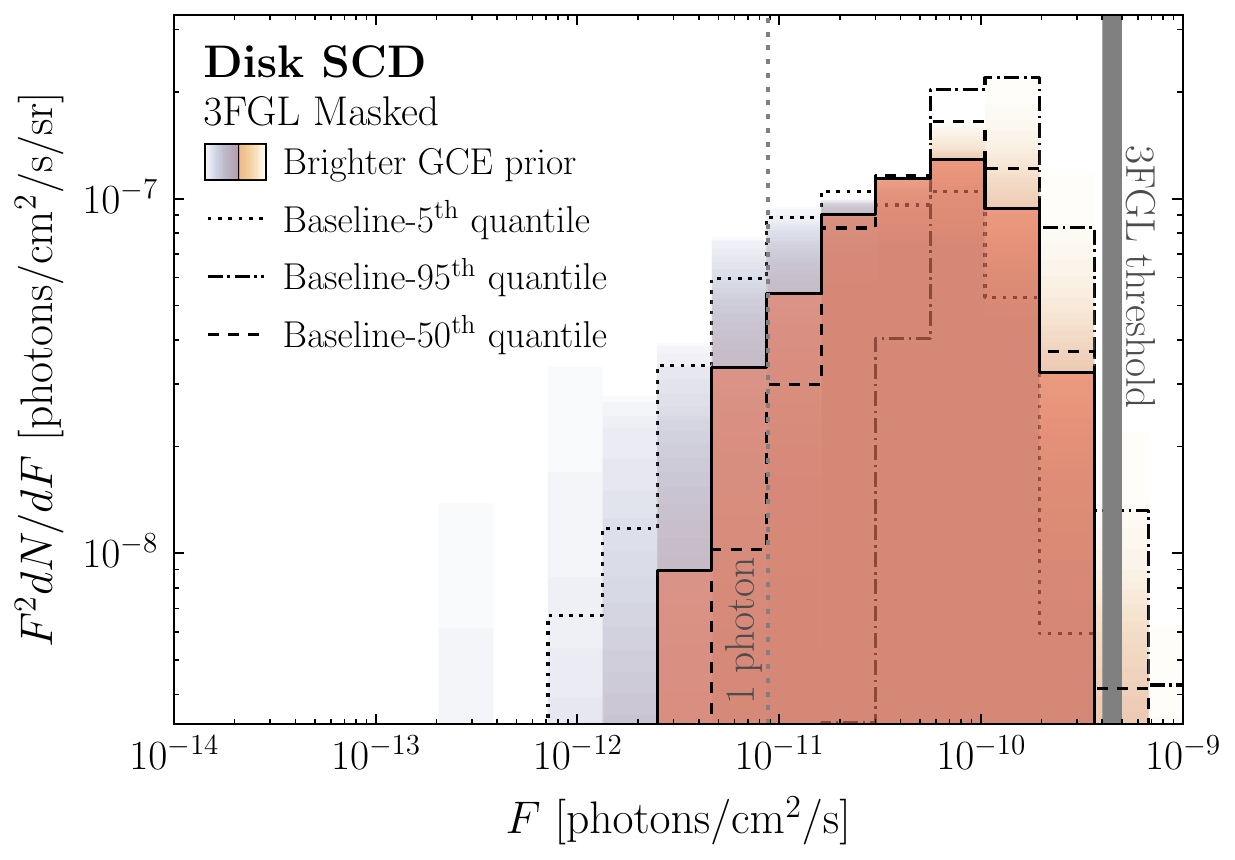}
\vspace{-0.2cm}
\caption{A comparison of how our results for the GCE and disk SCD are impacted by our choice of lower flux prior.
If we increase the GCE flux prior so that the mean now only goes as low as $\simeq 10^{-12}$\,photons/cm$^2$/s (cf. our default used in Figs.~\ref{fig:Results} and \ref{fig:SCD-Disk} which goes an order of magnitude dimmer), we see the GCE simply moves up until it is jutting against the prior, whereas the disk results are largely unchanged.
}
\label{fig:brighterGCE}
\end{figure*}

By default we simulate GCE sources down to a level where they are exceptionally dim, as outlined in Sec.~\ref{ssec:priors}.
In particular, the mean SCD can be as dim as $\simeq 10^{-13}\,$photons/cm$^2$/s.
This choice was partly motivated by the fact that originally in our analysis we only simulated sources down to a mean of $\simeq 10^{-12}\,$photons/cm$^2$/s (as used in Ref.~\cite{List:2021aer}); however, as we saw the results on the Fermi data were peaked down nearly as low as we simulated, we decided to simulate even dimmer sources to test the impact of this.

In Fig.~\ref{fig:brighterGCE} we show the impact of that change and we also plot our default conclusions from Figs.~\ref{fig:Results} and \ref{fig:SCD-Disk} for comparison.
The disk remains largely unchanged.
For the GCE the SCD increases in brightness, although it simply moves up to just above the new prior minimum; the peak remains below the 1 photon threshold.
Based on these results one could consider simulating dimmer sources still.
Nevertheless, by $10^{-13}$\,photons/cm$^2$/s we are considering sources that are only expected to generate 0.01 photons on average; they are exceptionally Poisson-like, and the probability that such a source emits $\geq 2$ detected counts is $5 \times 10^{-5}$.
More interesting would be to consider how to include in our CNN training the physical fact that Poisson emission becomes degenerate with point sources, rather than to simply simulate more indistinguishable maps.

\subsection{Varying the Modality of Simulated Sources}

\begin{figure*}[!t]
\centering
\includegraphics[width=.45\linewidth]{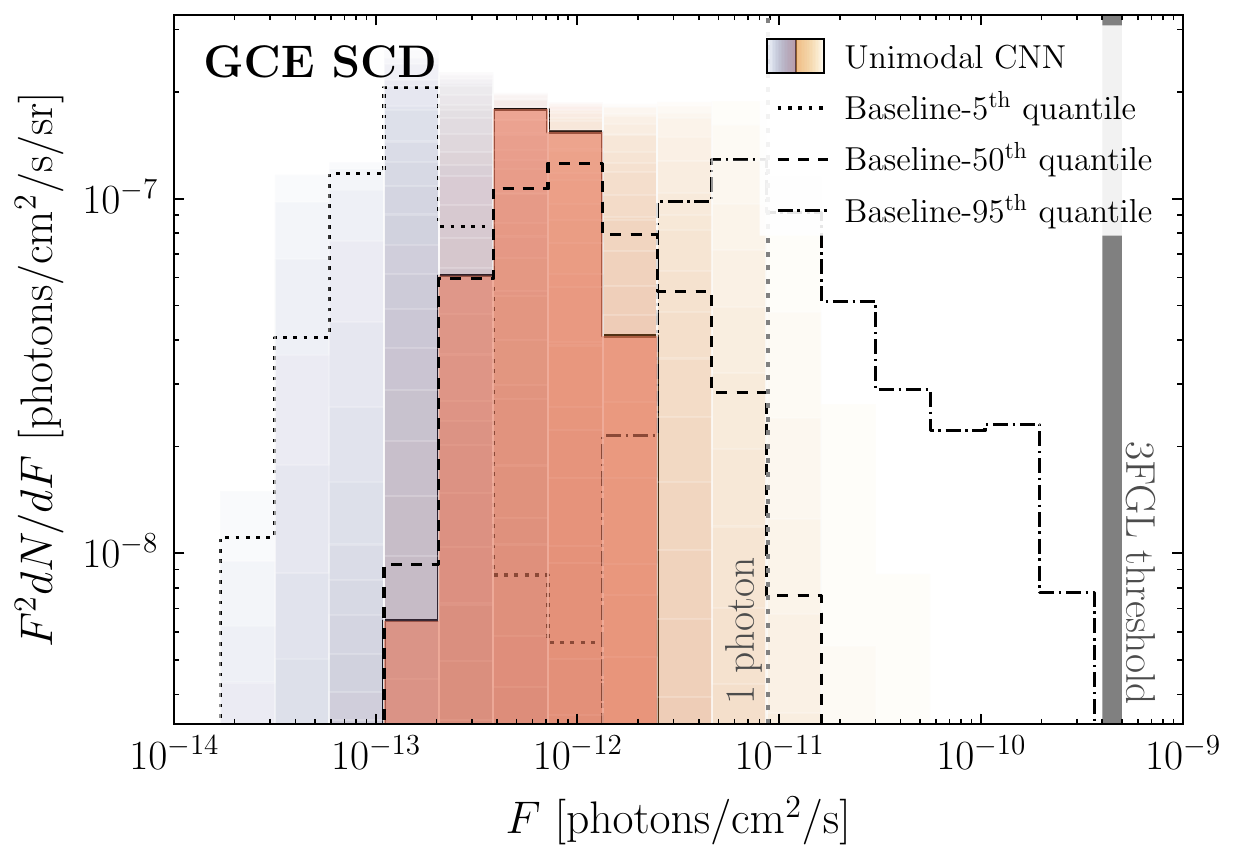}
\hspace{0.5cm}
\includegraphics[width=.45\linewidth]{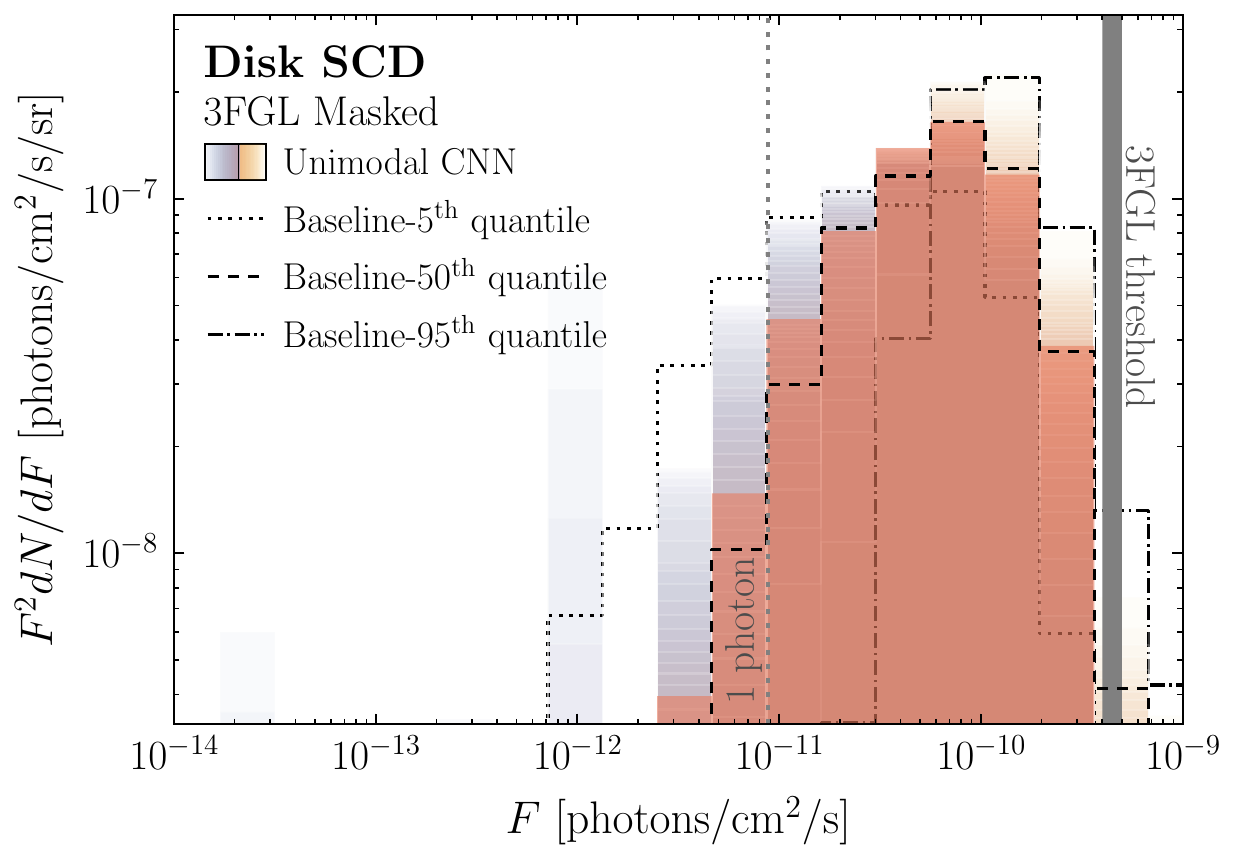}
\vspace{-0.2cm}
\caption{SCDs inferred when the CNN is trained on unimodal GCE SCDs only, as opposed to our default bimodal approach.
As expected, the CNN trained on unimodal SCDs makes narrower predictions.
}
\label{fig:unimodal}
\end{figure*}

As outlined in Sec.~\ref{ssec:priors}, by default the GCE is simulated from a bimodal SCD.
The underlying motivation is to test the hypothesis that the GCE could have a (potentially resolvable) point-source contribution as well as a Poisson-like component that could be attributed to DM.
As each mode is generated to have a flux that could be as low as zero, this also means that in training the CNN will have seen maps that are effectively unimodal too.
In this section we explore how much this bimodal choice impacted our findings (cf. Fig.~\ref{fig:old-edep-results} for a clear example where the CNN returns a bimodal result).
The results of the unimodal analysis are shown in Fig.~\ref{fig:unimodal}.
As can be seen, the unimodal distribution is thinner and has less support up to higher fluxes, but the distributions found in each case are broadly consistent.
The impact on the disk SCD is negligible.

\end{document}